\documentclass[12pt,aps,floats,showpacs,showkeys,preprintnumbers,nofootinbib]{revtex4}
\pdfoutput=1

\usepackage{amsmath,amssymb,bbm}
\usepackage{epsfig}
\usepackage{array}
\usepackage{comment}
\usepackage[pdftex,colorlinks=false,urlcolor=blue,linkcolor=blue]{hyperref}
\usepackage{epstopdf}
\newcommand{\PreserveBackslash}[1]{\let\temp=\\#1\let\\=\temp}
\newcolumntype{C}[1]{>{\PreserveBackslash\centering}p{#1}}
\newcolumntype{R}[1]{>{\PreserveBackslash\raggedleft}p{#1}}
\newcolumntype{L}[1]{>{\PreserveBackslash\raggedright}p{#1}}
\newcommand{\sgn}{\mathop{\mathrm{sgn}}}

\def\tanb{\tan\beta}
\def\cotb{\cot\beta}
\def\sinb{\sin\beta}
\def\cosb{\cos\beta}
\def\sina{\sin\alpha}
\def\cosa{\cos\alpha}

\def\sbma{\sin(\beta-\alpha)}
\def\cbma{\cos(\beta-\alpha)}
\def\gam{\gamma}
\def\lam{\lambda}

\def\mh{m_h}
\def\mH{m_H}
\def\mA{m_A}
\def\ha{A}
\def\mha{m_\ha}
\def\hpm{H^\pm}
\def\hp{H^+}
\def\hm{H^-}
\def\hl{h}
\def\hh{H}
\def\mhl{m_{\hl}}
\def\mhh{m_{\hh}}
\def\mhat{\hat m_{12}}
\def\wp{W^+}
\def\wm{W^-}
\def\wpm{W^\pm}

\def\mhpm{m_{\hpm}}
\def\mot{m_{12}}

\def\CU{C_U}
\def\CD{C_D}

\def\CV{C_V}
\def\CG{C_g}
\def\CP{C_\gamma}
\def\cu{\CU}
\def\cd{\CD}
\def\cv{\CV}
\def\cg{\CG}
\def\cp{\CP}

\def\gev{~{\rm GeV}}
\def\tev{~{\rm TeV}}
\def\fbi{~{\rm fb}^{-1}}
\def\pb{~{\rm pb}}
\def\fb{~{\rm fb}}
\def\br{{\rm BR}}
\def\anti{\overline}
\def\sig{\sigma}
\def\rts{\sqrt s}

\def\eg{{\it e.g.}}
\def\ie{{\it i.e.}}

\def\hsm{{H_{\rm SM}}}

\def\muggf{\mu_{\rm ggF+ttH}}
\def\muvbf{\mu_{\rm VBF+VH}}

\def\thdmc{{\tt 2HDMC}}

\def\beq{\begin{equation}}
\def\eeq{\end{equation}}
\def\bea{\begin{eqnarray}}
\def\eea{\end{eqnarray}}
\def\bit{\begin{itemize}}
\def\eit{\end{itemize}}
\def\ben{\begin{enumerate}}
\def\een{\end{enumerate}}
\def\rgghgamgam{\mu_{gg}^h(\gam\gam)}
\def\rvbfhgamgam{\mu_{\rm VBF}^h(\gam\gam)}

\def\rgghzz{\mu_{gg}^h(ZZ)}

\def\plhcfd{postLHC8-FDOK}

\def\lsim{\mathrel{\raise.3ex\hbox{$<$\kern-.75em\lower1ex\hbox{$\sim$}}}}
\def\gsim{\mathrel{\raise.3ex\hbox{$>$\kern-.75em\lower1ex\hbox{$\sim$}}}}

\begin{document}
\title{Constraints on and future prospects for Two-Higgs-Doublet Models in light of the LHC Higgs signal}

\vspace*{-1cm}
\begin{flushright}
LPSC14081\\[-1mm]
UCD-2014-03\\[5mm]
\end{flushright}

\author{B\'eranger Dumont$^{1}$}
\email[]{dumont@lpsc.in2p3.fr}
\author{John F.~Gunion$^2$}
\email[]{jfgunion@ucdavis.edu}
\author{Yun~Jiang$^2$}
\email[]{yunjiang@ucdavis.edu}
\author{Sabine Kraml$^{1}$}
\email[]{sabine.kraml@lpsc.in2p3.fr}

\affiliation{(1) \,Laboratoire de Physique Subatomique et de Cosmologie, Universit\'e Grenoble-Alpes,
CNRS/IN2P3, 53 Avenue des Martyrs, F-38026 Grenoble, France}
\affiliation{(2) \,Department of Physics, University of California, Davis, CA 95616, USA}

\begin{abstract}
We analyze the Two-Higgs-Doublet Models (2HDMs) of Type~I and Type~II for consistency with
the latest measurements of the $\sim 125.5\gev$ Higgs-like signal at the LHC. To this end, we perform scans of the 2HDM parameter space taking into account all relevant pre-LHC constraints as well as the most recent limits coming from searches for heavy Higgs-like states at the LHC. The current status of the 2HDMs of Type~I and Type~II is discussed assuming that the observed $125.5\gev$ state is one of the two $CP$-even Higgs bosons, either the lighter $h$ or the heavier $H$. Implications for future experiments, including expectations regarding other lighter or heavier Higgs bosons are given.  The possible importance of heavier Higgs bosons feeding the signals for the $125.5\gev$ state is also evaluated.
\end{abstract}


\maketitle

\section{Introduction}

Following the discovery by the ATLAS and CMS experiments at the LHC~\cite{Aad:2012tfa,Chatrchyan:2012ufa}  of a Higgs-like particle, additional measurements of its properties  using  the full data sets at $\sqrt{s} = 7$~and 8 TeV reveal that the observed state with a mass near $125.5\gev$ is quite Standard Model (SM)-like~\cite{ATLASnew,CMS:new,Aad:2013wqa}. It is thus clear that models with an extended Higgs sector will  be significantly constrained by the data. 

In particular, it is interesting to consider the
simplest such extensions of the SM, namely Two-Higgs-Doublet Models (2HDMs).
For comprehensive reviews see,
\eg,~\cite{Gunion:1989we,Gunion:2002zf,Branco:2011iw}.
These models have attracted a lot of attention recently.  A large number of
papers~\cite{Chiang:2013ixa,Grinstein:2013npa,Coleppa:2013dya,Eberhardt:2013uba,Chang:2013ona,Cheung:2013rva,Celis:2013ixa,Wang:2013sha,Baglio:2014nea,Inoue:2014nva}
performed fits of current data for the 125.5 GeV Higgs-like state (as
per the status post Moriond 2013) within the context of 2HDMs, and
investigated the consequent phenomenology of the other Higgs states present in the models. Among these papers, \cite{Grinstein:2013npa,Coleppa:2013dya,Chang:2013ona} consider 2HDMs with a conserved $Z_2$ symmetry,
\cite{Craig:2013hca,Barger:2013ofa,Celis:2013ixa,Wang:2013sha} focus on the case of an aligned 2HDM,
\cite{Cheung:2013rva,Inoue:2014nva} investigate the possibility of
$CP$-violation in the Higgs sector,
and \cite{Baglio:2014nea} concentrates on the question of the
triple-Higgs coupling.
The general conditions for the alignment limit, in which the lightest
$CP$-even Higgs boson of a 2HDM mimics the Standard Model Higgs, without
decoupling the other scalars were studied in \cite{Carena:2013ooa}.
Moreover, the prospects for future LHC running and/or for other future
colliders in view of the current data were investigated in
\cite{Chen:2013rba,Craig:2013hca,Barger:2013ofa,Kanemura:2014dea,Wang:2014lta}.
The possibility of $CP$ violation in 2HDMs also implies an important
link of such models to electroweak baryogenesis, a topic that was revisited
recently in
\cite{Dorsch:2013wja,Shu:2013uua,Ahmadvand:2013sna,Dorsch:2014qja}.


 The present work  goes beyond what was done in the above-referenced studies
of 2HDMs in that we provide a
very comprehensive and complete analysis of the status of the
$CP$-conserving 2HDMs of Type~I and Type~II, considering both the cases
where
the observed Higgs particle at the LHC is the lighter $CP$-even state $h$
or the heavier $CP$-even state $H$.~\footnote{Given that the observed state clearly has $ZZ,WW$ couplings that are not far from SM-like, it cannot be identified with the $\ha$ which has no $VV$ tree-level couplings. In this paper, we also do not consider cases in which the observed state is a mixture of two or more nearly degenerate 2HDM states.} (The possibility of the heavier $H$ being identified with the $125.5\gev$ state was also considered in~\cite{Coleppa:2013dya,Chang:2013ona,Wang:2013sha,Kanemura:2014dea,Wang:2014lta}.)
In particular, we employ all the
latest results for the signal strength measurements from LHC8, include
a consistent treatment of feed down (FD) from the production of heavier
Higgs states, and discuss the prospects for LHC14.

In scanning the 2HDM parameter space, we use the parameter set consisting of the physical Higgs masses, $\mh$, $\mH$, $\mhpm$ and $\mA$, the $Z_2$ soft-breaking parameter $m_{12}$, the $CP$-even Higgs mixing parameter $\alpha$ and the 
ratio of the two vacuum expectation values $\tanb=v_2/v_1$.  
We impose all relevant constraints from precision electroweak data, from stability, unitarity and perturbativity of the potential, as well as from $B$ physics and from the direct searches at LEP. For this we closely follow the approach of~\cite{Drozd:2012vf}. Points are retained only when these ``preLHC'' constraints are all satisfied.  
Once the preLHC constraints have been applied, we require that the rates for channels involving heavier Higgs bosons all lie below the existing 95\% confidence level (C.L.) limits coming from the LHC.  

The next step, and the most important one for this work, is to impose the restrictions on the 2HDM parameter space from the measured Higgs signal. The Higgs measurements are conventionally phrased in terms of ``signal strengths'', {\it i.e.}\ the ratios to the SM predictions for different production channels, $X$, and different decay modes, $Y$,
\beq
\mu_X(Y)={\sigma (X)\br(Y)\over \sigma_{\rm SM}(X) \br_{\rm SM}(Y)}\,,
\eeq 
where the numerator and denominator are evaluated for the same Higgs mass. 
The production modes for the Higgs boson considered are ggF (gluon fusion, also denoted as $gg$), 
VBF (vector boson fusion), ttH (associated production with $t\bar t$) and 
VH (associated production with a vector boson). The relevant decay modes are those    
into $Y=\gam\gam, VV, b\anti b, \rm{and~}\tau\tau$ (where $VV \equiv ZZ, WW$).  
In practice, we employ the signal strength likelihoods in the $\muggf$ versus $\muvbf$ planes for each of the final states $Y$ as determined in \cite{Belanger:2013xza}, 
combining all publicly available information from ATLAS and CMS.\footnote{Combining VBF and VH is motivated 
in models where the couplings of the Higgs to $WW$ and to $ZZ$ are scaled equally, as is the case  in any 2HDM because of custodial symmetry. Combining ggF and ttH is more a matter of convenience, partially motivated by the fact that the current LHC measurements do not probe ggF and ttH in any given final state at the same time: $H \to b\bar{b}$ is probed via ttH, not ggF, whereas all the other final states are probed quite precisely via ggF  and with much poorer precision via ttH.} 
The  $[\muggf,\muvbf]$ approach has been systematically adopted by the experimental collaborations. It has the advantage of taking into account 
correlations not accounted for when individual $X\to H \to Y$ channels are treated separately.
We will require that the Higgs rates for all channels fall within  the 95\%~C.L. regions in the $[\muggf,\muvbf]$ plane.  
Points which satisfy the preLHC constraints, the heavy Higgs limits and the Higgs fitting constraints  will be labelled as ``postLHC8" points.

There is a further issue arising from the fact that there are various
ways in which the $125.5\gev$ Higgs boson can be produced as a result of feed down from the production of heavier states -- in the case of $\mh=125.5\gev$ this includes the $H,A,\hpm$~\cite{Arhrib:2013oia} while if $\mH=125.5\gev$ only the $\ha$ and $\hpm$ can feed the $125.5\gev$ signal.  If such FD processes occur at a significant rate, the fit to the Higgs measurements using only direct $h$ or $H$ production processes may no longer be valid. For most of our plots, we will show only those postLHC8 points for which the production rate from FD  will not distort the fits to the $125.5\gev$ resonance.  Such points are called ``FDOK.'' The detailed feed down limits employed for a point to be FDOK will be given later. 
These FD processes may be tested by a variety of means.  For example, for the important FD sources of  $A\to Zh$ and $H\to hh$, the final state mass can be reconstructed and $\mA$ and $\mH$ will be determined should the rates be significant; see the current limits from CMS in \cite{HIG13025}. Then, data points lying within the relevant mass windows can be separated off. For the $H\to hh$ case, the decay products from the second $h$ are visible for most $h$ decays and events with this extra final state ``activity" can be separated off.  More complicated FD chains will have even more extra particles and constraints that will allow their separation.  Feed down must also be considered in the $\mH=125.5\gev$ case --- the process $gg\to A\to ZH$ adds events to the $ZH$ final state beyond those from $Z^*\to ZH$; we will comment on this possibility in Section \ref{hhsection}.
 
The rest of the paper is organized as follows. 
In Section~\ref{proceduredetails} we give details on the scan ranges and the constraints incorporated. 
The case of the lighter $CP$-even Higgs $h$ being the observed state near $125.5\gev$ is discussed in 
Section~\ref{hlsection}, and the case of the heavier  $CP$-even Higgs $H$ in Section~\ref{hhsection}. 
Section~\ref{conclusions} contains our conclusions. 
Details on the feed down of a heavier Higgs to the $125.5\gev$ state are discussed in Appendix~\ref{app:feeddown}. 
Appendix~\ref{nondecoup} gives details regarding the nondecoupling charged-Higgs contributions to the 
Higgs-$\gam\gam$ coupling and the relationship to wrong-sign Higgs Yukawa couplings for down-type quarks.

\section{ Procedural Details }
\label{proceduredetails}
In this section we provide some details regarding the parameter scans, the fitting of the signal strengths, 
and the incorporation of limits related to the Higgs bosons that are heavier than the $125.5\gev$ state.

\subsection{Scan ranges and procedures} \label{sec:scan}

As in~\cite{Drozd:2012vf},
we employ a modified version of the code \thdmc~\cite{Eriksson:2009ws,Eriksson:2010zzb} for our numerical calculations. 
All relevant contributions to loop-induced processes are taken into account, in particular those with heavy quarks ($t$ and $b$), $W^\pm$ and $H^\pm$.
A number of different input sets can be used in the \thdmc\ context.  We have chosen to use the ``physical basis" in which the inputs are the physical Higgs masses ($\mh,\mH,\mA,\mhpm$), the  
vacuum expectation value ratio ($\tanb$), and the $CP$-even Higgs mixing angle, $\alpha$, supplemented by $\mot^2$.  The additional parameters  $\lam_6$ and $\lam_7$ are assumed to be zero as a result of a $Z_2$ symmetry being imposed on the dimension-4 operators
under which $H_1\to H_1$ and $H_2\to -H_2$. $\mot^2\neq0$ is allowed as a ``soft'' breaking of the $Z_2$ symmetry.  With the above inputs, $\lam_{1,2,3,4,5}$ as well as $m_{11}^2$ and $m_{22}^2$ are determined (the latter two via the minimization conditions for a  minimum of the vacuum)~\cite{Gunion:2002zf}. 
We scan over the following ranges:\footnote{The upper and lower bounds on $\tanb$ are chosen to ensure that the bottom and top Yukawa couplings, respectively, lie within the perturbative region. Unlike the $Z_2$ symmetric 2HDM which constrains $\tanb \lesssim 7$~\cite{Chen:2013rba}, high $\tanb$ values are allowed when the $Z_2$ symmetry is softly broken. A safe upper limit, as adopted here, is $\tanb\leq 60$.}
\bea
 & \alpha\in[-\pi/2,+\pi/2]\,, \quad \tanb\in[0.5,60]\,, \quad m_{12}^2\in[-(2\tev)^2,(2\tev)^2]\,, & \cr
 & \mA\in[5\gev,2\tev]\,, \quad \mhpm\in[m^*,2\tev] \,, & 
\eea
where $m^*$ is the lowest value of $\mhpm$ allowed by LEP direct production limits and $B$ physics constraints. The LEP limits on the $H^\pm$ are satisfied by requiring $\mhpm\geq 90\gev$. The lower bounds from $B$ physics are shown as a function of $\tanb$ in Fig.~15 of \cite{Branco:2011iw} in the case of the Type~II model (roughly $m^*\sim 300\gev$ in this case) and in Fig.~18 of \cite{Branco:2011iw} in the case of the Type~I model.

\begin{table}[t]
\begin{center}
\begin{tabular}{|c|c|c|c|c|c|}
\hline
\ & Type~I and Type~II  & \multicolumn{2}{c|}  {Type~I} & \multicolumn{2}{c|}{Type~II} \cr
\hline
Higgs & $C_V$ & $C_U$  & $C_D$ & $C_U$ & $C_D$  \cr
\hline
 $h$ & $\sin(\beta-\alpha)$ & $\cosa/ \sinb$ & $\cosa/ \sinb$  &  $\cosa/\sinb$ & $-{\sina/\cosb}$   \cr
\hline
 $H$ & $\cos(\beta-\alpha)$ & $\sina/ \sinb$ &  $\sina/ \sinb$ &  $\sina/ \sinb$ & $\cosa/\cosb$ \cr
\hline
 $A$ & 0 & $\cotb$ & $-\cotb$ & $\cotb$  & $\tanb$ \cr
\hline 
\end{tabular}
\end{center}
\vspace{-.15in}
\caption{Tree-level vector boson couplings $C_V$ ($V=W,Z$) and fermionic couplings $C_{F}$ ($F=U,D$)
normalized to their SM values for the Type~I and Type~II 2HDMs. }
\label{tab:couplings}
\end{table}

The couplings, normalized to their SM values, of the Higgs bosons to vector bosons ($C_V$) 
and to up- and down-type fermions ($C_U$ and $C_D$) are functions of $\alpha$ and $\beta$ 
as given in Table~\ref{tab:couplings}; see {\it e.g.}~\cite{Gunion:1989we} for details. The Type~I and Type~II models are distinguished only by the pattern of their fermionic couplings. 
We note that the range of $\alpha$ employed guarantees that the top quark Yukawa coupling is always positive in our convention.  For the most part, in particular for the case of $\mh\sim 125.5\gev$, one also finds that $\sin(\beta-\alpha)>0$.

For the remaining physical Higgs masses, we consider  
\beq
  \mh\in[123\gev,128\gev]\,,\quad \mH\in\ ]128\gev,2\tev] \,,
\eeq  
for the case that $h$ is the observed state near 125.5~GeV, or 
\beq
   \mH\in[123\gev,128\gev]\,, \quad \mh\in[10\gev,123\gev[ \,,
\eeq   
for the case that $H$ is the observed state near 125.5~GeV. 
The window of $125.5\pm 2.5\gev$ is adopted to account for theoretical uncertainties.
However, we do not consider the cases where the $A$ and/or the other $CP$-even Higgs are close
to $125.5\gev$ and possibly contribute to the observed signal.
Thus, for $\mh\sim 125.5\gev$ ($\mH\sim 125.5\gev$)  we require that $\mH$ ($\mh$) and $\mA$ not be within the   $[123,128]\gev$ window nor within a $\mh\pm4\gev$  ($\mH\pm4\gev$) mass window where $\mh$ ($\mH$) is the particular mass value generated within the $[123,128]\gev$ range.

We should note that our scans were performed in a manner that provides adequate point density in all regions of the various plots and predictions of interest.  This means that, besides very broad (usually flat) scans in the input parameters mentioned above, we also used scans designed to focus on ``hard-to-reach" regions of parameter space of particular interest and importance.

\subsection{Limits imposed by nonobservation of Higgs bosons other than the ${\bf 125.5\gev}$ state} \label{sec:HAlimits}

As noted in the Introduction, it is necessary to take into account LHC exclusion limits for Higgs bosons that are heavier than $125.5\gev$. 
 In the case that the $\hl$ is identified with the $\sim 125.5\gev$ state the relevant channels are $gg\to H\to 4\ell,2\ell2\nu$,
$gg\to H,A\to \tau\tau$ and $gg\to b\anti b H,b\anti b A\to b\anti b\tau\tau$ at the LHC.  When these limits have been applied the points will be denoted by the phrase  ``$H/A$ limits." In the case that $\mhh\sim 125.5\gev$ is assumed, the only LHC limits that apply are those on $gg\to A\to\tau\tau$ and $gg\to b\anti b A\to b\anti b\tau\tau$.  Points that remain acceptable are denoted by ``$A$ limits".  Direct search limits on the $\hpm$ at the LHC do not impact the parameter space once the $B$ physics limits described below are imposed. In addition, the impact of the CMS search for $H \to hh$ and $A \to Zh$~\cite{HIG13025} and its relation with FD will be discussed in Sections III B and C.

For $H/A \to \tau\tau$, we employ the recent CMS limits based on the 8 TeV data~\cite{htautau},
which are presented separately for the bbH ($b\bar{b}$ associated production of the Higgs) and ggF production modes. In taking these limits into account, it will be important to note that in Type~II models the coupling of the $H$ and $A$ to down-type fermions can be dramatically enhanced at large $\tan \beta$ compared to the SM expectation  and that this enhancement will influence both bbH and ggF production.  When $\mha$ and $\mhh$ are within 15\% of one another, which is the approximate resolution in the invariant mass of a pair of $\tau$ leptons, we will add their  signals together.

Turning to $H \to ZZ$, we employ the latest ATLAS and CMS searches for heavy Higgs-like states in the $H \to ZZ \to 4\ell$ channel~\cite{ATLAS:2013nma,CMS-PAS-HIG-13-002}~\footnote{Note that the CMS results on $H \to ZZ$ are presented after combination of the $4\ell$ (where $\ell = e,\mu$) and $2\ell2\tau$ channels. However, the limit is almost uniquely driven by the $4\ell$ channels as can be seen from the result based on $H \to ZZ \to 2\ell2\tau$ only, which is available as supplementary material on the TWiki page~\cite{cms2l2tau}.}
and the CMS search in the $H \to ZZ \to 2\ell2\nu$ channel~\cite{CMS-PAS-HIG-13-014}. In the context of 2HDMs, there are two important considerations associated with using the limits as presented by the ATLAS and CMS collaborations.
First, the $VV$ couplings shown in Table~\ref{tab:couplings} imply $(C^h_V)^2+(C^H_V)^2=1$ for the  coupling strengths relative to the SM Higgs. Thus, if the $h$ is the $125.5\gev$ state, the $H$ will have a small coupling to $W,Z$ due to the fact that the $h$ must be very SM-like in order to describe the data at $\sim125.5$~GeV, as shown, \eg, in Fig.~10 of \cite{Belanger:2013xza}. Thus, only ggF production is relevant for $H$. However, only ATLAS presents constraints on a high mass Higgs arising purely from the ggF initial state with the full statistics at $7+8$~TeV, for $m_H > 200$~GeV~\cite{ATLAS:2013nma}. 
All the other results, \ie\ ({\it i})~ATLAS $H \to 4\ell$ for $m_H \in [130,180]$~GeV, ({\it ii})\ CMS $H \to 4\ell$ and ({\it iii})\ CMS $H \to 2\ell2\nu$ are implemented in our analysis under the assumption that the experimental search is fully inclusive. The limit is thus rescaled by a factor $\sigma^{\rm tot}_{\hsm}/\sigma_{gg \to \hsm}$.
Second, the width of the $H$ in the 2HDM 
can be much smaller than the large SM Higgs widths assumed in the ATLAS and CMS analyses. We correct for the width difference by rescaling the observed limits on $\sigma\times \br$ by the factor $f=\sqrt{\Gamma_H^2+(4\gev)^2\over \Gamma_\hsm^2+(4\gev)^2}$, where $4\gev$ is the experimental resolution in the $4\ell$ final state~\cite{korytov:privcom}.

\subsection{Constraints from the signal strength measurements at 125.5 GeV}
\label{sec:signalstrength}

For each scan point that passes the constraints explained above, we compute the 
predictions for $\mu_{\rm ggF+ttH}(Y)$ and $\mu_{\rm VBF+VH}(Y)$ for the main decay modes $Y$ ($\gam\gam, VV, b\anti b, \rm{and~}\tau\tau$) in terms of the reduced 
couplings $\cu$, $\cd$, and $\cv$, see Table~\ref{tab:couplings}.   
To this end, we also need the loop-induced $\gamma\gamma$ and $gg$ couplings of the Higgs boson with mass around 125.5~GeV; for these we employ the full 1-loop amplitudes in \thdmc\ (including the contribution from the charged Higgs bosons in the $\gam\gam$ case), where SM contributions are scaled according to the values of $\cu$, $\cd$ and $\cv$. 

To combine the information
provided by ATLAS, CMS and the Tevatron experiments on the 
$\gam\gam$, $ZZ^{(*)}$, $WW^{(*)}$, $b\bar{b}$ and $\tau\tau$  final states 
including the error correlations
among the (VBF+VH) and (ggF+ttH) production modes, 
we follow the approach of \cite{Belanger:2013xza}. 
Concretely, we fit the likelihood from the 68\%~C.L. contour provided by the experiments  for each decay mode $Y$  
in the $\mu_{\rm ggF+ttH}(Y)$ versus $\mu_{\rm VBF+VH}(Y)$ plane, using a Gaussian approximation. 
For each experiment, $- 2 \log L_Y = \chi_Y^2$ can then be expressed as
\begin{align} \label{eq:bestlikegaussian}
\chi^2_Y &= (\boldsymbol{\mu}_Y - \hat{\boldsymbol{\mu}}_Y)^T
\begin{pmatrix} \sigma_{{\rm ggF},Y}^2 & \rho_Y \sigma_{{\rm ggF},Y} \sigma_{{\rm VBF},Y} \\ \rho_Y \sigma_{{\rm ggF},Y}\sigma_{{\rm VBF},Y} & \sigma_{{\rm VBF},Y}^2 \end{pmatrix}^{-1}
(\boldsymbol{\mu}_Y - \hat{\boldsymbol{\mu}}_Y) \\
&= (\boldsymbol{\mu}_Y - \hat{\boldsymbol{\mu}}_Y)^T
\begin{pmatrix} a_Y & b_Y \\ b_Y & c_Y \end{pmatrix}
(\boldsymbol{\mu}_Y - \hat{\boldsymbol{\mu}}_Y) \nonumber \\
& = a_Y(\mu_{{\rm ggF},Y}-\hat{\mu}_{{\rm ggF},Y})^2
+2b_Y(\mu_{{\rm ggF},Y}-\hat{\mu}_{{\rm ggF},Y})
(\mu_{{\rm VBF},Y}-\hat{\mu}_{{\rm VBF},Y})
+c_Y(\mu_{{\rm VBF},Y}-\hat{\mu}_{{\rm VBF},Y})^2 \,, \nonumber 
\end{align}
where the indices ggF and VBF stand for (ggF+ttH) and (VBF+VH), respectively, 
and $\hat{\mu}_{{\rm ggF},Y}$ and $\hat{\mu}_{{\rm VBF},Y}$ denote the
best-fit points obtained from the measurements~\cite{Belanger:2013xza}.
The two-dimensional (2D) covariance matrix is explicitly shown in the first line of Eq.~\eqref{eq:bestlikegaussian}, with $\rho_Y$ corresponding to the correlation between the measurement of (ggF+ttH) and (VBF+VH).
From a digitized version of the 68\% C.L. contour, it is possible to fit simultaneously the parameters $a_Y$, $b_Y$, $c_Y$, $\hat{\mu}_{{\rm ggF},Y}$ and $\hat{\mu}_{{\rm VBF},Y}$. A combination of ATLAS and CMS can then be made for each decay mode $Y$ and expressed again in terms of $a_Y$, $b_Y$, $c_Y$, $\hat{\mu}_{{\rm ggF},Y}$ and $\hat{\mu}_{{\rm VBF},Y}$.

Adding up the individual $\chi_Y^2$, we thus obtain a ``combined likelihood,'' which can be used 
in a simple, generic way to 
constrain nonstandard Higgs sectors and new contributions to the loop-induced processes, provided they have the same Lagrangian structure as the SM. 
In this paper, for each scan point that passes the constraints of Sections~\ref{sec:scan} 
and \ref{sec:HAlimits}, we demand that each $\chi^2_Y$ (for ATLAS and CMS combined)
 be smaller than $6.18$, which corresponds to 95\% C.L. in two dimensions. These points are labelled as  ``postLHC8.''

\clearpage
\section{\bf \boldmath $\mh\sim 125.5\gev$ scenarios}
\label{hlsection}

In this section we focus on the case that the observed $\sim 125.5\gev$ state is the $\hl$.  
 
\subsection{Current constraints}

To cover the case $m_h=125.5\pm2.5$~GeV, we scan over  $\mhh$, $\mha$ and $\mhpm$ as discussed 
earlier in section~\ref{proceduredetails}.
As regards Yukawa couplings, the scan range of $|\alpha|\leq \pi/2$ implies that $\cu^h=\cd^h>0$ for Type~I, whereas for Type~II $\cd^h<0$ is possible when $\sina>0$.  Note that with this scanning range for $\alpha$, $\sbma<0$ is in principle possible, but is excluded by LHC data in the case of $\mh=125.5\gev$.  Thus, for both Type~I and Type~II models, $\cv^h$ and $\cu^h$ are always positive.

\begin{figure}[tb]
\begin{center}
\includegraphics[width=0.58\textwidth]{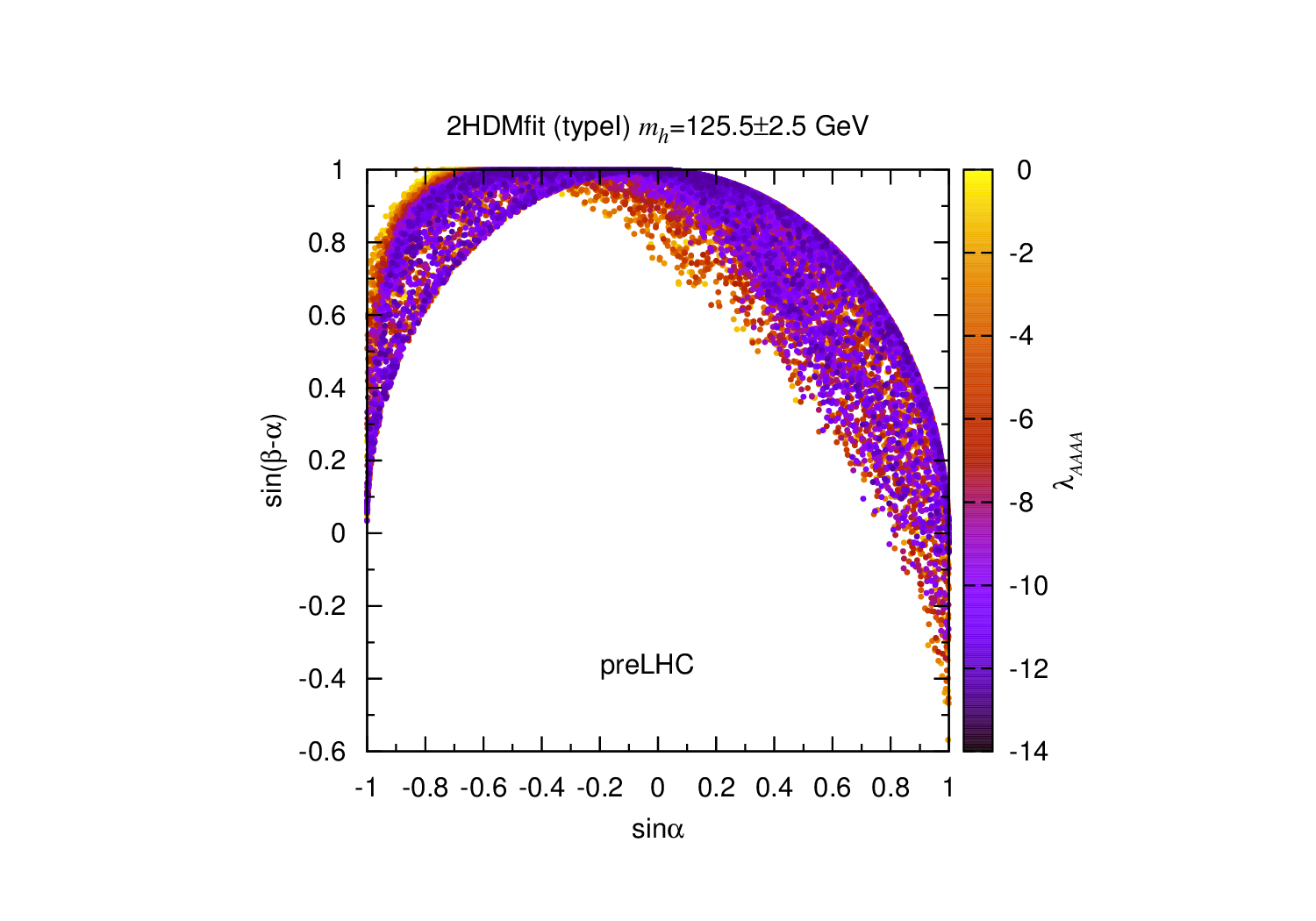}\hspace{-28mm}\includegraphics[width=0.58\textwidth]{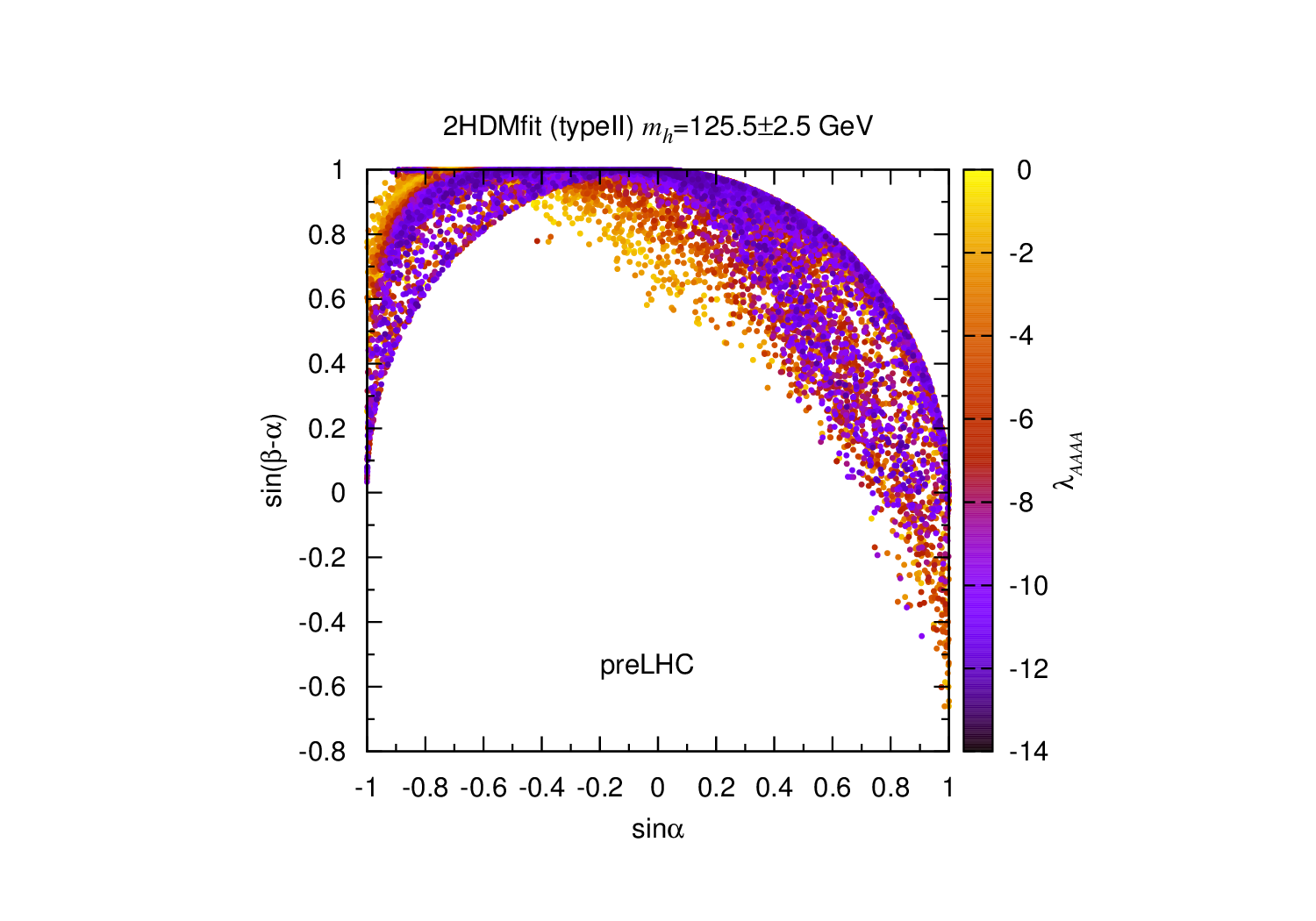}
\end{center}\vspace*{-10mm}
\caption{Values for the quartic coupling $\lam_{AAAA}$ for  2HDMs of Type~I (left) and Type~II (right) in the $\sbma$ versus $\sina$ plane for $\mh\sim 125.5\gev$ shown as ``temperature plots''. For all points, the full set of preLHC constraints is satisfied, including all quartic couplings having absolute values below $4\pi$.
}
\label{laaaa}
\end{figure}

As regards the possibilities for $\sina$, which determines the sign of $\cd^h$ in Type~II, there are important constraints from perturbativity of the quartic Higgs couplings (even in the case of the Type~I model). Often the strongest constraint is associated with the $\lam_{AAAA}$ quartic coupling of four $A$ Higgs bosons.  Figure~\ref{laaaa} shows the values of $\lam_{AAAA}$ that arise after the preLHC conditions listed in the Introduction are satisfied. From the figure we see that perturbativity of this quartic coupling creates a boundary of maximal $\sbma$ values for $\sina>0$ and of minimal $\sbma$ values for $\sina\lsim -0.3$.  The other boundaries  arise as a result of constraining other quartic couplings to their perturbative domain.  Note in particular that in both Type~I and Type~II the maximal value of $\sbma$ decreases as $\sina$ increases starting from $\sina\sim 0$, whereas $\sbma\sim 1$ is possible for a broad range of $\sina<0$ values.  This will impact many phenomenological results.  In particular, even though the $gg\to h$ fusion production rate is insensitive (not very sensitive) to the sign of $\sina$ for Type~I (Type~II), the $\gam\gam$ partial width is -- as $\sbma$ declines, the $W$-loop contribution to the $h\gam\gam$ coupling decreases, resulting in a decrease in $\br(\hl\to\gam\gam)$. Hence, the rate for $gg\to \hl\to\gam\gam$ quickly falls below the level acceptable for LHC precision Higgs results. This is also the case for $H \to WW^*$ originating from VBF or VH: while probed with poorer precision, the $\sbma$ factor associated with the $HVV$ vertex is present in both production and decay.

\begin{figure}[tb]
\begin{center}
\includegraphics[width=0.49\textwidth]{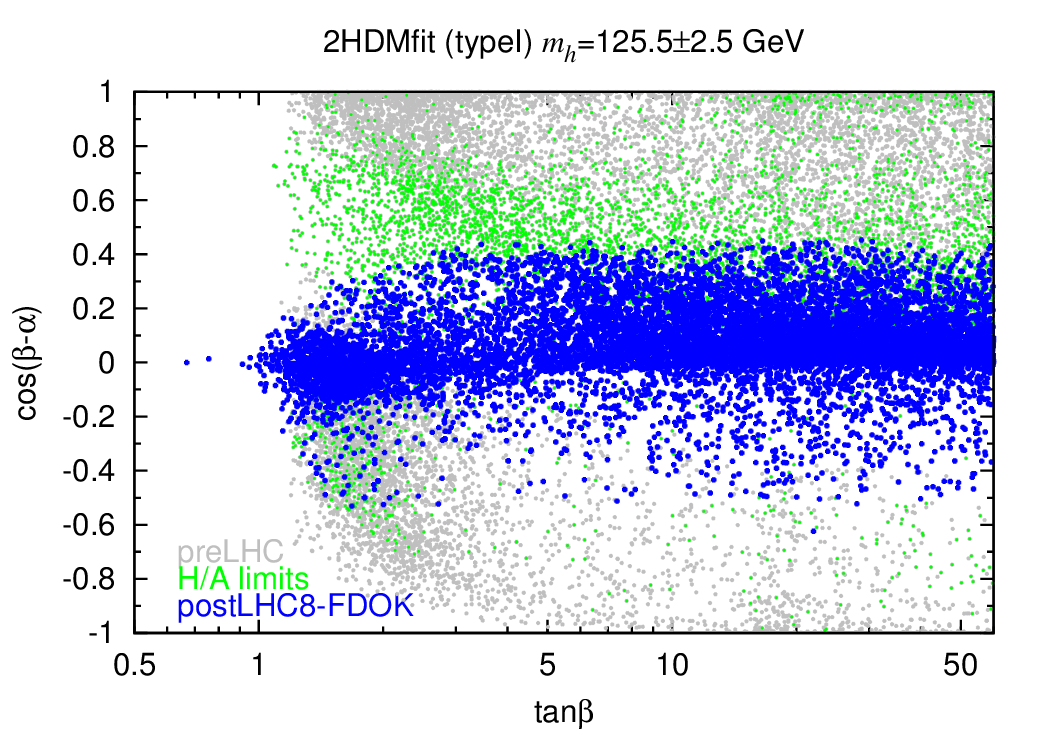}
\includegraphics[width=0.49\textwidth]{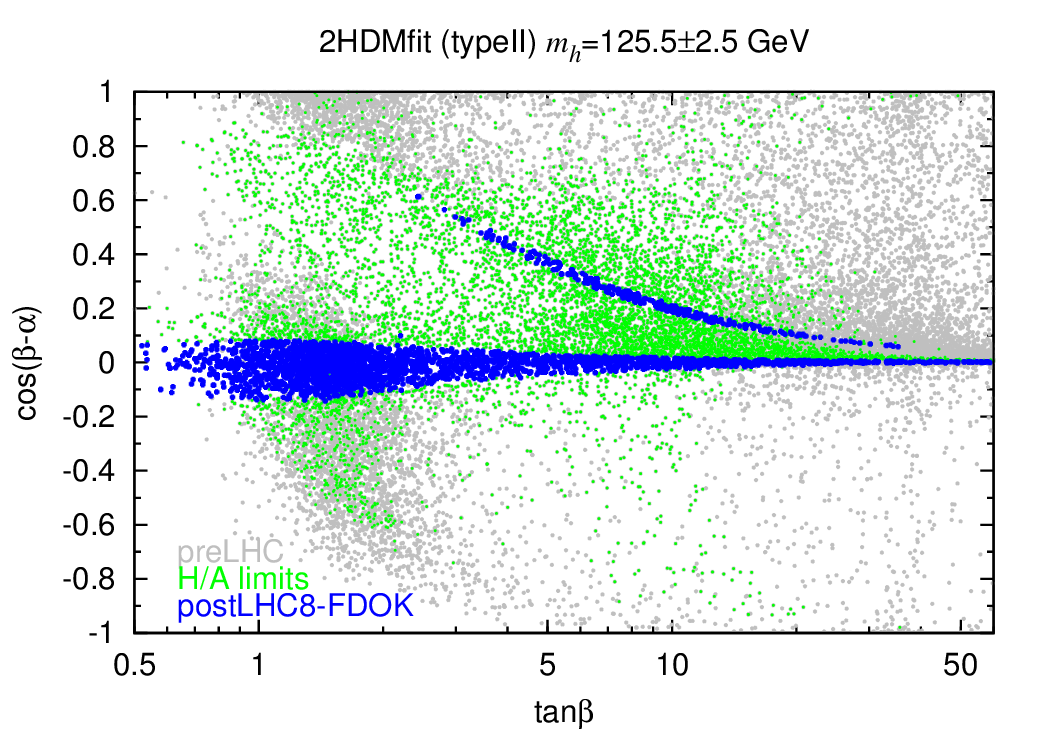}
\end{center}\vspace*{-5mm}
\caption{Constraints  in the $\cbma$ versus $\tanb$ plane for $\mh\sim 125.5\gev$. Grey points satisfy preLHC constraints, while green points satisfy in addition the LHC limits on  $\hh$ and $\ha$ production.  Blue points fall moreover within the 7+8 TeV $95\%$ C.L. ellipses in the $[\muggf(Y),\muvbf(Y)]$ plane for each of the final states considered ($Y=\gam\gam,VV,b\bar b,\tau\tau$), and the amount of FD from $H$ or $A$ production is small.
}
\label{bmavstb}
\end{figure}

Let us now turn to the constraints on the model parameter space that originate from requiring that  the 
signal strengths of the observed Higgs are matched at 95\% C.L. in each final state. 
We begin with the plots of Fig.~\ref{bmavstb} showing points in the $\cbma$ vs.\ $\tanb$ plane. Recall that if the $h$ is SM-like, then  $\sbma$ will be large and $\cbma$ will be constrained to smaller values. The color scheme is the following:
\vspace*{-.1in}
\bit
\item grey points are those that survive the preLHC constraints;\vspace*{-.1in}
\item green points are those for which all LHC limits related to the heavier Higgs bosons are obeyed at 95\% C.L. --- we employ the label ``H/A limits" for such points;\vspace*{-.1in}
\item blue points are those for which, {\it in addition}, the $h$ predictions fall within the 95\% C.L. regions in the $[\muggf,\muvbf]$ plane for {\it all} final state channels ($\gam\gam,VV,b\anti b$, and $\tau\tau$) --- we term points at this level postLHC8 --- {\it and} in addition the effects of feed down on the 95\% C.L. ellipses is small, which we call ``postLHC8-FDOK.'' 
\eit
\vspace*{-.1in}
We will discuss FD in more detail later, giving the precise criteria required for a point to be ``FDOK".  For now, let us note the following features of the plots.
Looking at the blue points that survive at the \plhcfd\ level, we observe that for $\mh\sim 125.5\gev$ in Type~I models $|\!\cbma |$ cannot be too large, especially if $\tanb \sim 1$. In the Type~II models, either $|\!\cbma|$ can be quite close to 0 or it can fall in a second branch where fairly large positive $\cbma\gsim 0.3$ is allowed if $\tanb\lsim 7$.  It turns out that this branch is associated with $\sin(\beta+\alpha) \approx 1$ and $\sin\alpha>0$ [for which the $b$-quark Yukawa coupling has the opposite sign relative to the $\sin(\beta-\alpha)\to 1$ limit]. This ``wrong-sign" Yukawa coupling, $\cd^h\sim -1$ is the focus of \cite{Ferreira:2014naa}.

\begin{figure}[t]
\begin{center}
\includegraphics[width=0.49\textwidth]{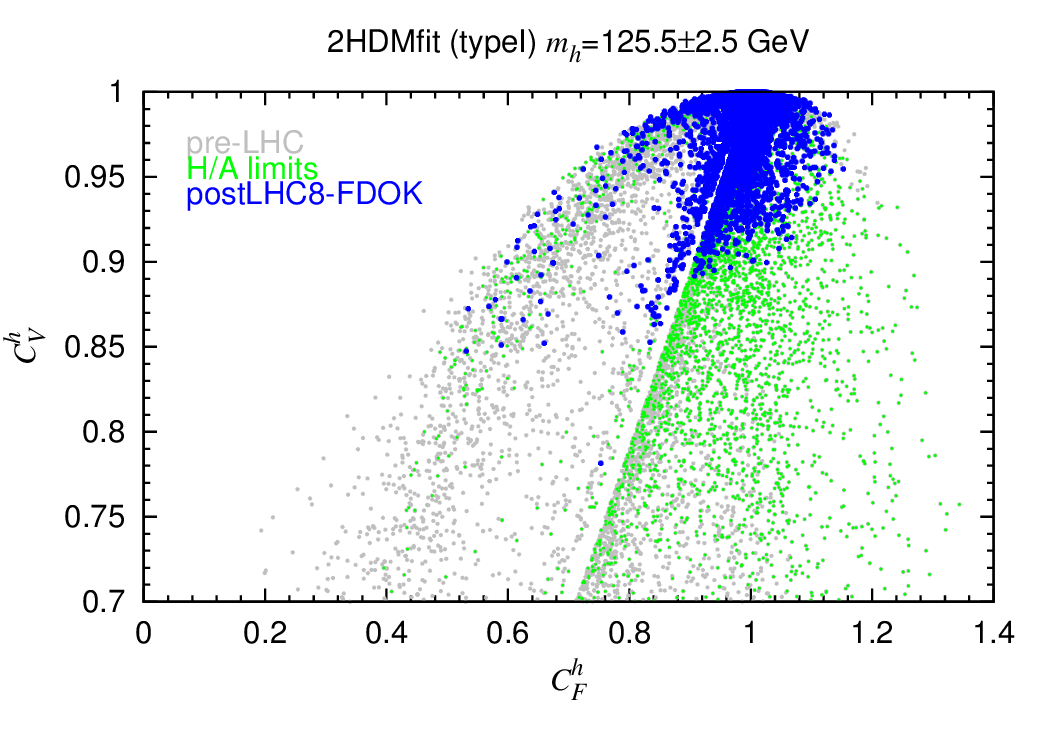}
\includegraphics[width=0.49\textwidth]{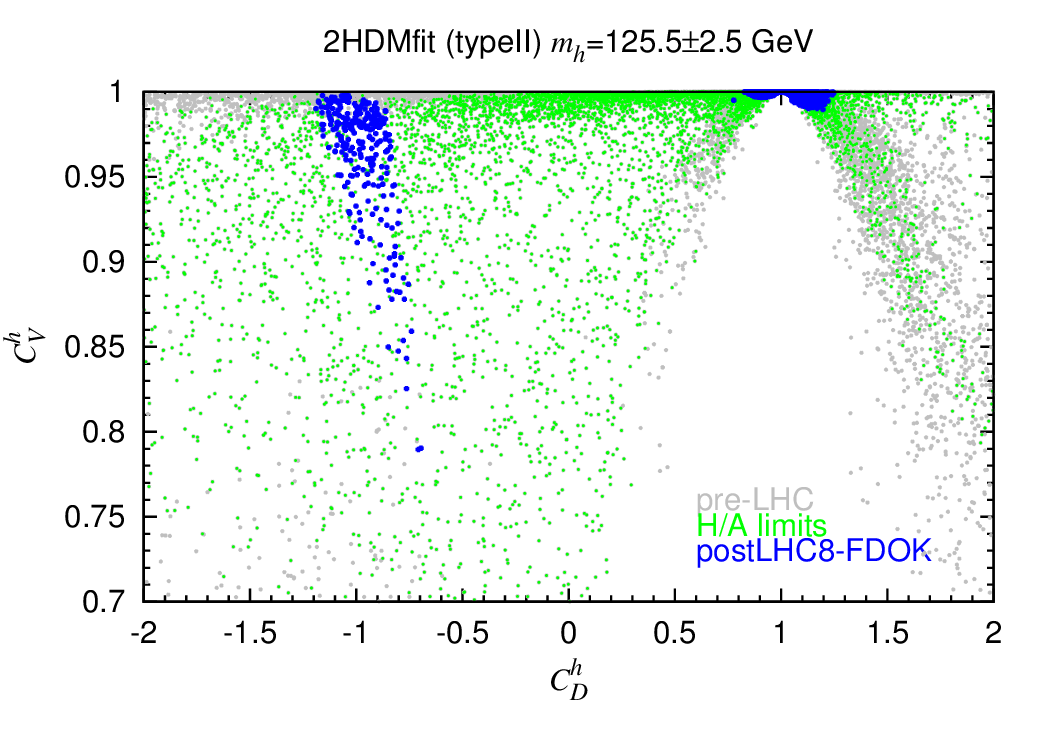}\\
\end{center}\vspace*{-5mm}
\caption{$\cv$ versus $C_F=\cu=\cd$ for Type~I (left) and $\cv$ versus $\cd$ for Type~II (right). Color scheme as in Fig.~\ref{bmavstb}. We have restricted the plots to $\cv \geq 0.7$ so as to most clearly display the postLHC8-FDOK points.}\label{cfcuvscv}
\end{figure}

Insight into the underlying couplings, $\cv^h$, $\cu^h$ and $\cd^h$ is provided by Fig.~\ref{cfcuvscv}.  There, we see that for both Type~I and Type~II, $\cv^h\sim +1$ is required for a decent fit to the Higgs data.  Further, $\cu^h\sim +1$ is needed in order to describe the observed $\gam\gam$ final state rates (\ie\ a SM-like cancellation between the $W$ and $t$ loops contributing to the $h\gam\gam$ coupling is required).  In Type~I, $\cd^h=\cu^h$ and therefore both must also be close to $+1$. However, this is not required in the Type~II models. In fact,  the second branch apparent in Fig.~\ref{bmavstb}  corresponds to the 
$\cd^h\sim -1$ region of the right-hand Type~II plot of Fig.~\ref{cfcuvscv} --- note that the magnitude, $|\cd^h|\sim 1$, is approximately fixed by the need for 
acceptable fits to the $b\anti b$ and $\tau\tau$ final state rates.  

\begin{figure}[t]
\begin{center}
\includegraphics[width=0.49\textwidth]{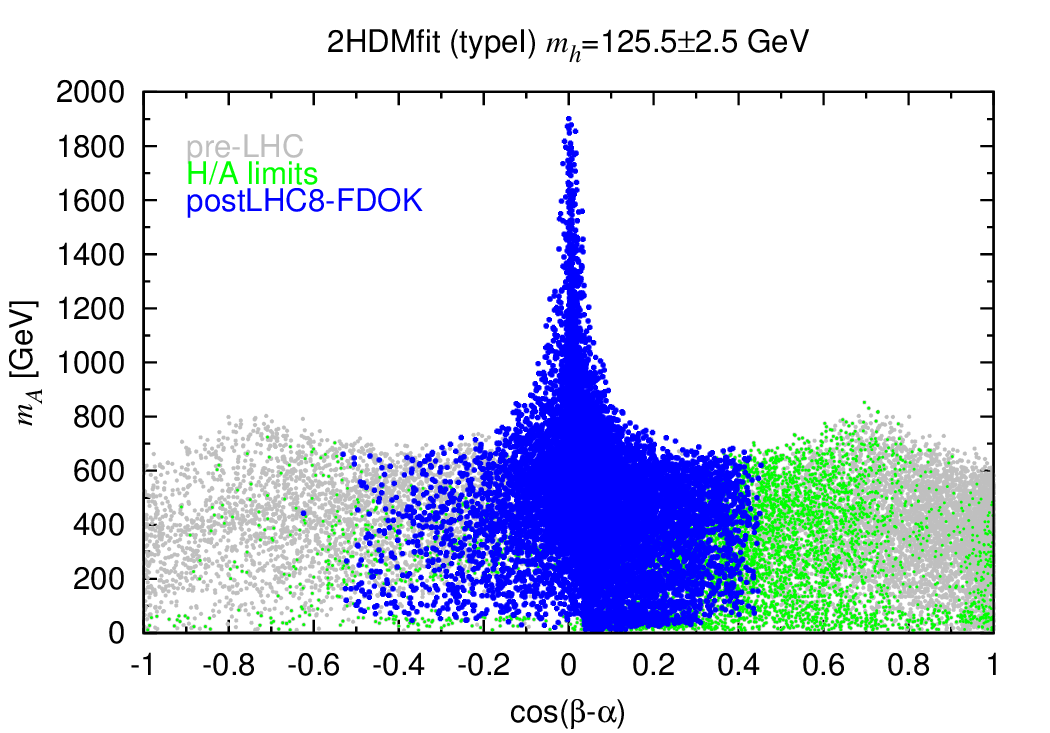}
\includegraphics[width=0.49\textwidth]{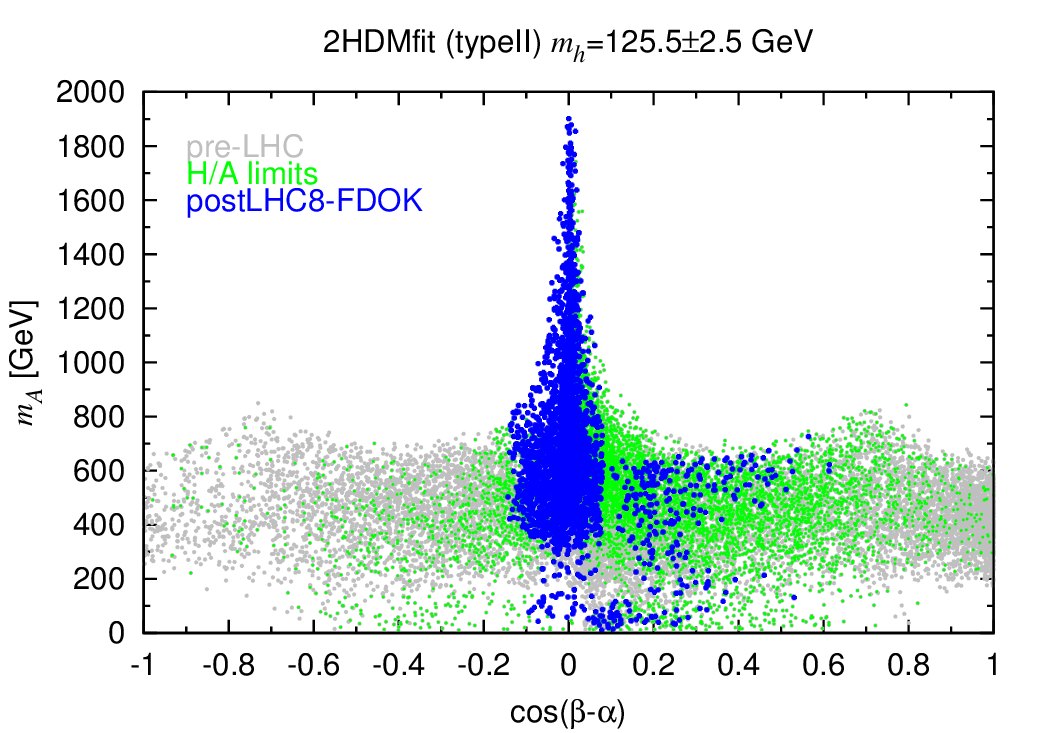}\\
\end{center}\vspace*{-5mm}
\caption{Constraints  in the  $\mA$ versus $\cbma$ plane for $\mh\sim 125.5\gev$. Color scheme as in Fig.~\ref{bmavstb}. The corresponding results in the $m_H$ vs.\ $\cbma$ and $m_{\hpm}$ vs.\ $\cbma$ planes give essentially the same picture as for $\mA$, except that $\mH$ and $\mhpm$ do not go below $125\gev$ and $100\gev$ ($300\gev$) for Type~I (Type~II), respectively.}
\label{mHvscba}
\end{figure}

Also of interest for the $\mh\sim 125.5\gev$ case 
is the range of the heavy Higgs masses as  a function of $\cbma$, shown in Fig.~\ref{mHvscba}. 
Clearly, once $\mA$ is above about $800\gev$ we are deep into the small $|\!\cbma|$ decoupling region, whereas for masses below $\sim 800\gev$ there is considerable spread in the allowed $|\!\cbma|$ values, in particular in Type~I. Thus, if an $A$ (or $H$ or $\hpm$) is found above $\sim 800\gev$, the 2HDMs require that the $h$ is very SM-like, but if $\mA$ is found to be lower in mass, then the $h$ need not be so SM-like.  Conversely, if the $h$ is found to have very SM-like $VV$ coupling, \ie\ if $|\!\cbma|\approx 0$, 
then $\mA$, $\mhh$ and $\mhpm$ could each take a large range of  values.
The larger $\cbma>0$ points in the Type~II case are the same as the $\cd^h\sim -1$ points of Fig.~\ref{cfcuvscv}.  That they cannot occur at high $\mH$ is associated with nondecoupling perturbativity limits, see \cite{Ferreira:2014naa}.

\begin{figure}[t]
\begin{center}
\includegraphics[width=0.49\textwidth]{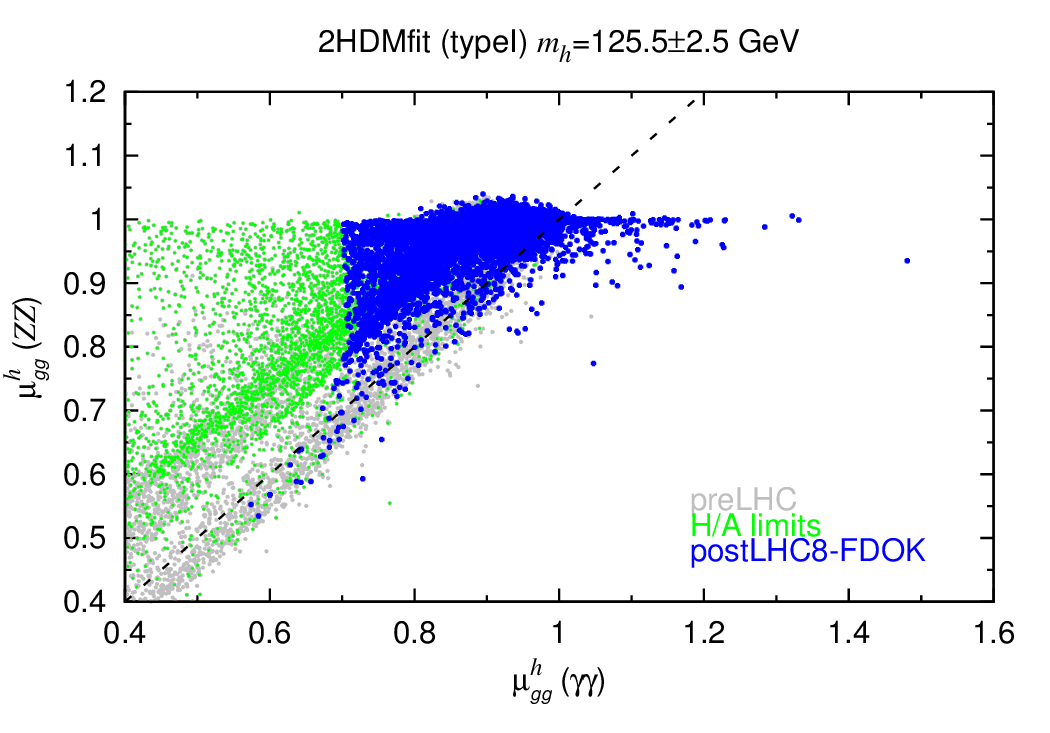}
\includegraphics[width=0.49\textwidth]{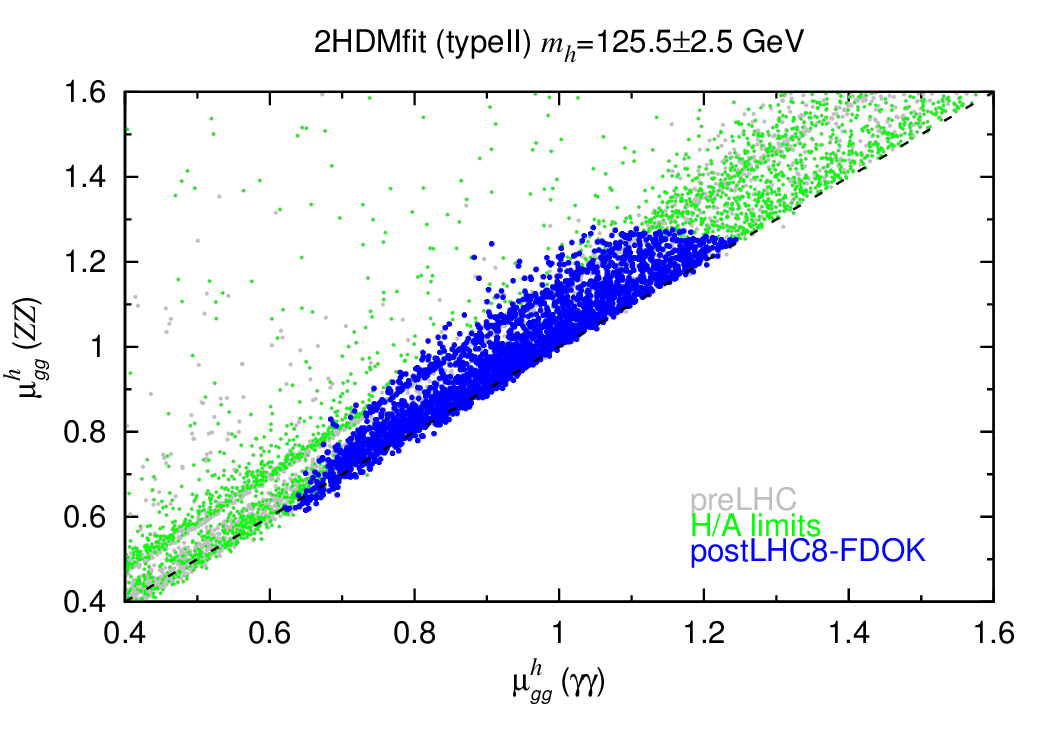}\\
\includegraphics[width=0.49\textwidth]{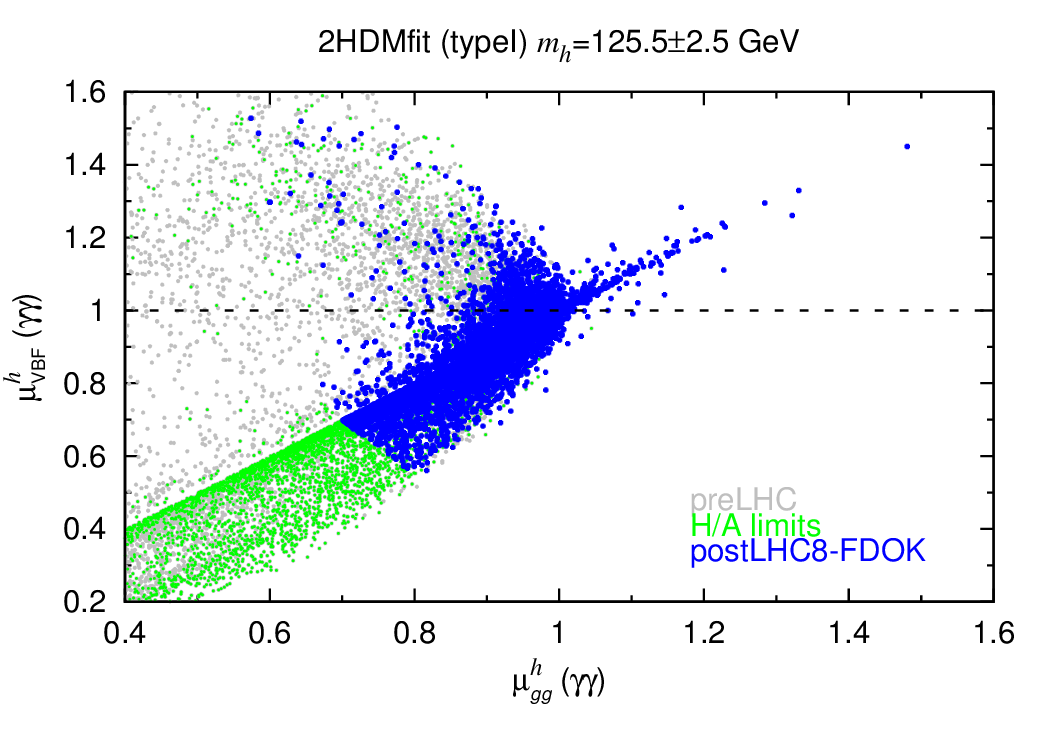}
\includegraphics[width=0.49\textwidth]{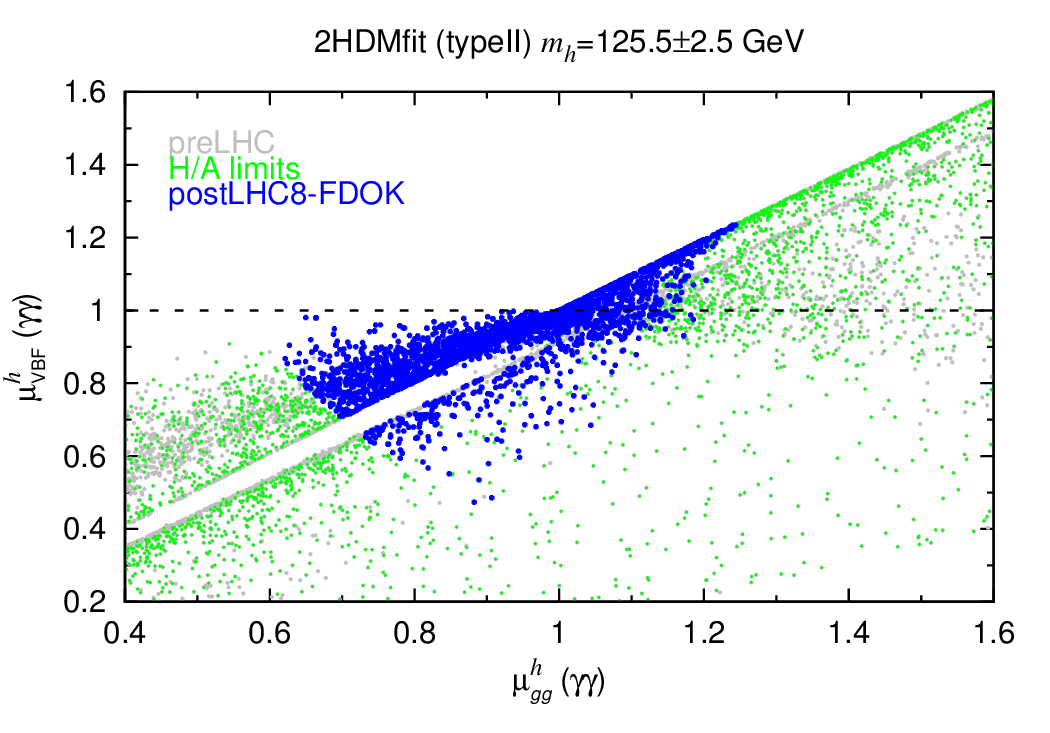}
\end{center}\vspace*{-5mm}
\caption{Correlations of signal strengths for the $\mh\sim 125.5\gev$ scenario following the color scheme of Fig.~\ref{bmavstb};  in the upper row $\rgghzz$  vs.\ $\rgghgamgam$, in the lower row $\rvbfhgamgam$ vs.\ $\rgghgamgam$. 
}
\label{gamgamvszz}
\end{figure}

To illuminate the precision with which individual channels are being fit, we show in Fig.~\ref{gamgamvszz} 
the signal strengths for $\rgghzz$ vs.\ $\rgghgamgam$ (upper row) 
as well as for $\rvbfhgamgam$ vs.\ $\rgghgamgam$ (lower row).
From this figure, it is apparent that requiring the  $\mu$ values to lie within $\pm 10\%$ of unity would have a strong impact.  Even $\pm 20\%$ measurements will remove many parameter choices.   
Note also that with sufficiently precise measurements of $\rgghzz$ and  $\rgghgamgam$ there is a chance to distinguish Type~I from Type~II models; for most of the blue points if $\rgghgamgam$ is $> 1$, then $\rgghzz/\rgghgamgam<1$ for Type~I, whereas for Type~II $\rgghzz/\rgghgamgam>1$ always. 
Likewise, there are complementary correlations between the ggF and VBF modes, as illustrated for the $\gam\gam$ final state in the lower row of Fig.~\ref{gamgamvszz}.  In particular, if $\rvbfhgamgam>1$, then $\rgghgamgam<1$ is required in Type~I, whereas just the opposite statement applies in Type~II. In general, the cross correlations between different production$\times$decay modes carry interesting information because of the dependences in particular of the $hgg$ and $h\gam\gam$ couplings on $C_U$, $C_V$ (and in Type~II for large $\tan\beta$ also on $C_D$) and thus can be useful for distinguishing scenarios. 

\begin{figure}[t]
\begin{center}
\includegraphics[width=0.47\textwidth]{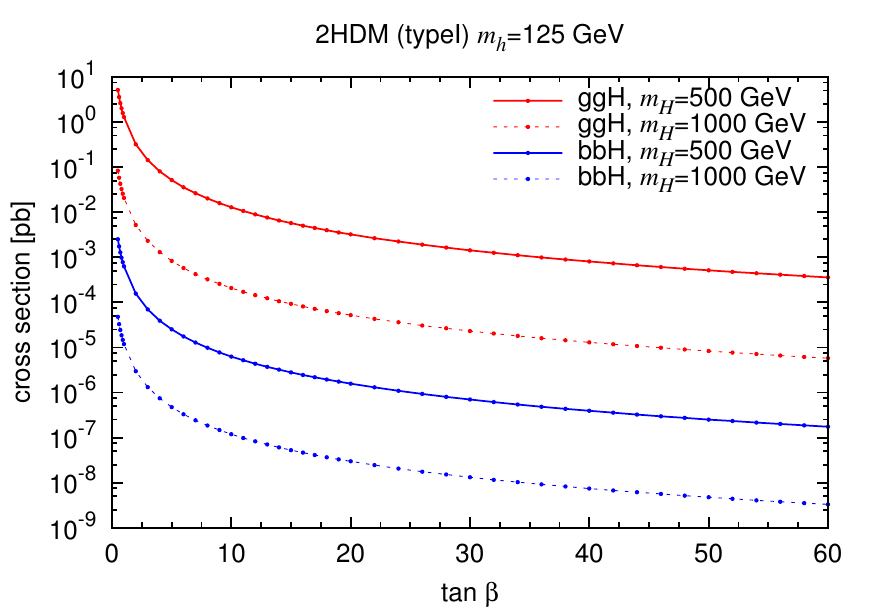}
\includegraphics[width=0.47\textwidth]{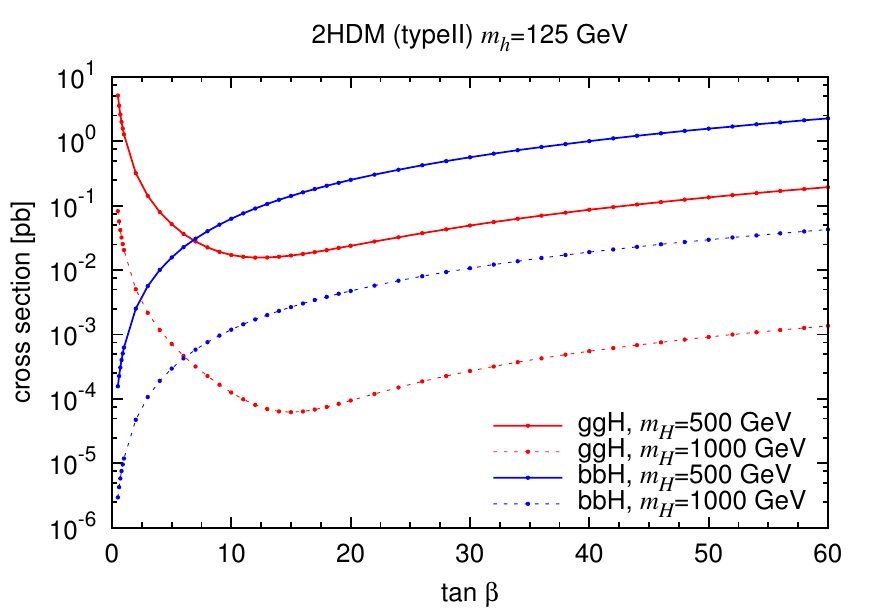}\\
\end{center}\vspace*{-5mm}
\caption{Cross sections (in pb) for $gg\to H$ and $bbH$ production at $\sqrt{s}=8\tev$ as a function of $\tanb$ for $\mH=500\gev$ and $1\tev$. The results shown are for the SM limit of $\sbma=1$. }
\label{bbHtoggH}
\end{figure}

With regard to the $H/A$ limits coming from heavy Higgs bosons, we find that they have significant impact.  In the case of Type~I, the 
limits coming from $gg\to H,A\to \tau\tau$ are always stronger (even at high $\tanb$) than those coming from $gg\to b\anti b H, b\anti bA$ with $H,A\to \tau\tau$.  
 This is because,  in Type~I models, all the fermionic couplings are the same and it is only a question of the $\tanb$-independent ratio of the $gg\to H,A$ cross section to the $gg\to b\anti b H,b\anti b A$ cross section. This ratio is always quite a bit larger than 1 (by typically a factor of at least 100) at any given mass. For Type~II models, we note that the down-type coupling is enhanced by $\tanb$ and affects both production modes, in the  $gg\to H,A$ case by enhancing the $b$-quark contribution to the one-loop coupling. To illustrate the comparison between these two production modes, we plot for two representative masses ($500\gev$ and $1\tev$) the two cross sections  vs.\ $\tanb$ in Fig.~\ref{bbHtoggH} with $\alpha$ chosen so that $\sbma=1$. In the case of Type~II, we observe that at low $\tanb$ it is $\sigma(gg\to H)$ that is biggest while at high $\tanb$ it is $\sigma(bbH)$ that is biggest by a factor of $\sim 10$ to $100$.

\subsection{Implications for the future}

\begin{figure}[t]
\begin{center}
\hspace*{-.2in}\includegraphics[width=0.49\textwidth]{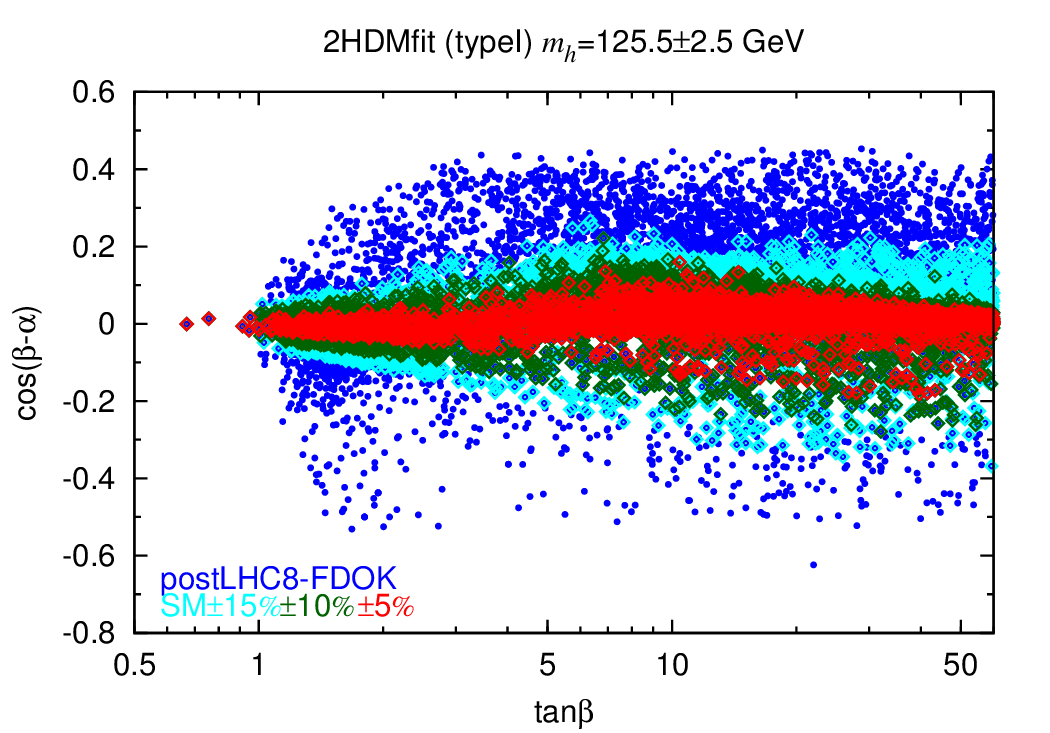}
\includegraphics[width=0.49\textwidth]{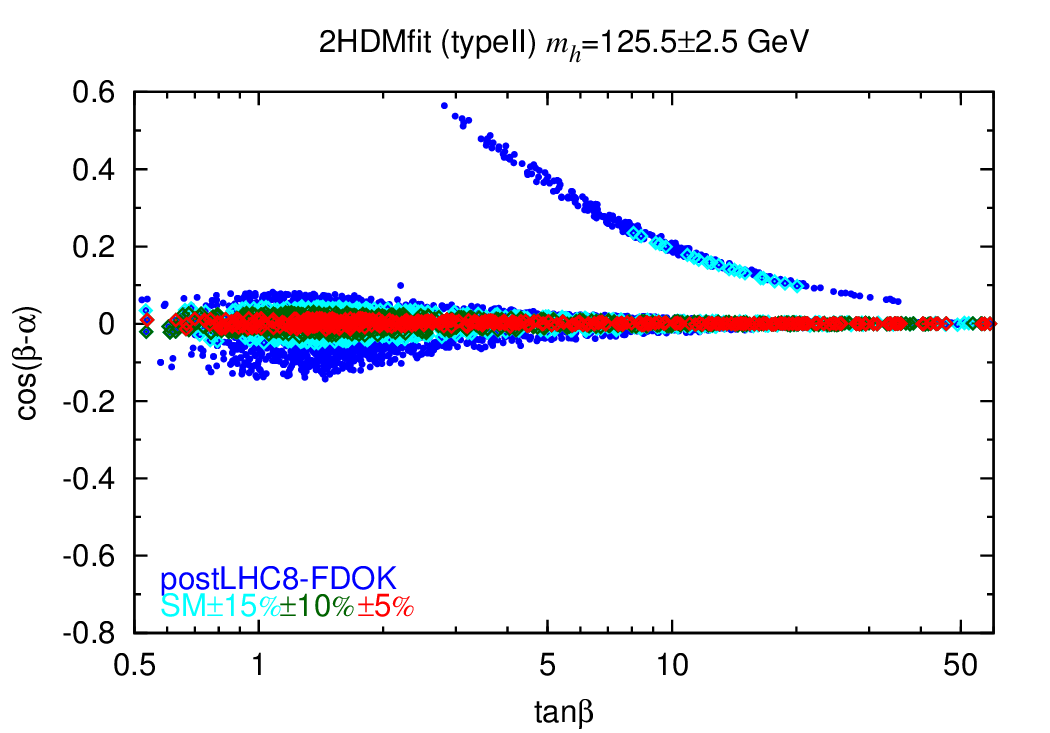}
\end{center}\vspace*{-5mm}
\caption{The postLHC8-FDOK points in the $\cbma$ vs.\ $\tanb$  plane for the $\mh\sim 125.5\gev$ scenario comparing current $h$ fits (blue) to the case that  the rates for all the channels listed in Eq.~(\ref{xxyychannels})  are within $\pm 15\%$ (cyan), 
$\pm 10\%$ (green) or $\pm 5 \%$ (red) of the SM Higgs prediction. FDOK is also required for these latter points.
} 
\label{cbmavstbfuture}
\end{figure}

At this point, we turn to a consideration of what future measurements at LHC 13/14 or a linear collider might be most revealing. Let us first quantify the extent to which future higher precision measurements   at the next LHC run might be able to restrict the model parameter space. Typical results are illustrated in Fig.~\ref{cbmavstbfuture}. To make clear the impact of increased precision in the future, we will show points that survive if the observed values of  $\mu_{X}^h(Y)$ {\it all}   lie  within $P\%$ of the SM prediction for the following channels $(X,Y)$:\footnote{It is important to note that the $({\rm VBF},\tau\tau)=({\rm VH},bb)$ channels have exactly the same scaling factor in  2HDMs. We mention them together since they are experimentally very different channels and although individually they may not be measurable with a certain level of accuracy, in combination they should be able to determine the common $\mu$ to the specified accuracy.} 
 \beq 
 (gg,\gam\gam),  ~ (gg,ZZ),   ~ (gg,\tau\tau),  ~ ({\rm VBF},\gam\gam),  ~ ({\rm VBF},ZZ),  ~ ({\rm VBF},\tau\tau)=({\rm VH},bb),  ~ ({\rm ttH},bb)\,.
 \label{xxyychannels}
 \eeq 
Here, we will consider $P=\pm15\%$, $\pm10\%$ and $\pm5\%$.  
For this we use the shorthand notation SM$\pm 15\%$, SM$\pm10\%$ and SM$\pm5\%$, respectively.
Not unexpectedly, as increasingly precise agreement with the SM is imposed in the various channels, one is quickly pushed to small $|\!\cbma|$, but $\tanb$ remains unrestricted.  Note that even SM$\pm10\%$ on each of the individual $\mu$'s  will have eliminated the ``wrong-sign" down-quark Yukawa region (which corresponds to $\sin\alpha>0$ or $\cd^h<0$) of the Type~II model. 

\begin{figure}[t]
\begin{center}
\includegraphics[width=0.49\textwidth]{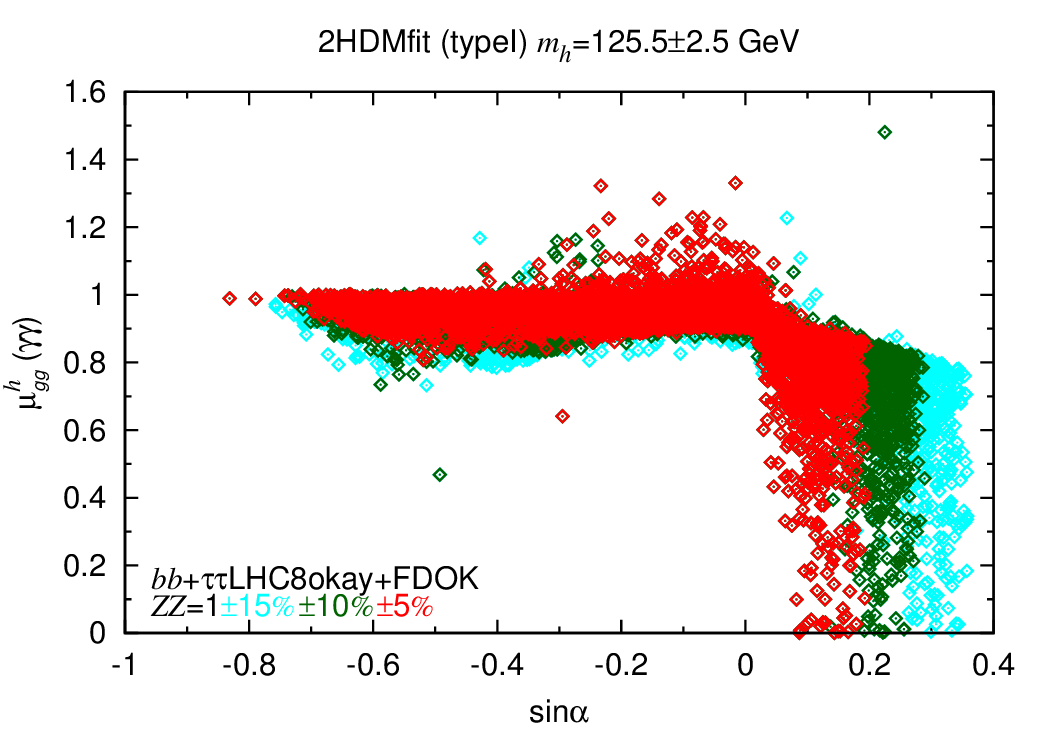}
\includegraphics[width=0.49\textwidth]{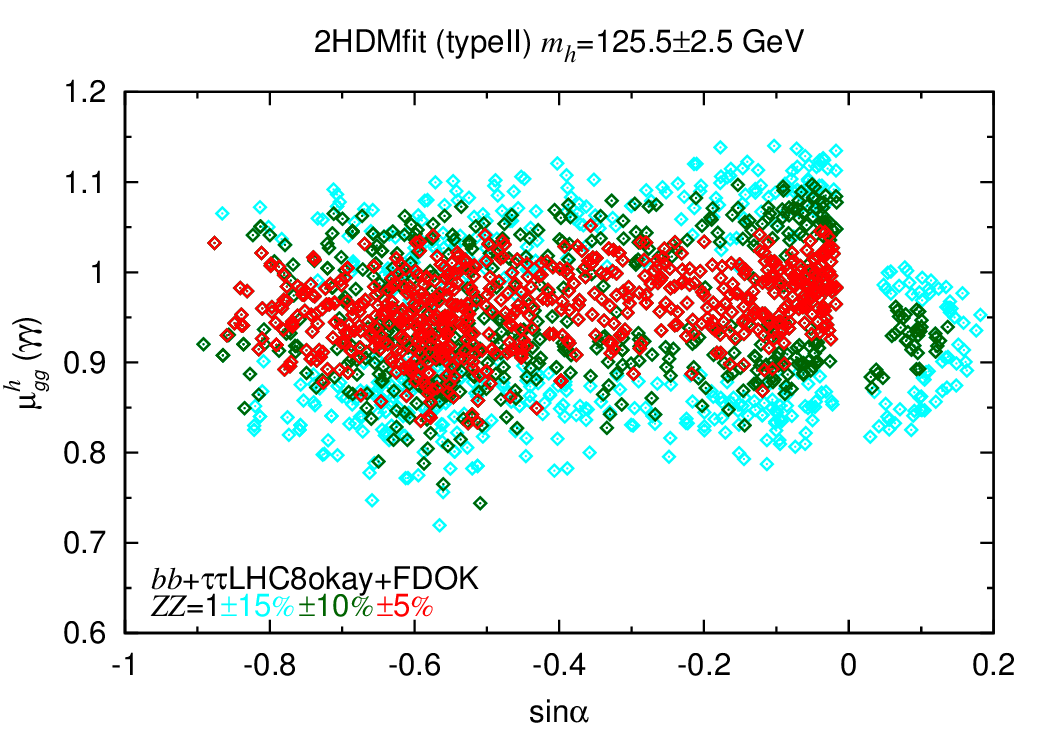}
\end{center}\vspace*{-5mm}
\caption{The postLHC8-FDOK points in the $\rgghgamgam$ vs.\ $\sin\alpha$ plane for the $\mh\sim 125.5\gev$ scenario, requiring that the $gg\to ZZ$ and $VV\to ZZ$ rates are both within $\pm 15\%$ (cyan), $\pm 10\%$ (green) and  $\pm 5\%$ (red) of the SM predictions. The individual rates for the $\tau\tau$ and $bb$ final states are consistent with current 95\% CL limits. 
}
\label{sinaplots}
\end{figure}

\noindent
The reason for this is clarified by Fig.~\ref{sinaplots}.  
Here, we  require that the rates for the $bb$ and $\tau\tau$ final states are consistent at 95\% C.L. with current data and then examine the implications for $gg\to h \to \gam\gam$ when  requiring that the $gg\to h\to ZZ$ and $VV\to h\to ZZ$ rates be progressively closer to the SM prediction. We see that once SM$\pm 10\%$ is required for the $ZZ$ final state, then $\rgghgamgam$ is at least $5\%$ below the SM in the $\sin\alpha>0$ region.  As explained in \cite{Ferreira:2014naa} this is because for $\sin\alpha>0$ the charged-Higgs loop contribution to the $h\gam\gam$ coupling does not decouple and causes a decrease in $\Gamma(h\to\gam\gam)$ of order $10\%$.  This decrease is only partially compensated by a $6\%$ increase in the cross section for $gg\to h$ that arises due to the fact that the interference between the top and bottom loops has a sign that is opposite the normal sign in this $\sin\alpha>0$ region.  As regards the Type~I model, the sign of the up-type and down-type Yukawa couplings is independent of the sign of $\sina$. Nonetheless, $\sina>0$ values are disfavored since the $\rgghgamgam$ rate reduces as $\sina$ increases past zero. This is in fact also true in Type~II models.  As noted in relation to Fig.~\ref{laaaa}, this is a result of requiring that all quartic Higgs couplings remain perturbative, defined by having absolute values below $4\pi$.

\begin{figure}[t]
\begin{center}
\includegraphics[width=0.49\textwidth]{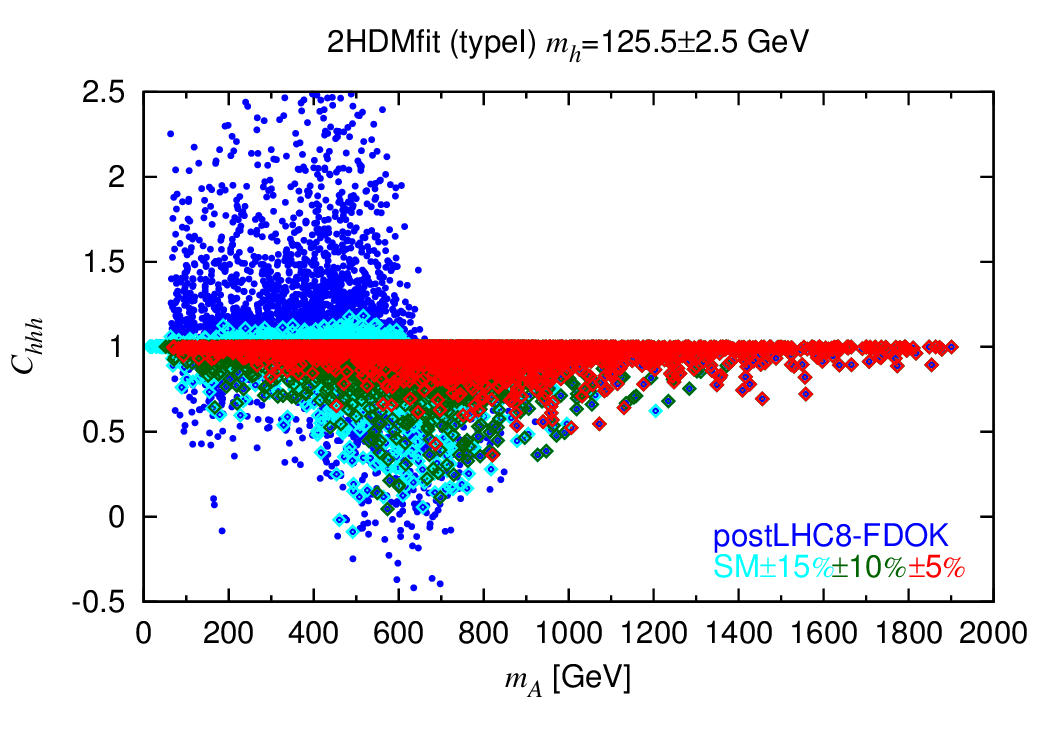}
\includegraphics[width=0.49\textwidth]{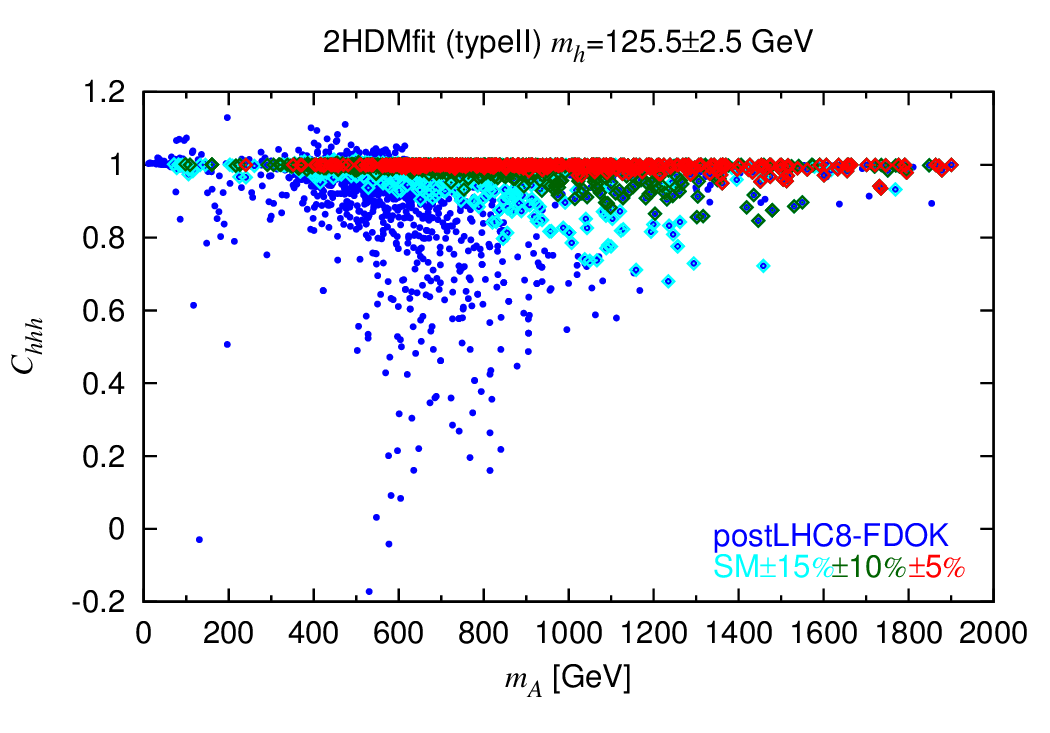}
\end{center}\vspace*{-5mm}
\caption{The postLHC8-FDOK points in the $C_{hhh}$ vs.\ $\mA$  plane for the $\mh\sim 125.5\gev$ scenario comparing current $h$ fits  to the case where future measurements show that  the rates for all the channels listed in Eq.~(\ref{xxyychannels})   are within $\pm 15\%,\pm 10\%,\pm 5 \%$ of the SM Higgs prediction; FDOK is required in all cases. Color scheme is as for Fig.~\ref{cbmavstbfuture}.}
\label{hhhcoup}
\end{figure}

One quantity of particular interest for a SM-like $h$ is the triple-Higgs coupling strength $\lam_{hhh}$.  We plot the current and possible future expectations in Fig.~\ref{hhhcoup} for $C_{hhh}$ (defined as the value of $\lam_{hhh}$ relative to the SM value). We observe that if the $\mu_X(Y)$ measurements were to have excursions from the SM predictions at the currently allowed 95\% level extreme, then measurement of a large deviation from $C_{hhh}=1$ would be quite likely (also see \cite{Efrati:2014uta}). For example, at the high-luminosity LHC14, with $L=3000\fbi$ one can measure $\lam_{hhh}$ to the 50\% level \cite{Dawson:2013bba}, and given the limited constraints on the model implied by current Higgs data, deviations from $C_{hhh}=1$ of this order, indeed up to 100\% or more,  are possible. However, if future LHC measurements imply increasingly smaller deviations from $\mu_X(Y)=1$ in the various channels, then observing a deviation from $C_{hhh}=1$ becomes increasingly difficult, even at the ILC.  For example, from \cite{Dawson:2013bba} 
we find that the predicted precision 
on $\lam_{hhh}$ for ILC1000 with $L=500-1000\fbi$ of 21\% and for ILC1000 with $L=1600-2500 \fbi$ is of order 13\%.  At CLIC3000 with $L=2000\fbi$ the accuracy achievable would be about 10\%.  

Comparing to the deviations shown in Fig.~\ref{hhhcoup}, we see that in Type~I a determination of $\mu_X(Y)$ rates at the level of SM$\pm 10\%$ still allows $C_{hhh}$ as small as $\sim 0$, while SM$\pm 5\%$ allows $C_{hhh}$ as small as 0.3, either of which will be observable for any of the listed machines and integrated $L$ values. In contrast, for Type~II, even SM$\pm 15\%$ would already imply that $C_{hhh}$ must lie below $1$. 
This agrees with the conclusion reached in  \cite{Baglio:2014nea} where it is stated that current $68\%$~C.L. ($1\sigma$) limits (which are very close to our SM$\pm 15\%$ constraint) imply $C_{hhh}\leq 1$ for Type~II. We note further that the smallest $C_{hhh}$ for SM$\pm 10\%$ is $\sim 0.9$, while for SM$\pm 5\%$ it is $\sim 0.95$.   The former would require CLIC3000 while the latter would be beyond the reach of any of the above $e^+e^-$ colliders. Thus, it is clear that future LHC Higgs data could have a very significant impact on the prospects for seeing an interesting deviation from  $C_{hhh}=1$ at ILC/CLIC. As an aside, we note from Fig.~\ref{hhhcoup} that for Type~II  (but not Type~I) models SM$\pm 5\%$ is only possible for $\mha\gsim 250-300\gev$ depending on $\tanb$.

With this in mind, it is important to consider implications of the current and future $h$ fits for the heavier Higgs bosons.  Hopefully, one will retain a significant possibility of detecting the heavier Higgs bosons even if the $h$ is shown to be very SM-like.  To assess the situation, we consider only the high-rate gluon-fusion and $bb$ associated production processes for the $H$ and $A$.\footnote{For reasons of space, we will mainly present results for the inclusively summed ggF and $bb$ associated production of a given Higgs for $\rts=14\tev$.  Of course, it will be possible and of interest to separate these experimentally.} There are many final states of potential interest.  These include $H\to ZZ$, $H,A \to \tau\tau,\gam\gam,t\anti t$  as well as the $H\to hh$ and $A\to Zh$ final states.

\begin{figure}[t]
\begin{center}
\includegraphics[width=0.49\textwidth]{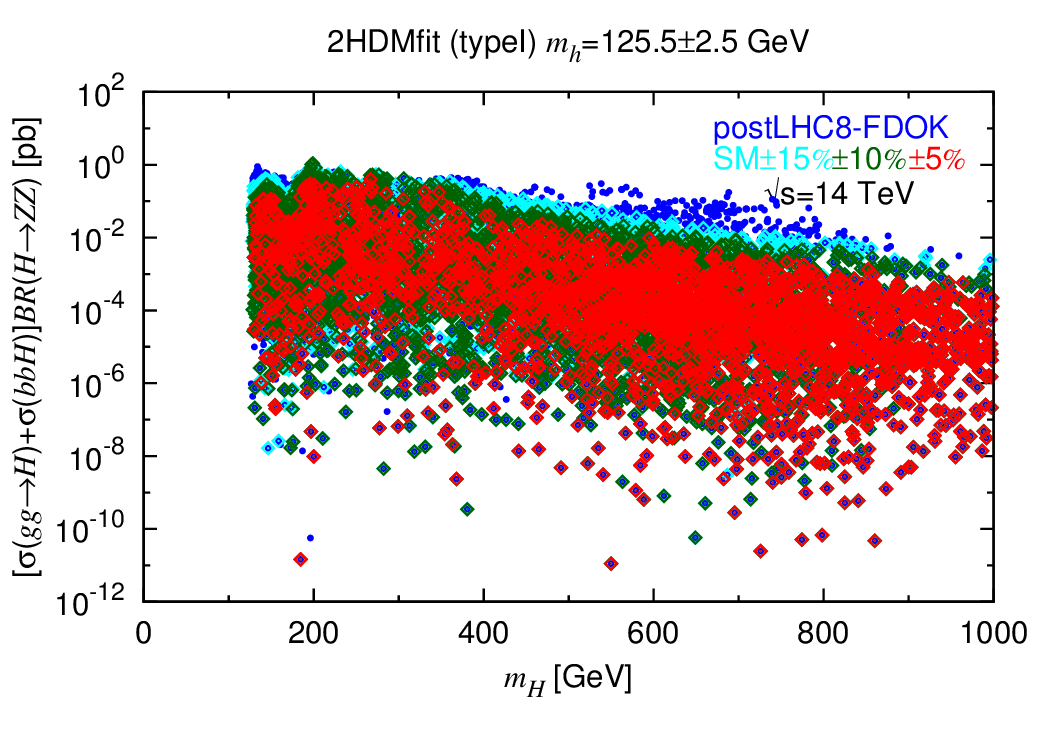}
\includegraphics[width=0.49\textwidth]{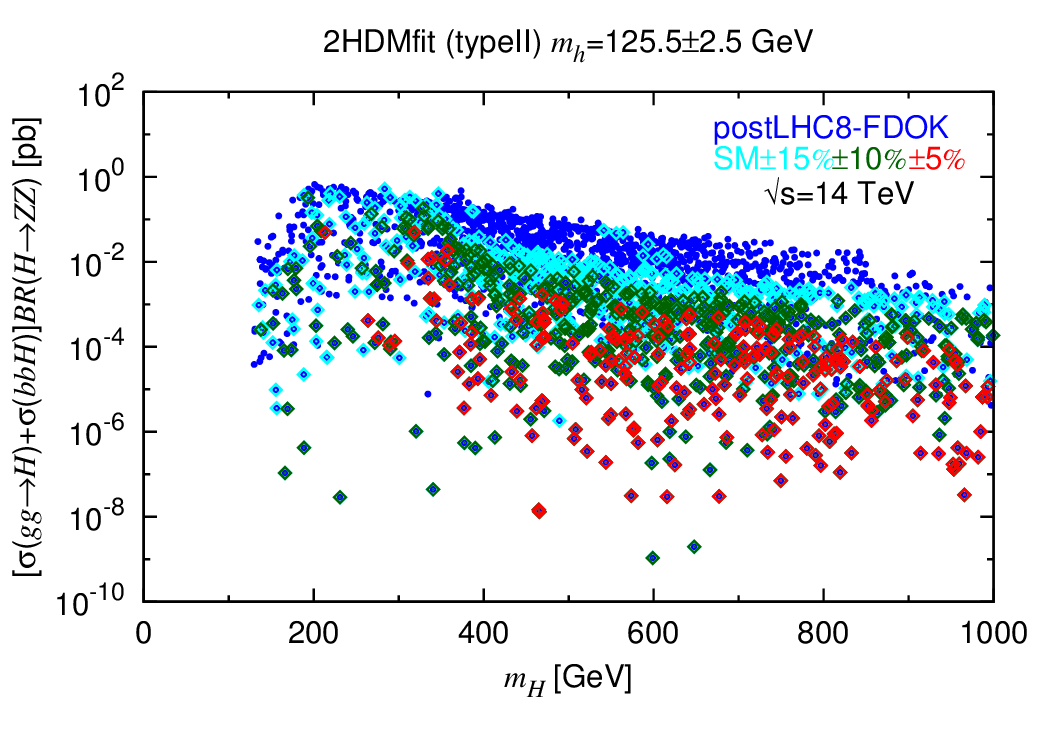}
\end{center}\vspace*{-5mm}
\caption{We plot $[\sigma(gg\to H)+\sigma(bbH)]\br(H\to ZZ)$ as functions of $\mH$, for Type~I (left) and Type~II (right) 2HDMs. Only FDOK points are shown. Implications of various levels of precision for future $h$ measurements are displayed. Color scheme is as for Fig.~\ref{cbmavstbfuture}.}
\label{ZZ14}
\end{figure}

Results for $H\to ZZ$ at $\rts=14\tev$ are shown in Fig.~\ref{ZZ14}.  We observe that substantial $\sigma \times \br$  values (as high as $\sim 1\pb$ at $\mH\sim 150\gev$ and $\sim 1\fb$ at $\mH\sim 1\tev$) are possible, but certainly not guaranteed.  In the case of the Type~II model, if the $h$ is determined to have SM-like rates within $\pm 10\%$ or, especially, $\pm 5\%$ then the maximum possible $\sigma \times \br$ is substantially reduced and the minimum allowed $\mH$ for $\pm 5\%$ is of order $200\gev$.

\begin{figure}[t]
\begin{center}\vspace*{-5mm}
\includegraphics[width=0.49\textwidth]{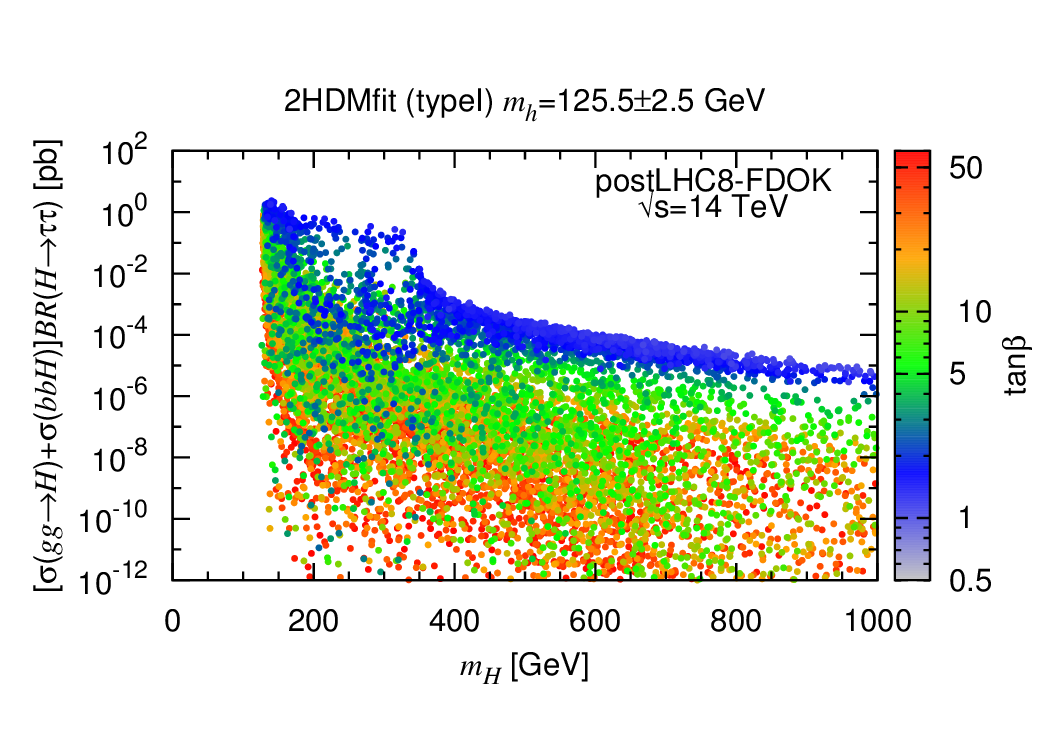}
\includegraphics[width=0.49\textwidth]{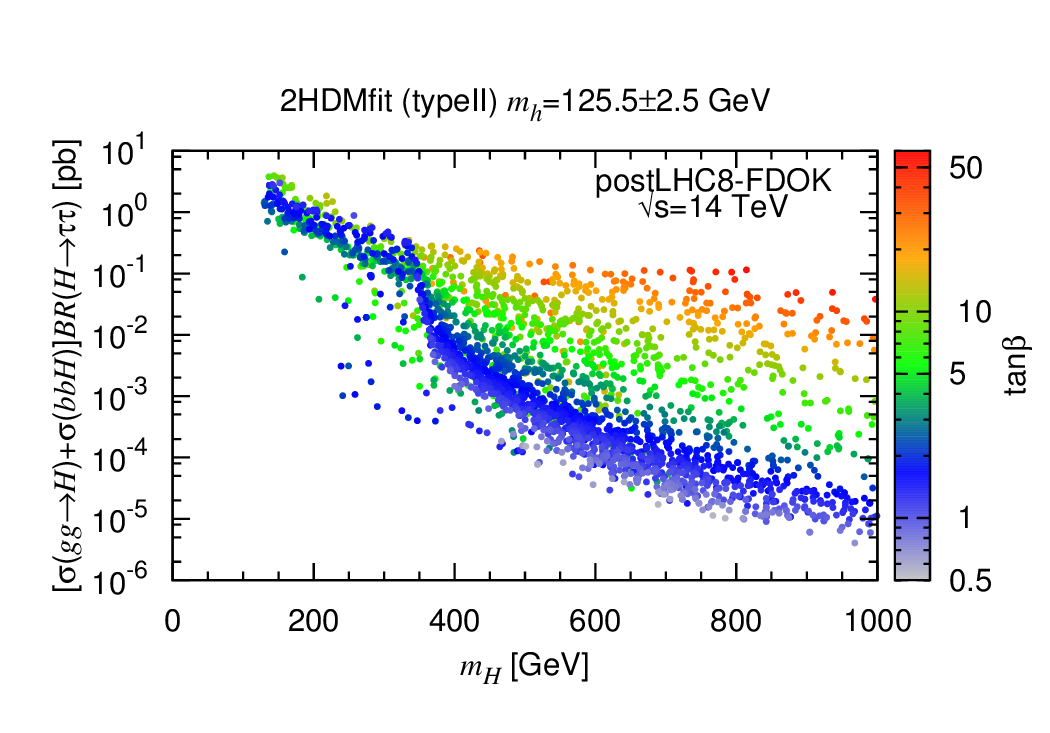}\\[-5mm]
\includegraphics[width=0.49\textwidth]{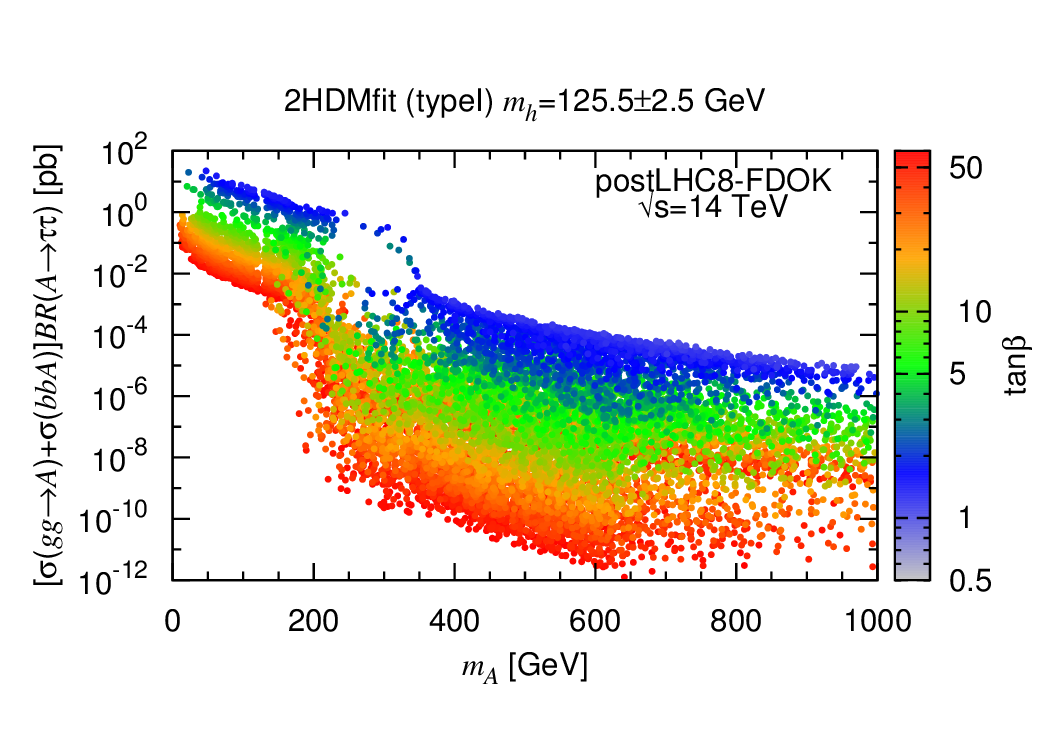}
\includegraphics[width=0.49\textwidth]{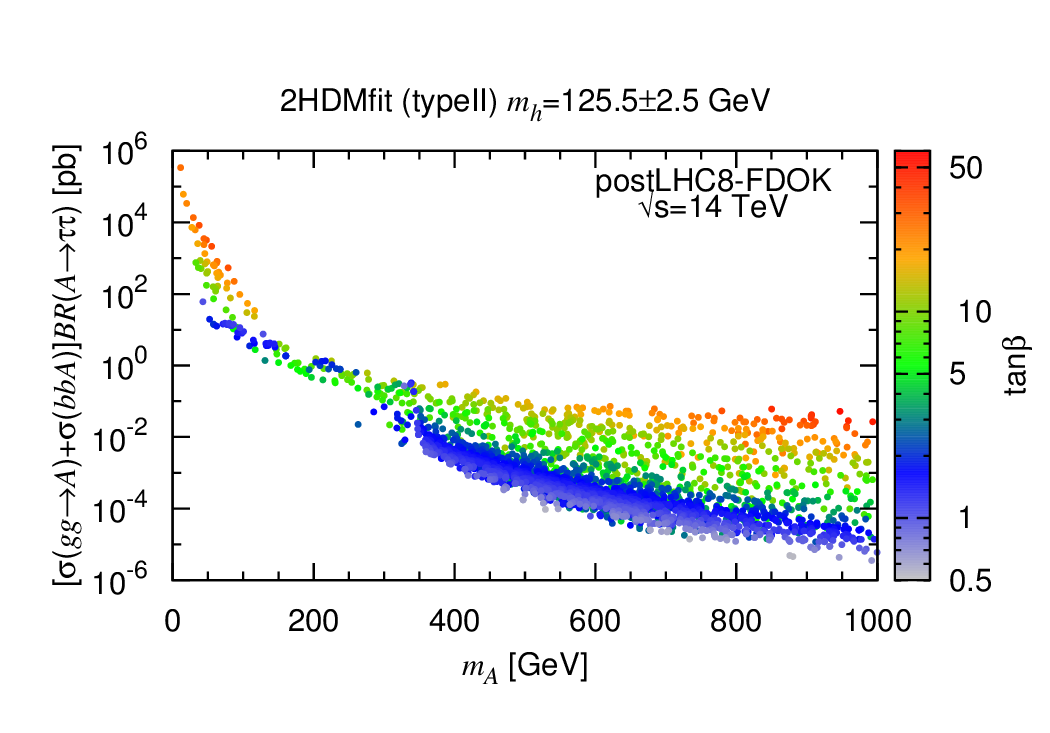}
\end{center}\vspace*{-8mm}
\caption{Scatter plots of $[\sigma(gg\to H)+\sigma(bbH)]\br(H\to \tau\tau)$ and $[\sigma(gg\to A)+\sigma(bbA)]\br(A\to \tau\tau)$, in pb, as functions of $\mH$ (top row) and $\mA$ (bottom row), respectively, for postLHC8-FDOK points with  $\mh\sim 125.5\gev$. 
The values of $\tan\beta$ are color-coded as indicated on the plots.}
\label{tautau14}
\end{figure}

Results for $gg + bb \to \hh,\ha$ production in the $\tau\tau$ final state are displayed in Fig.~\ref{tautau14} assuming $\rts=14\tev$. 
Overall, the range of possible cross sections is quite large, with maximum values of order 1 to 10 pb and minimum values below $10^{-10}\pb$ in the case of Type~I (although this range is somewhat narrowed on average as the $h$ is required to be more and more SM-like) and minimum values of order $10^{-4}-10^{-5}\pb$ in the case of Type~II. It is worth noting that
for lower values of $\mH$ and $\mA$, $[\sigma(gg\to H)+\sigma(bbH)]\br(H\to \tau\tau)$ and $[\sigma(gg\to A)+\sigma(bbA)]\br(A\to \tau\tau)$ are typically quite substantial in the Type~II case, but that few points survive below $\mH\sim 300\gev$ if the $125.5\gev$ state rates lie within $5\%$ of the SM Higgs predictions.  We comment on one particular feature of the plots, namely the fact that the $t\bar t$ threshold is not apparent for Type~II in the case of the $\ha$.  This is a direct consequence of the fact that the LHC8 constraints include the limits from \cite{htautau} on $gg\to A$ and $bbA$ with $A\to \tau\tau$.  The predicted 2HDM cross sections can significantly exceed these limits in the region below $2m_t$.  Since these limits are included in obtaining the postLHC8 results the $t\bar t$ threshold that would otherwise be apparent is not present.  Note that in the case of the $\hh$, the limits of \cite{htautau} do not have a strong impact because the predicted values of $\sigma(gg\to H)\br(H\to \tau\tau)$ are smaller due to the fact that $\hh\to ZZ$ decays are also present.

It is also interesting to consider the $\tan\beta$ dependence of the cross sections, indicated by a color code in Fig.~\ref{tautau14}.  As expected from the fermionic couplings in Table~\ref{tab:couplings}, this dependence is opposite in Type~I and Type~II. Concretely, in Type~I $[\sigma(gg\to H)+\sigma(bbH)]\br(H\to \tau\tau)$ and $[\sigma(gg\to A)+\sigma(bbA)]\br(A\to \tau\tau)$ increase as $\tan\beta$ gets smaller, while in Type~II  larger cross sections are obtained for larger $\tan\beta$. 
Note also that in Type~II  the $t\bar t$ threshold is visible for small $\tan\beta\lesssim 3$ but not for larger values. 

\begin{figure}[t]
\begin{center}\vspace*{-5mm}
\includegraphics[width=0.49\textwidth]{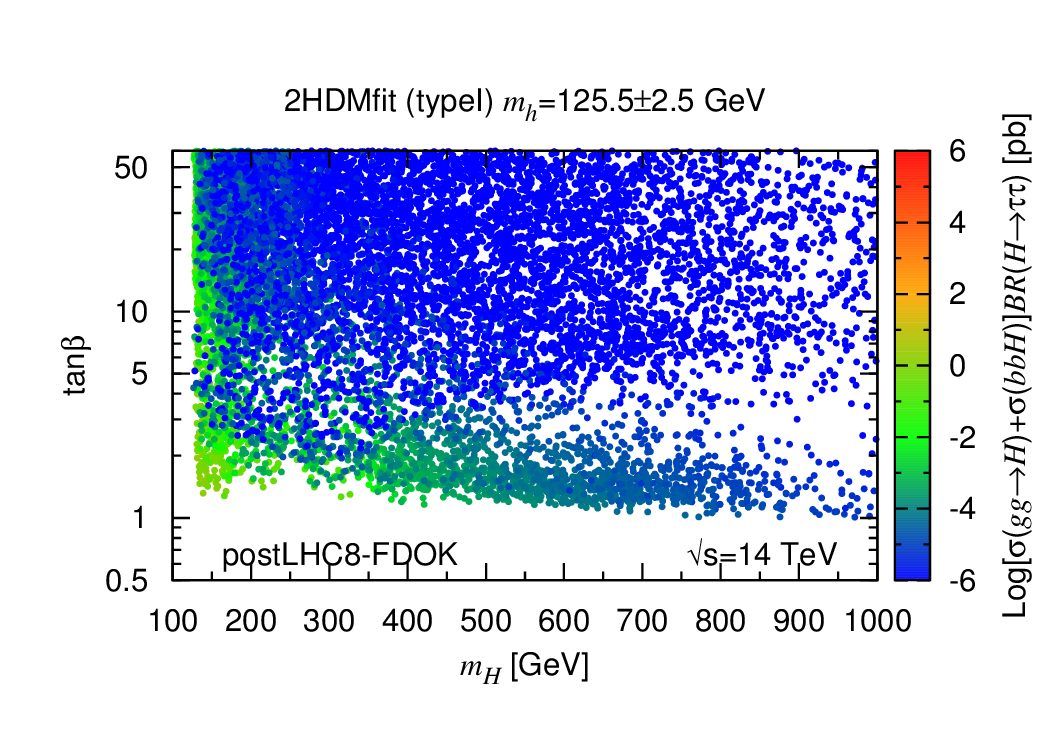}
\includegraphics[width=0.49\textwidth]{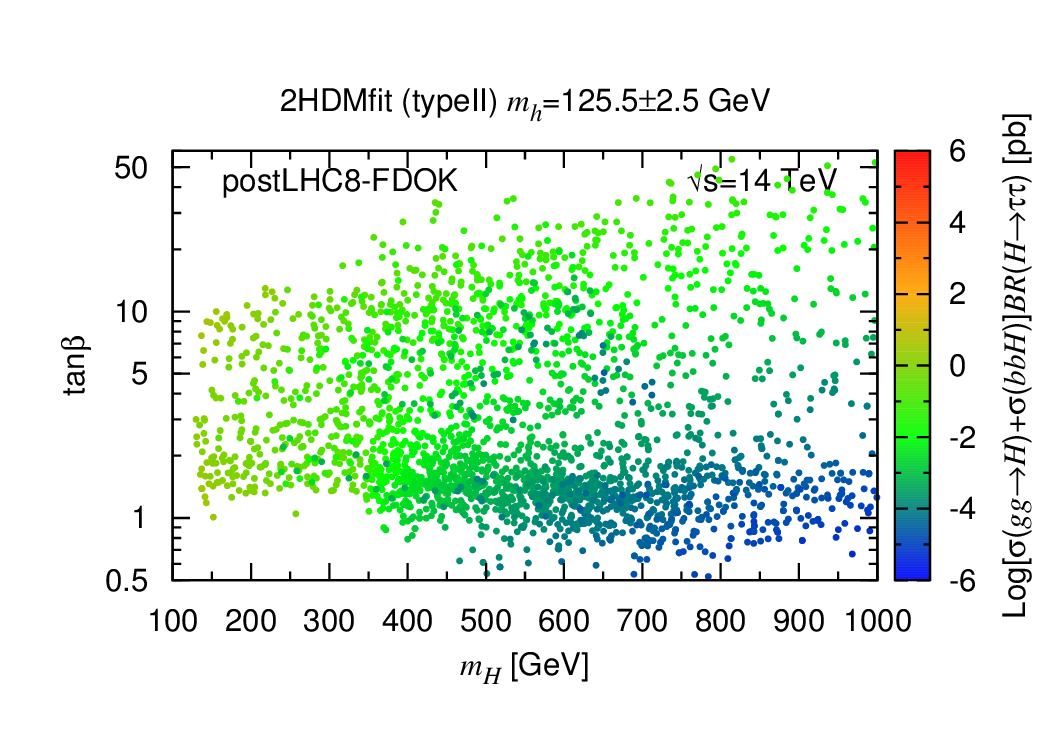}\\[-5mm]
\includegraphics[width=0.49\textwidth]{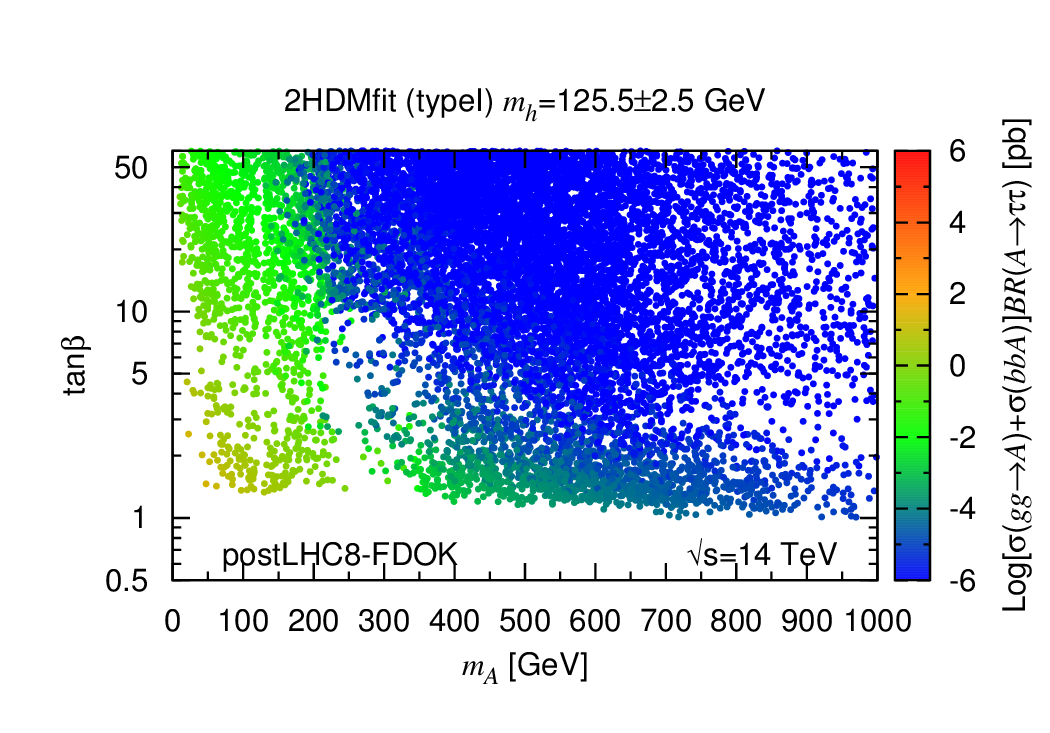}
\includegraphics[width=0.49\textwidth]{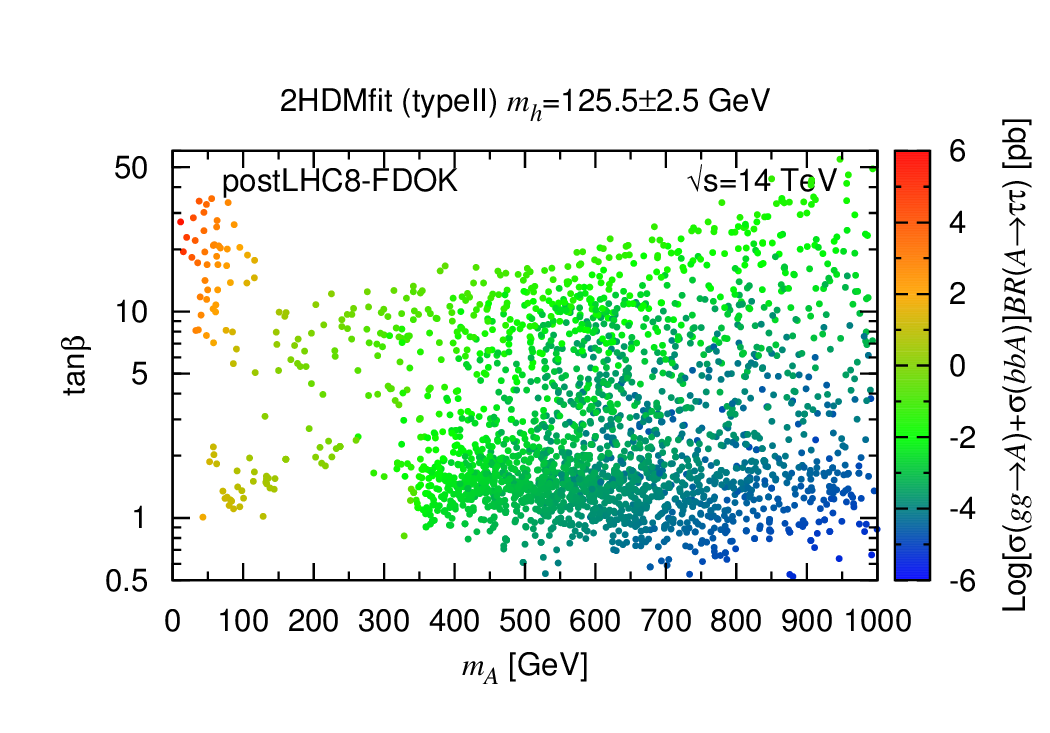}
\end{center}\vspace*{-8mm}
\caption{Same information as in Fig.~\ref{tautau14} but in the $\tan\beta$ vs.\ $\mH$ (top row) and $\tan\beta$ vs.\ $\mA$ (bottom row) planes with the 14~TeV cross sections color coded as indicated by the scales on the right of the plots. Only FDOK points are shown.
}
\label{tautau14alt}
\end{figure}

In the case of the Minimal Supersymmetric Standard Model (MSSM), which is a special case of a Type~II model, limits on $H,A\to \tau\tau$ are often presented in the $\tan\beta$ vs.\ $\mA$ plane. For the sake of comparison, we show in Fig.~\ref{tautau14alt} the $H,A\to \tau\tau$ rates (in pb) in the $\tan\beta$ vs.\ $\mH$ (top row) and $\tan\beta$ vs.\ $\mA$ (bottom row) planes for both Type~I and Type~II, plotting, however, only points with $\sigma>10^{-6}$~pb.  We note a very interesting difference with the MSSM case.  In the 2HDMs, it is possible to have small $\mha$ independent of the other Higgs masses. Further, small $\mha$ can escape LEP limits provided, in particular, that  $hA$ production is sufficiently suppressed. This is natural in the present case to the extent that $\sbma\sim 1$ for a SM-like $h$ given that the $ZAh$ coupling is $\propto \cbma$.  This is the origin of the points in the bottom right plots of Figs.~\ref{tautau14} and \ref{tautau14alt} with $\mha\lsim 100\gev$ and a very large cross section.  (Note that CMS and ATLAS 8 TeV constraints on the $\tanb$ vs.\ $\mha$ plane (not shown) do not exist below $\sim 90\gev$. Furthermore, we have explicitly checked that these limits do allow the few points shown with $\mha\gsim 90\gev$.)

As an aside, we note that in both the $\hh$ and $\ha$ cases the $\mu\mu$ final state rates are obtained by simply multiplying by the relevant ratio of branching ratios, $\br(H~{\rm or}~ A\to \mu\mu)/\br(\hh~{\rm or}~ \ha\to \tau\tau)$, which is essentially independent of $\tanb$ in either Type~I or Type~II with a value of order $3.5\times 10^{-3}$.  Looking at Fig.~\ref{tautau14}, it would appear that prospects for detecting the $\hh$ and $\ha$ in the $\mu\mu$ final state are significant for $\mha$ and $\mhh$ below the top threshold, especially in the case of $\ha\to\mu\mu$ in Type~II when $\mha\lsim 150\gev$.

\begin{figure}[t]
\begin{center}
\includegraphics[width=0.49\textwidth]{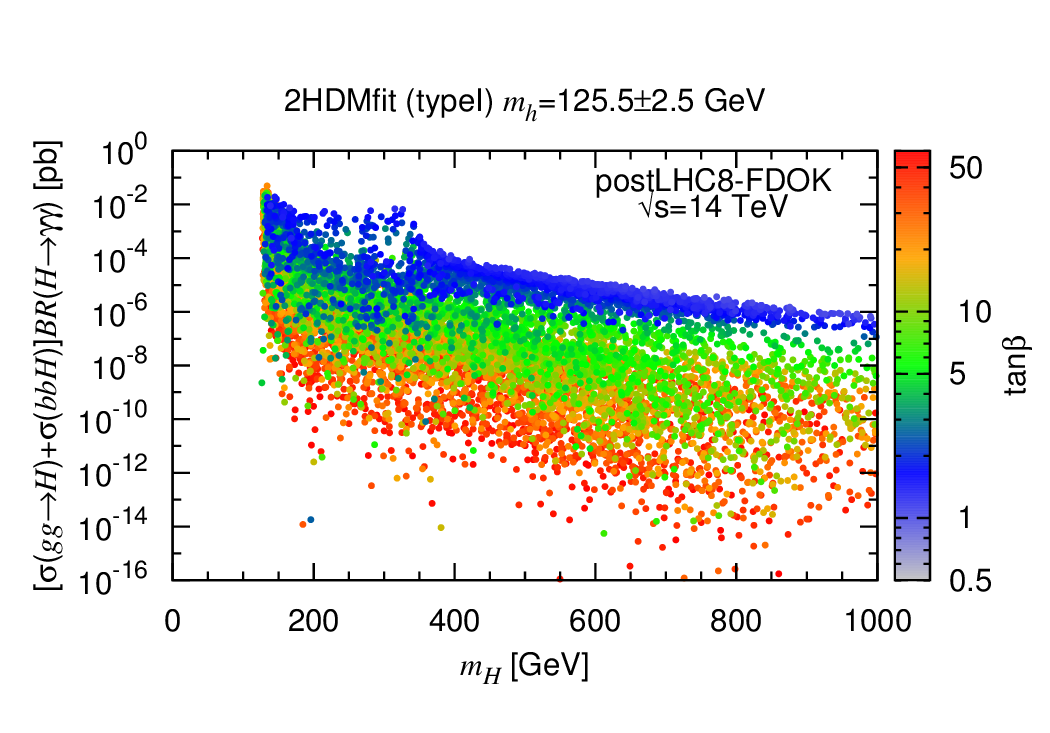}
\includegraphics[width=0.49\textwidth]{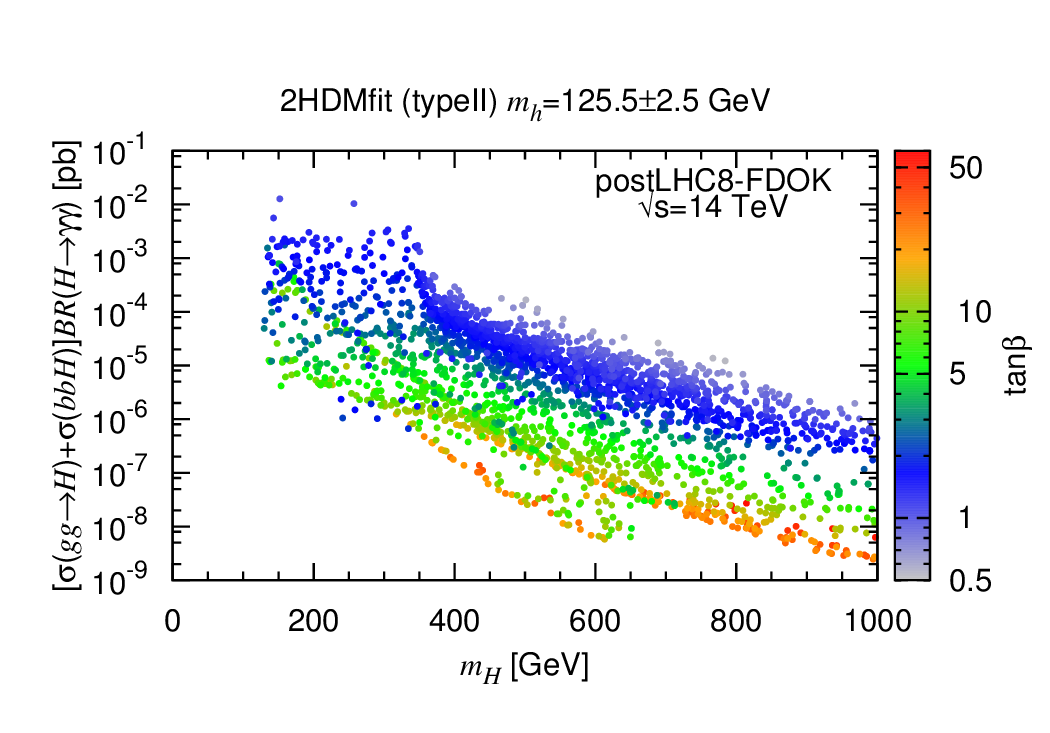}
\includegraphics[width=0.49\textwidth]{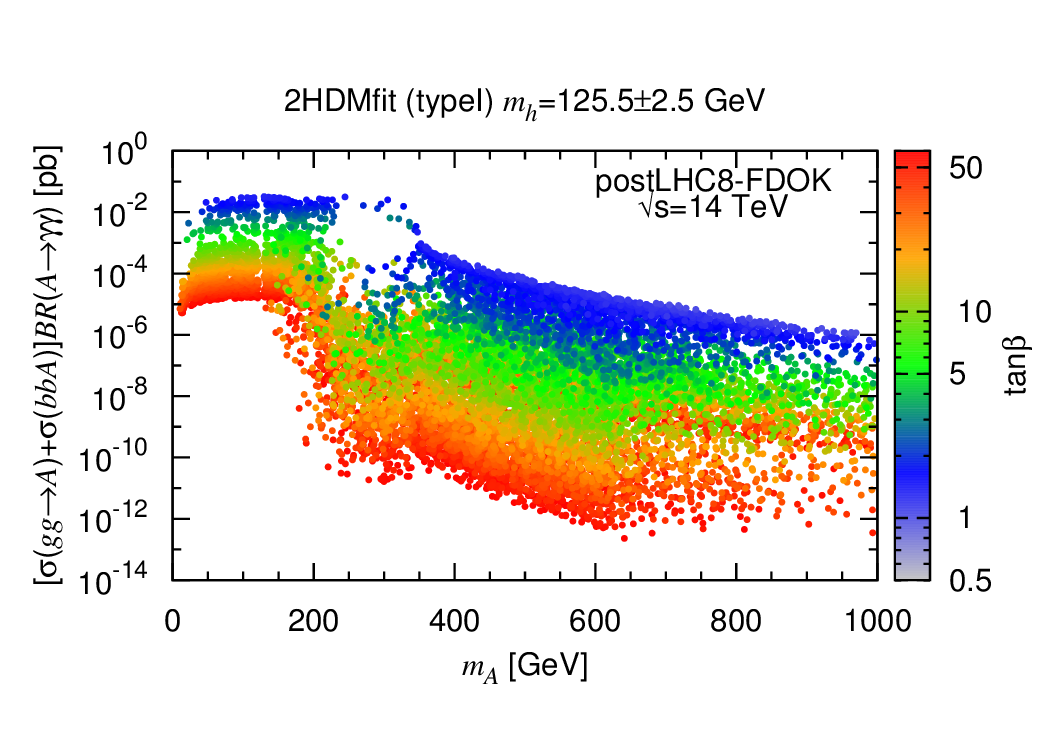}
\includegraphics[width=0.49\textwidth]{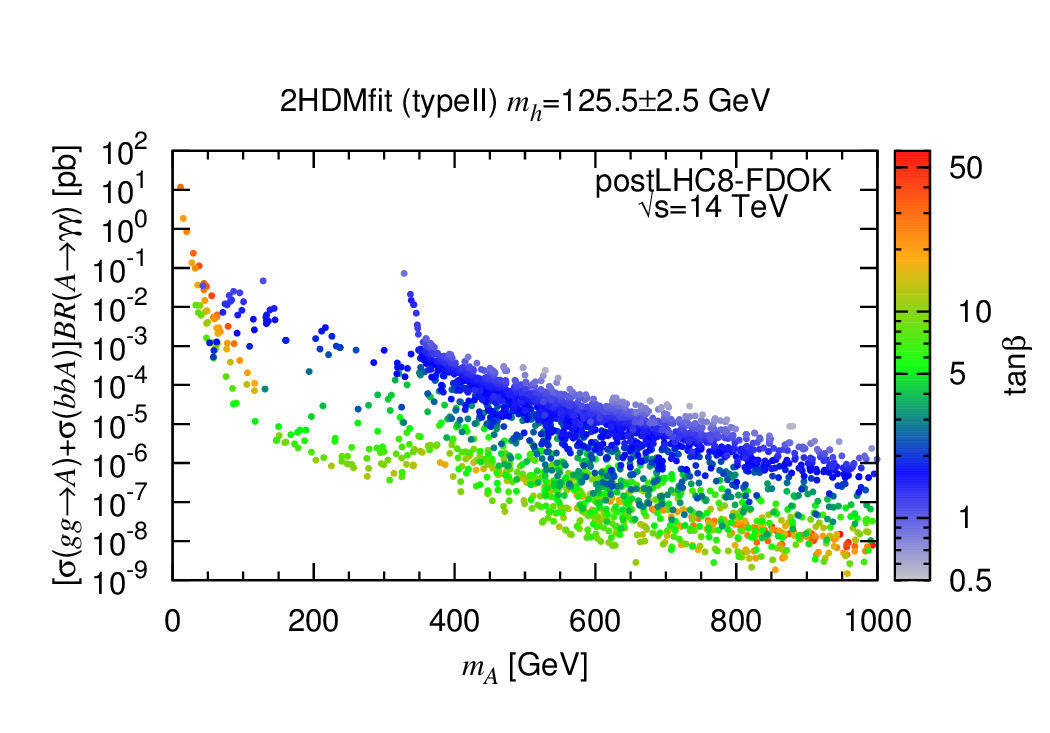}
\end{center}\vspace*{-5mm}
\caption{Scatterplots of $[\sigma(gg\to H)+\sigma(bbH)]\br(H\to \gam\gam)$ as function of $\mH$ (top row)  and 
$[\sigma(gg\to A)+\sigma(bbA)]\br(A\to \gam\gam)$ as function of  $\mA$ (bottom row), for postLHC8-FDOK points 
with  $\mh\sim 125.5\gev$. The values of $\tan\beta$ are color coded as indicated by the scales on the right of the plots.
}
\label{gamgam14}
\end{figure}

Corresponding results for the $H\to \gam\gam$ and $A\to\gam\gam$ final states are shown in Fig.~\ref{gamgam14}. Aside from $\mha\lsim 50\gev$ in Type~II, the largest $\sigma \times \br$'s are  of order 0.05 pb, with much lower values being more typical. If the $\gam\gam$ continuum background is sufficiently small,  $\sigma \times \br$ values as low as $10^{-3}-10^{-4}\pb$ might well be observable, although it must be kept in mind that the total width of the $H$ or $A$ will be of order several to a few tens of GeV.

In the preceding plots, we have not displayed the impact of future $h$ measurements that lie  within SM$\pm15\%$, SM$\pm 10\%$ and SM$\pm5\%$.  In the case of Type~I, agreement with the SM of  $\pm 10\%$ or better implies that $\mha\lsim 80\gev$ is excluded (whereas at the postLHC8 level very low $\mha$ is allowed). This is apparent from examining the reach in $\mha$ in Fig.~\ref{hhhcoup}. In the case of $\mhh$, which already must lie above $\sim 125.5\gev$, there is almost no impact as increasing agreement with the SM is required.   For Type~II, the impact is more varied.  In the case of the $\ha$, as one moves through SM$\pm 15\%,10\%,5\%$ fewer and fewer points are found at lower $\mha$, as can again be read from Fig.~\ref{hhhcoup}, but determining precise boundaries would require dedicated scanning.  In the case of the $\hh$, SM$\pm15\%$ and $\pm 10\%$ do not restrict $\mhh$ beyond the postLHC8 range, but heavier $\mhh$ is preferred by SM$\pm 5\%$.

\begin{figure}[t]
\begin{center}
\includegraphics[width=0.49\textwidth]{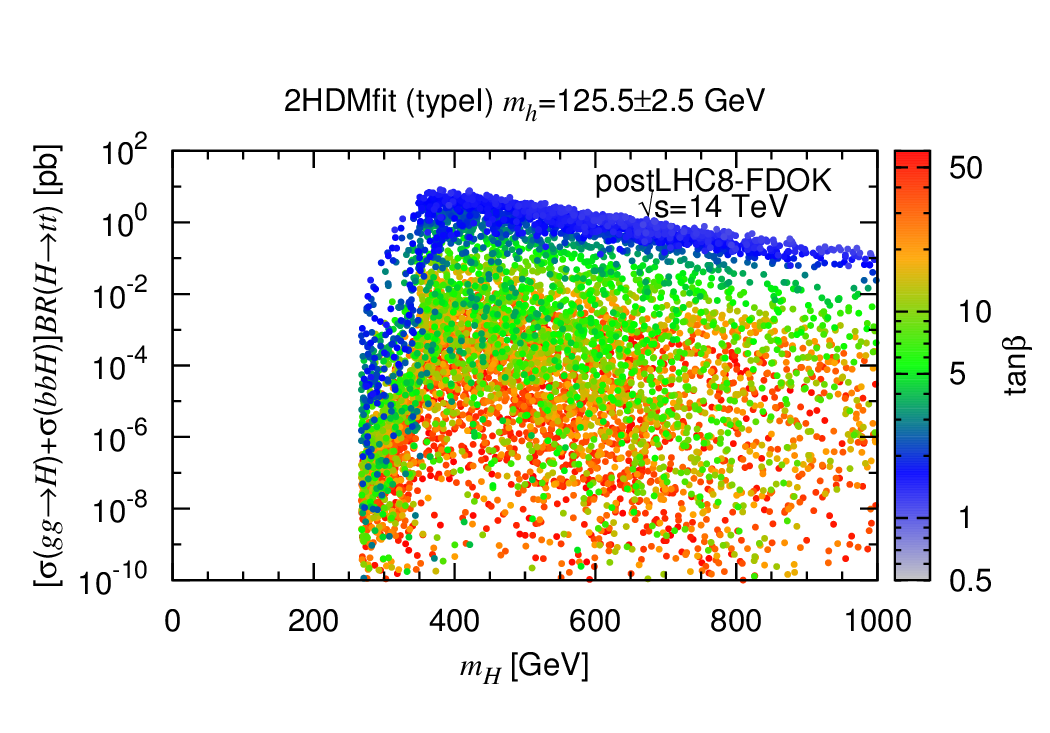}
\includegraphics[width=0.49\textwidth]{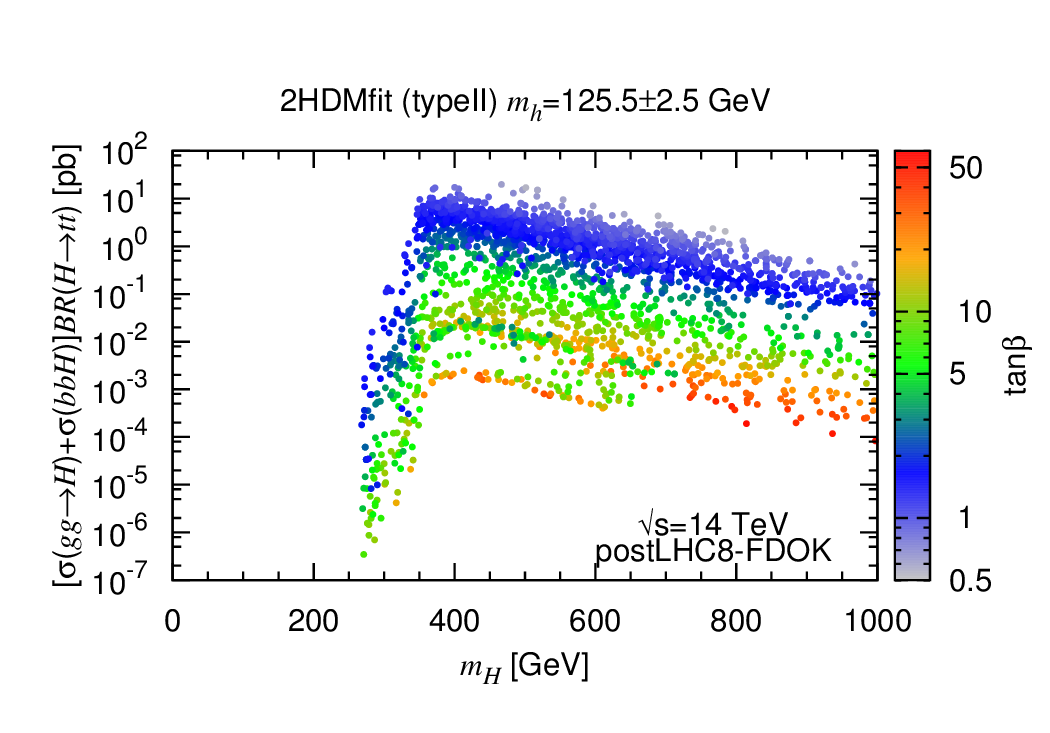}
\includegraphics[width=0.49\textwidth]{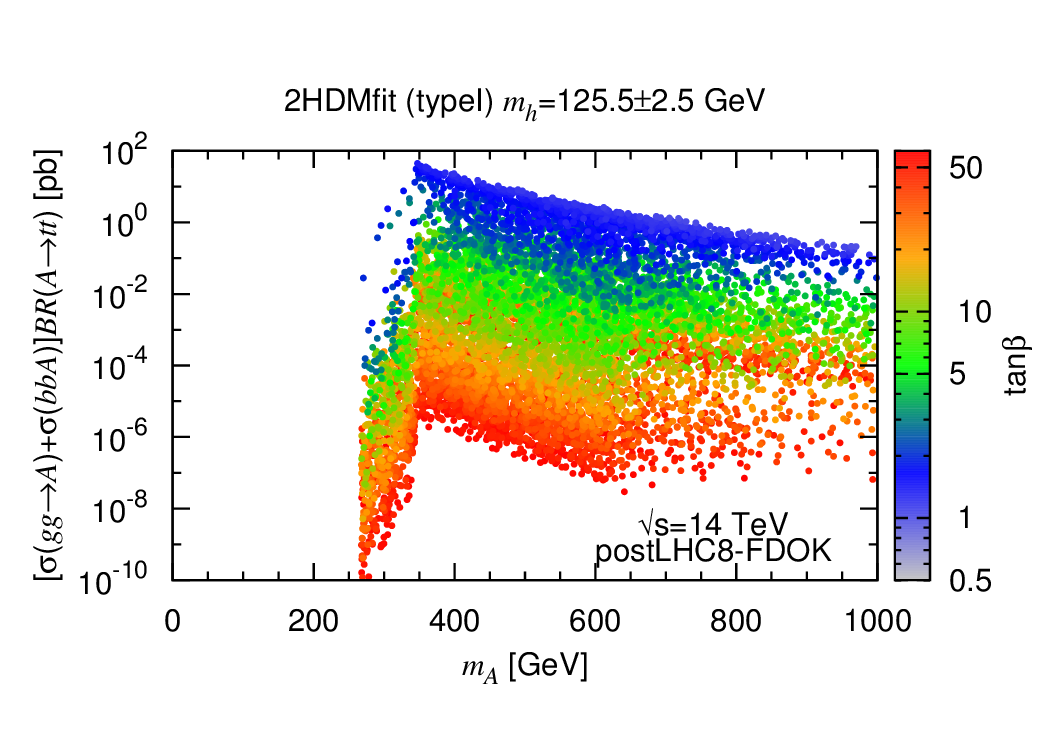}
\includegraphics[width=0.49\textwidth]{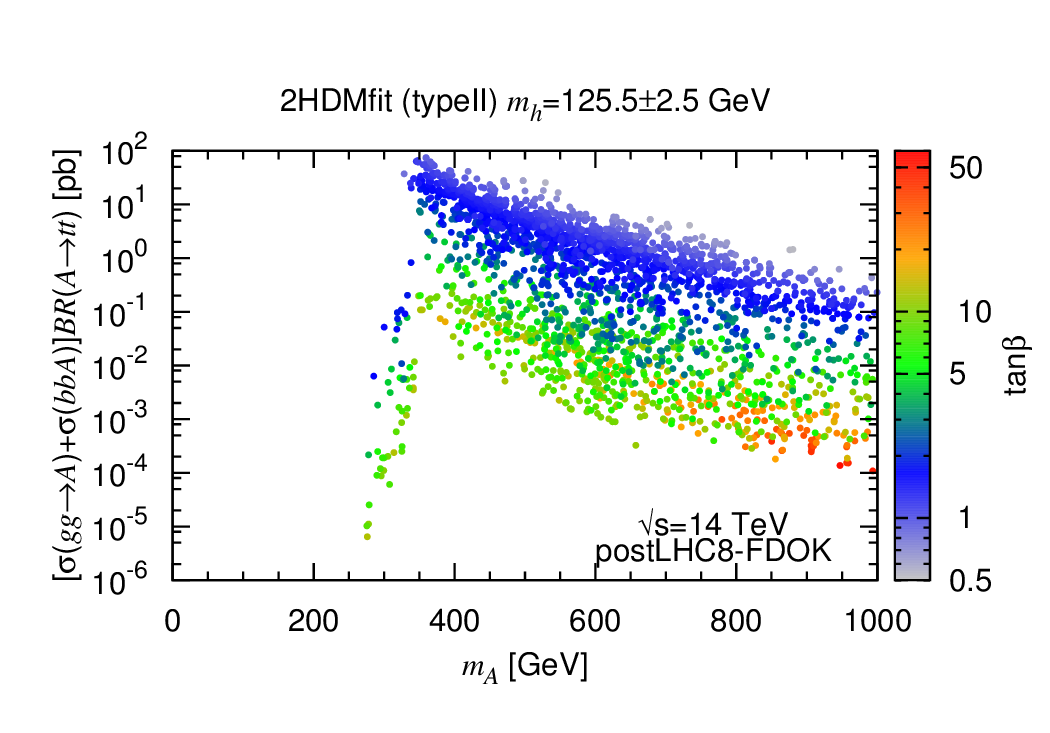}
\end{center}\vspace*{-5mm}
\caption{Scatter plots of $[\sigma(gg\to H)+\sigma(bbH)]\br(H\to t\bar t)$ as function of $\mH$ (top row)  and 
$[\sigma(gg\to A)+\sigma(bbA)]\br(A\to t\bar t)$ as function of  $\mA$ (bottom row) at $\sqrt{s}=14\tev$, 
for postLHC8-FDOK points with  $\mh\sim 125.5\gev$. 
The values of $\tan\beta$ are color coded as indicated by the scales on the right of the plots.
}
\label{tt14}
\end{figure}

Of course, once $\mha$ or $\mhh$ is above the $t\anti t$ threshold, the rates in the $t\bar t$ final state will be of great interest.  These are shown in Fig.~\ref{tt14}. Large $\sigma\times \br$ values 
are certainly possible, but so also are very small values, although in the case of Type~II the smallest values found at $\mhh$ or $\mha$ of order 1 TeV is $\sim 10^{-4}\pb$.  This latter might be detectable for full Run2 luminosity of $L=300\fbi$, and is certainly of great interest for the high-luminosity run of the LHC which might accumulate $L=3000\fbi$.

\begin{figure}[t]
\begin{center}
\includegraphics[width=0.49\textwidth]{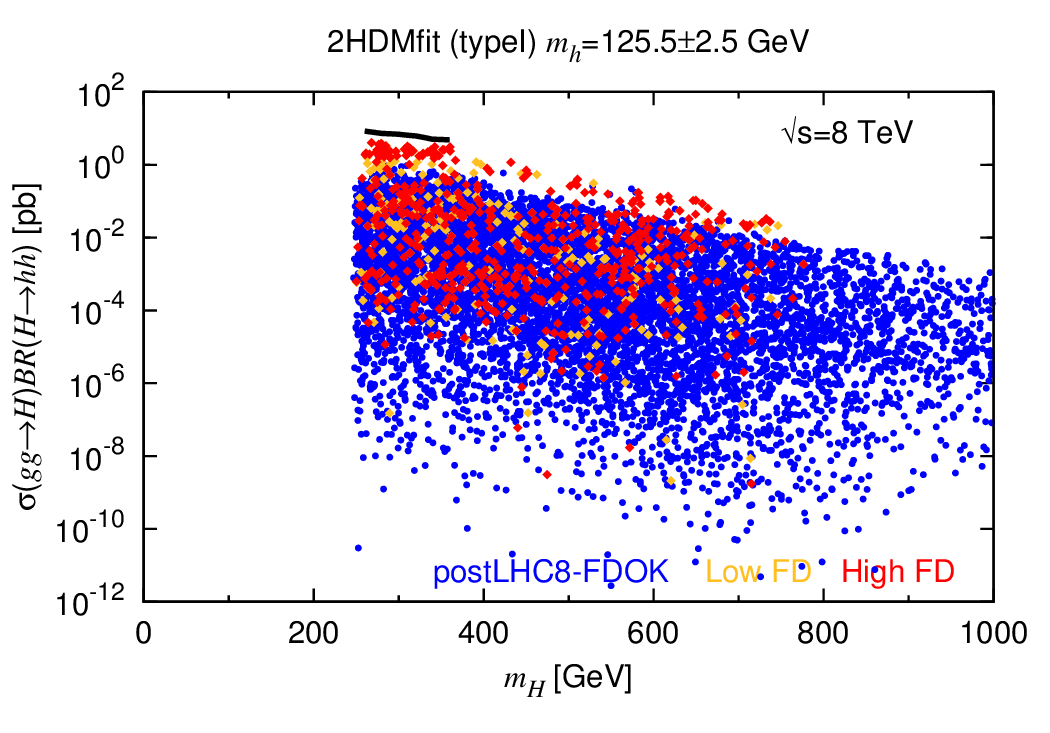}
\includegraphics[width=0.49\textwidth]{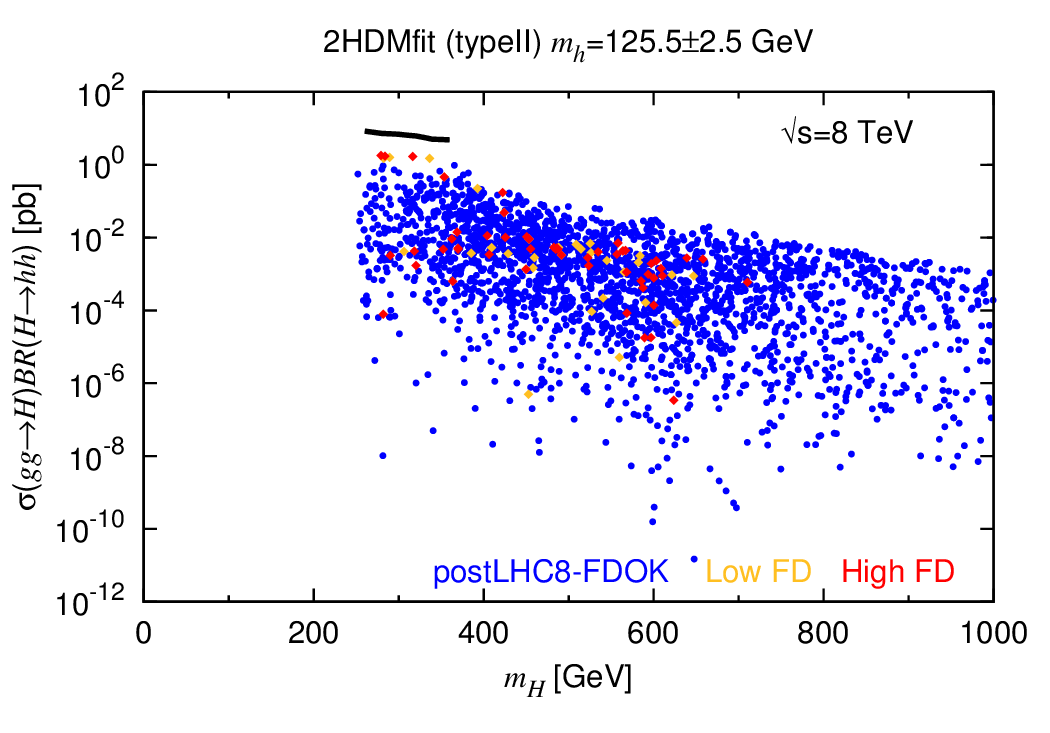}
\includegraphics[width=0.49\textwidth]{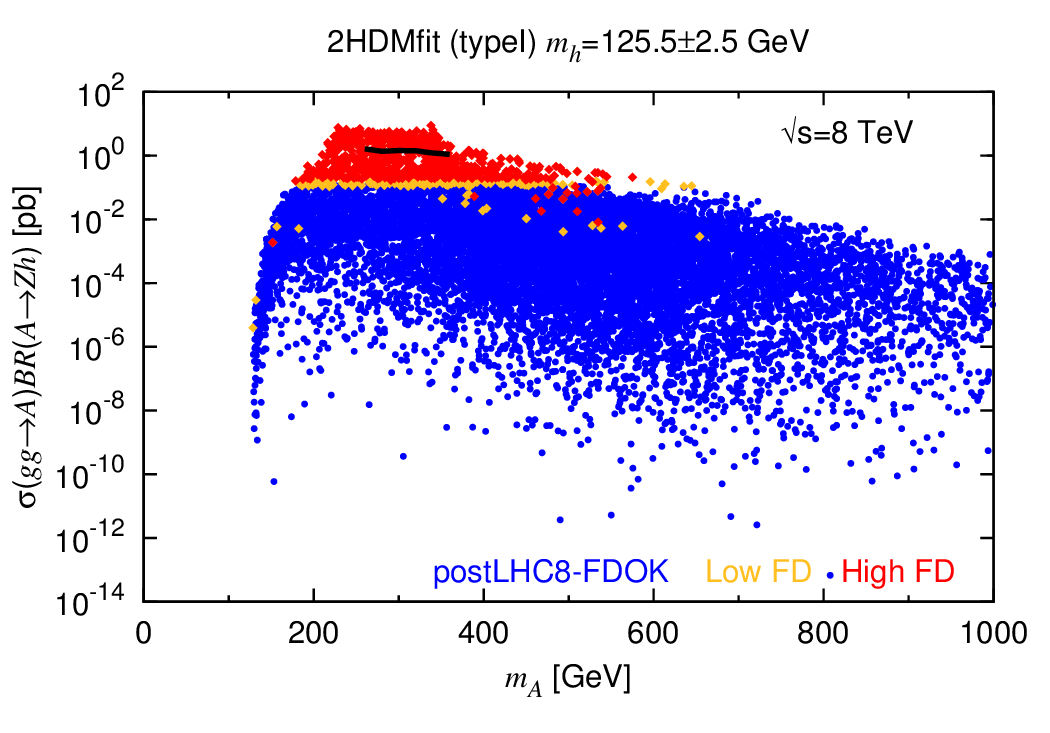}
\includegraphics[width=0.49\textwidth]{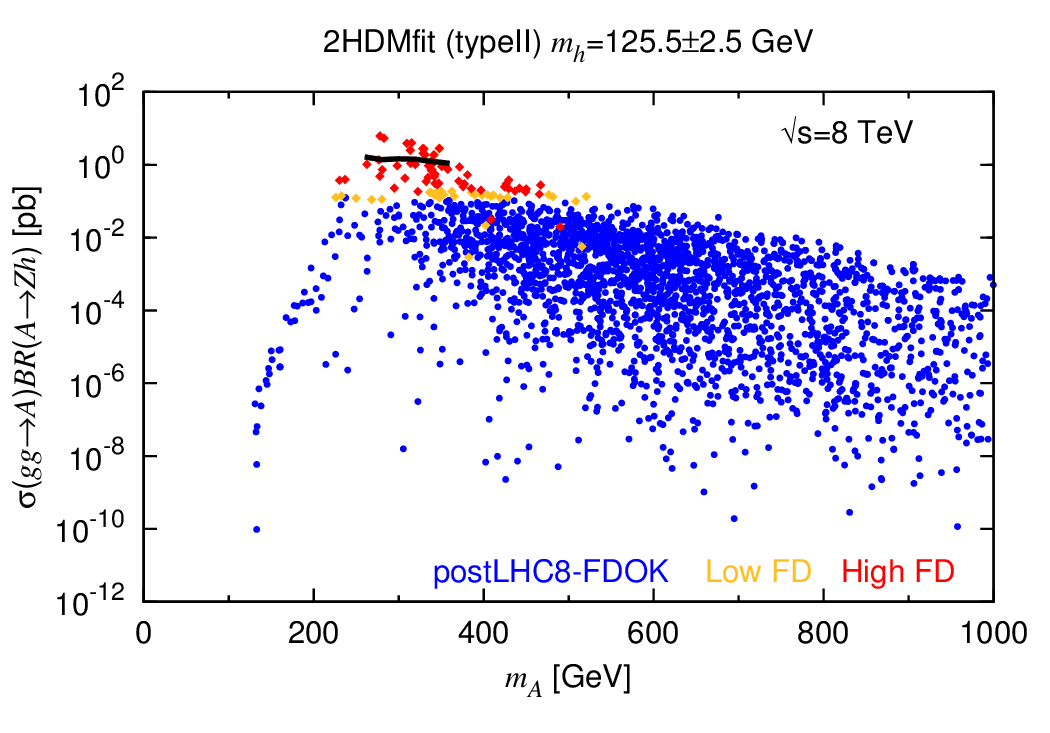}
\end{center}\vspace*{-5mm}
\caption{We plot $\sigma(gg\to H)\br(H\to hh)$ and $\sigma(gg\to A)\br(A\to Zh)$ as functions of $\mH$ and $\mA$, respectively, for Type~I (left) and Type~II (right) 2HDMs. Blue points fulfill all constraints, including the FDOK requirement. Also shown are the points that  have Low FD (in orange) and High FD (in red) as defined in Section~\ref{fdsection}.  The black lines show the current limits from the CMS analysis of Ref.~\cite{HIG13025}.}
\label{hhzh8}
\end{figure}

Perhaps most interesting are the rates for $H\to hh$ and $A\to Zh$.  First, we show in Fig.~\ref{hhzh8} the results for ggF at $\rts=8\tev$.  These results should be compared to the ggF limits obtained recently in~\cite{HIG13025}, which are shown as  black lines in the plots of Fig.~\ref{hhzh8}; see also projections at the 14~TeV LHC in~\cite{Coleppa:2014hxa}. Moreover, we show the points that have significant FD (as discussed in detail in the following subsection). We distinguish between points for which the amount of FD violates the FDOK criteria, but is still moderate in size (orange points, labelled ``Low FD''), as defined below in Section~\ref{fdsection}, and points with a high level of FD (red points, labelled ``High FD'').   In the case of $H\to hh$, none of the points are excluded by~\cite{HIG13025}, not even the High FD ones.
In the case of $\sig(gg\to A)\br(A\to Zh)$, on the other hand, the experimental limits exclude a significant fraction of the High FD points in the case of Type~I models. In the case of Type~II, only a few of the High FD points are excluded.
Nonetheless, the nearness of the black line limits to the High FD points indicates that  we should not be surprised if the above two processes are the dominant sources of feed down, as we shall describe in the next section. 

From Fig.~\ref{hhzh8} we see that the points with unacceptable FD levels are prominent in the $\mhh\sim 250-350\gev$ region and, especially,  in the $\mha\sim200-350\gev$ region. Looking back at, for example, Figs.~\ref{tautau14} and \ref{gamgam14}, which include only postLHC8-FDOK points, we observe corresponding ``holes" and depleted regions at low $\tanb$ in precisely these mass regions.  FD is largest at low $\tanb$ where  $\sigma(gg\to A,H)+\sigma(bbA,bbH)$ is largest, as is especially true in the case of Type~I models. In these figures, the surviving points that surround or outline the depleted regions are ones with very small $\cbma$ which implies very small $AZh$ coupling and therefore very small $\br(A\to Zh)$ ($A\to Zh$ being the feed down mechanism of primary importance, see Section~\ref{fdsection}).

It is interesting to note that to the extent that the $gg\to A\to Zh$ process contaminates direct $Zh$ production whereas feed down to the $Wh$ final state is substantially smaller (see later discussion), one might observe an apparent violation of custodial symmetry when extracting the $C_Z$ and $C_W$ effective coupling strengths independently of one another using $Zh$ and $Wh$ production, respectively. The fact that the direct limits in Fig.~\ref{hhzh8} are above the points that have large feed down indicates that apparent custodial symmetry violation might actually be a more sensitive probe of the presence of $H\to hh$ and, especially, $A\to Zh$   decays.

\begin{figure}[t!]
\begin{center}
\includegraphics[width=0.49\textwidth]{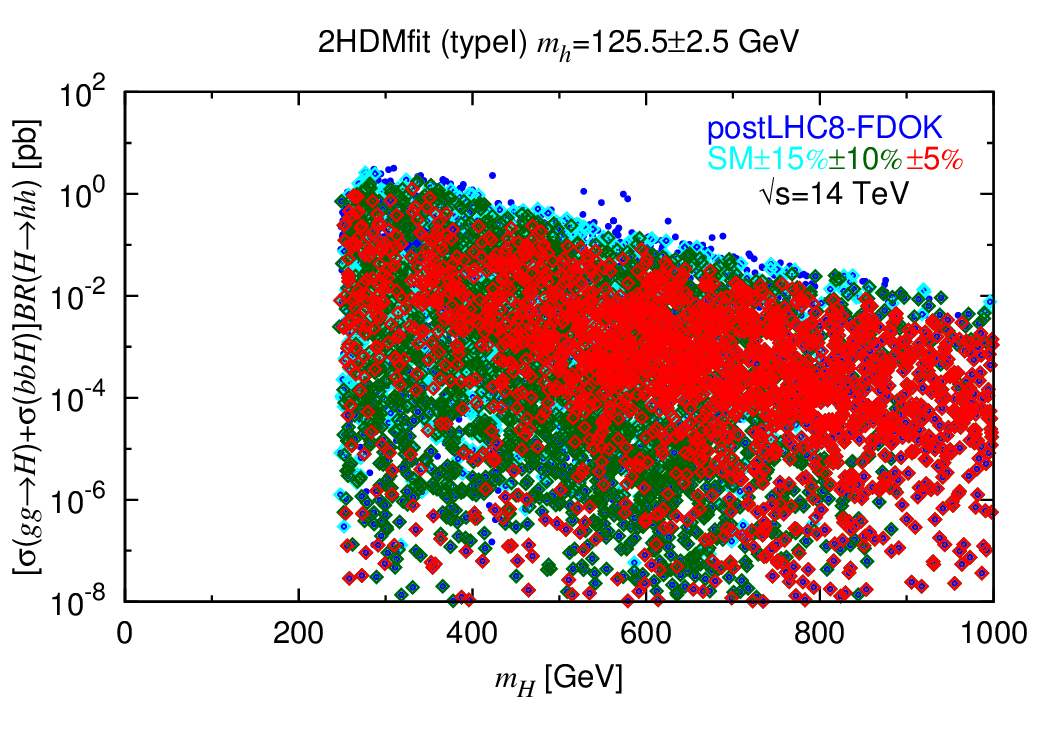}
\includegraphics[width=0.49\textwidth]{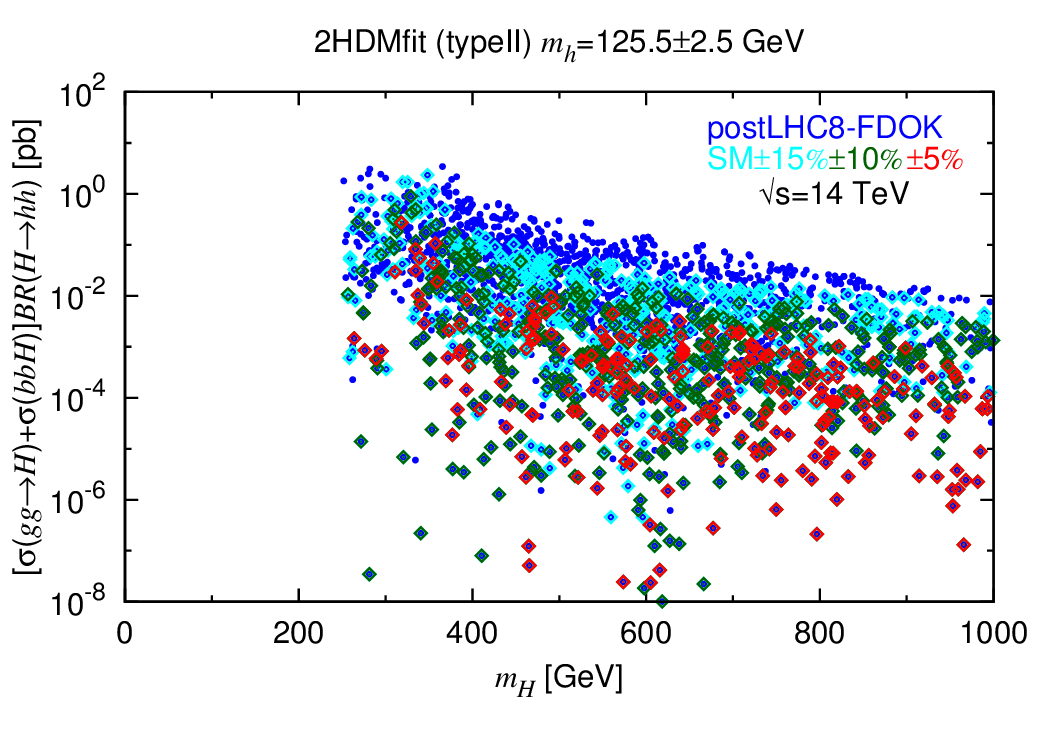}
\includegraphics[width=0.49\textwidth]{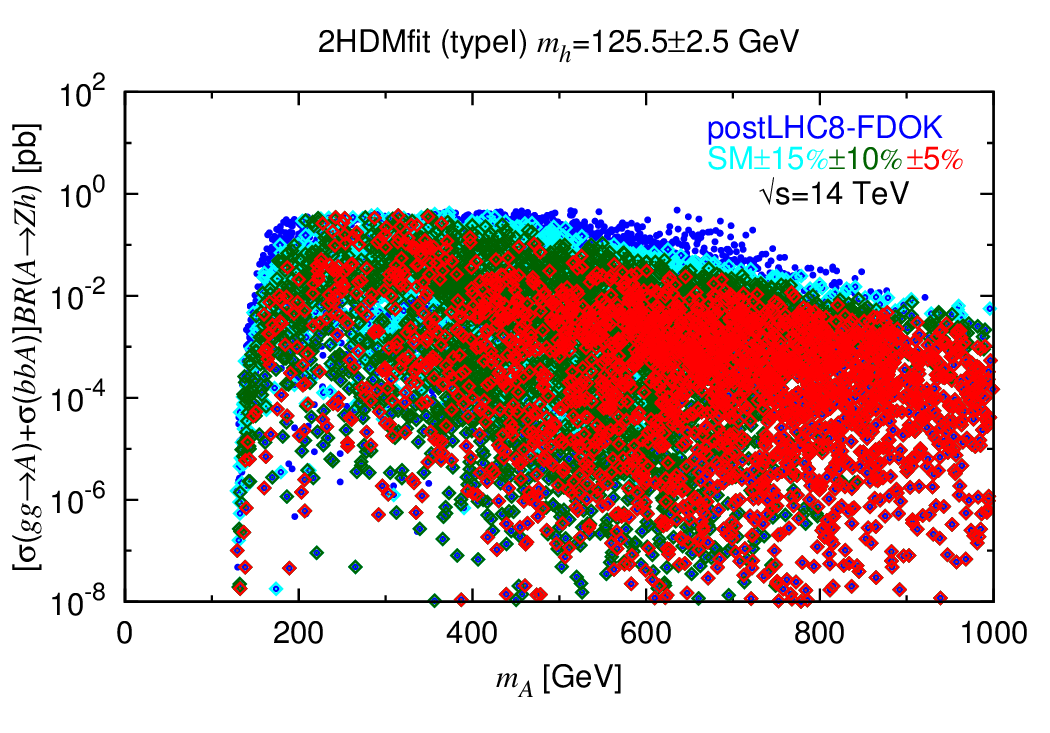}
\includegraphics[width=0.49\textwidth]{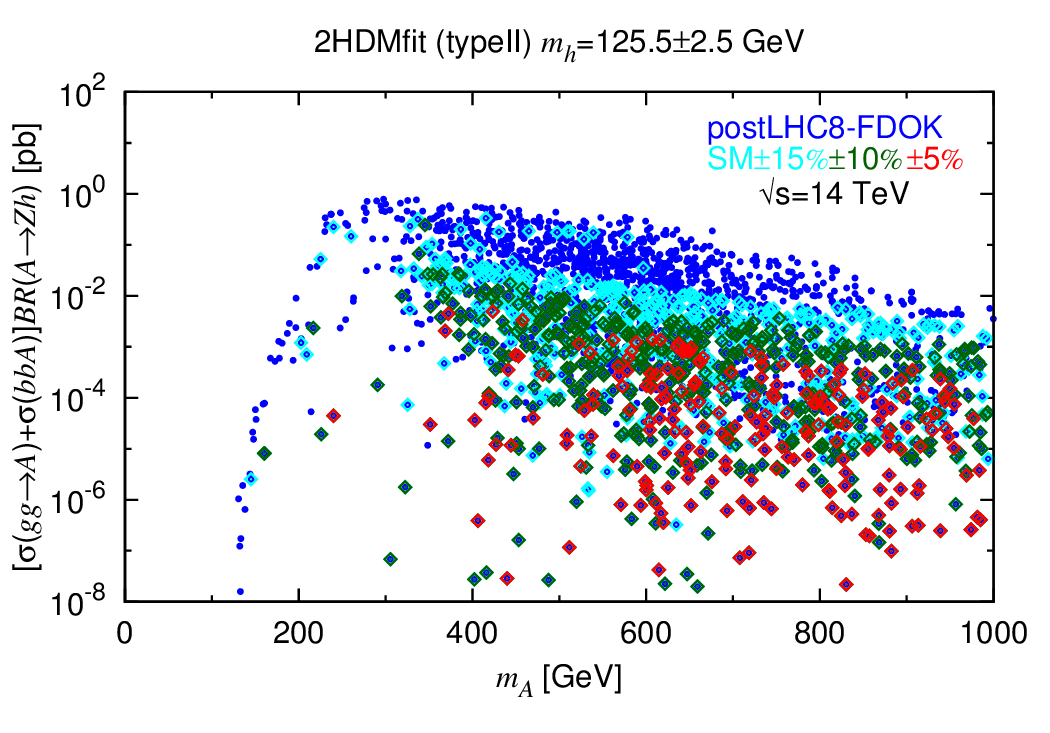}
\end{center}\vspace*{-5mm}
\caption{We plot $[\sigma(gg\to H)+\sigma(bbH)]\br(H\to hh)$  and $[\sigma(gg\to A)+\sigma(bbA)]\br(A\to Zh)$ as functions of $\mH$ and $\mA$, respectively, for Type~I (left) and Type~II (right) 2HDMs. For this figure, only FDOK points are shown. Implications of various levels of precision for future $h$ measurements are displayed. Color scheme as in Fig.~\ref{cbmavstbfuture}.  Values of $\sigma\times\br$ below $10^{-8}$ are not plotted. 
}
\label{hhzh14}
\end{figure}

In Fig.~\ref{hhzh14} we show results for the $hh$ and $Zh$ final states for $\rts=14\tev$, this time indicating the impact of SM$\pm15\%$, $\pm10\%$, $\pm5\%$ requirements. 
We note that if the $h$ measurements approach SM values, then this will limit only somewhat the maximum values achievable for the cross section in the $hh$ and $Zh$ final states in the case of Type~I models --- which means that there is a significant, although not large, probability of seeing the $H\to hh$ and $A\to Zh$ final states in gluon fusion and in associated production with $b$ quarks.  However, in the case of Type~II models, increasingly SM-like $h$ results imply much smaller cross sections than those shown (as allowed by current Higgs fitting for both $hh$ and $Zh$ final states).

\subsection{\bf \boldmath Feed down  of heavier Higgs to the $125.5\gev$ $h$} \label{fdsection}

Let us now turn to the issue of whether feed down from heavier Higgs decaying to the $\sim 125.5\gev$ Higgs  could invalidate our fitting. All the blue points plotted earlier are such that FD is neglectable in all channels of interest. The reason to be  concerned is that 
heavy Higgs bosons have a propensity for decaying to a vector boson plus a Higgs boson or to two Higgs bosons. The most direct cases are $H\to hh$ and $A\to Zh$, but there are also chains like $H\to AA$ followed by $A\to Zh$ or $H\to \hp\hm$ with $\hpm\to \wpm h$ and so forth.  The full formalism for the feed down calculations is given in Appendix~\ref{app:feeddown}.  For the present purposes, we will consider the most important FD sources and associated ratios
\bea
\mu^{\rm FD}_{{\rm ggF }h+{\rm bb}h}&\equiv&{\sum_{{\cal H}=H,A} \left( \sigma_{{\rm ggF}{\cal H}}+\sigma_{bb{\cal H}} \right) P_{\rm FD}({\cal H}\to h+X)\over \sigma_{{\rm ggF}h}+\sigma_{bbh}}\,,\\
\mu^{\rm FD}_{Zh}&\equiv&{\sigma_{{\rm ggF}A}\br(A\to Zh)\over \sigma_{Zh}}\,,
\eea
where $P_{\rm FD}({\cal H}\to h+X)$ is the net branching ratio to produce one (or more) $h$ in the ${\cal H}=H$ or $A$ decay chains --- see Appendix \ref{app:feeddown}.  Above, 
$\sigma_{{\rm ggF}{\cal H}}$ and $\sigma_{bb{\cal H}}$ refer to the cross sections for $gg\to {\cal H}$ and $bb{\cal H}$ associated production respectively, where $\cal H$ can in the present case be $\hh$ or $\ha$.

We emphasize that the amount of FD is computed  
without accounting for any reduced efficiency for accepting such events into the $125.5\gev$ signal as a result of the experimental cuts used to define the $gg\to h$, $bbh$ or $Z^*\to Zh$ channels. In practice, it could be that the actual FD after the experimental cuts currently employed to define the various channels is considerably smaller than this maximally conservative estimate.

\begin{figure}[t]
\begin{center}
\includegraphics[width=0.49\textwidth]{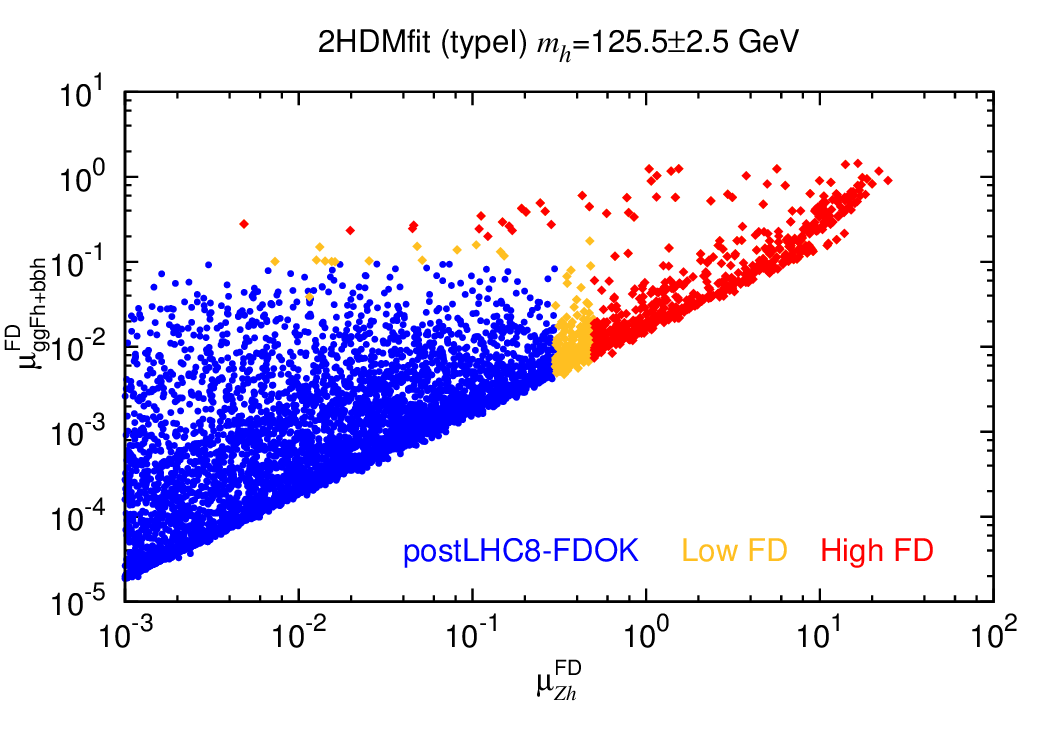}
\includegraphics[width=0.49\textwidth]{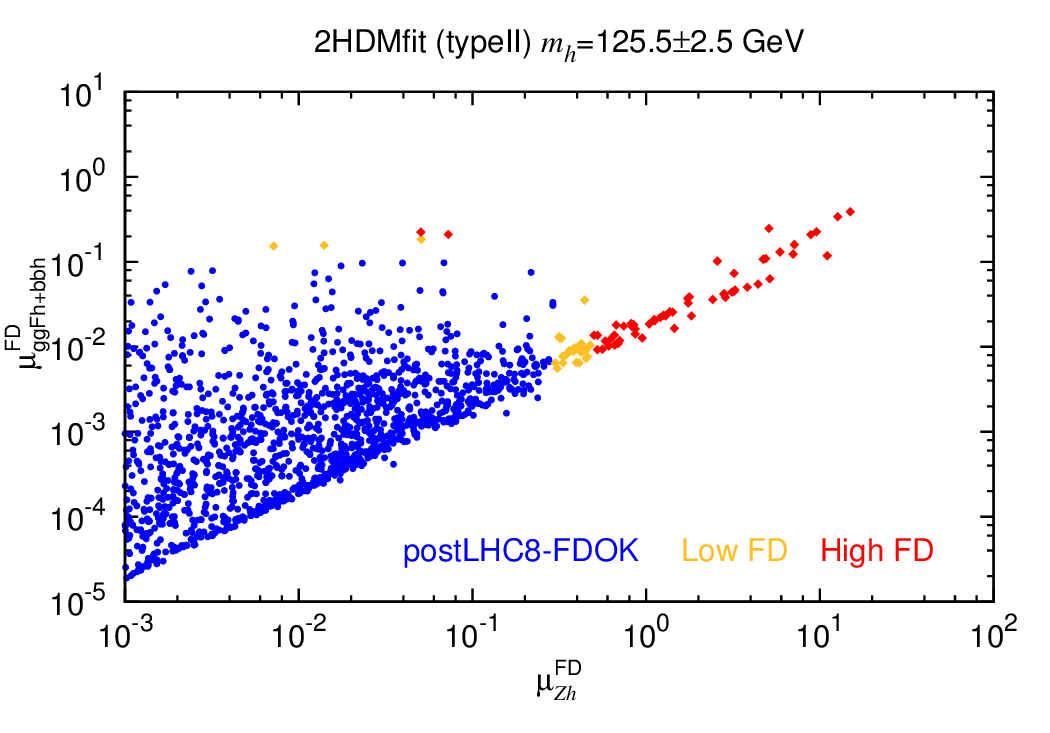}\\
\includegraphics[width=0.49\textwidth]{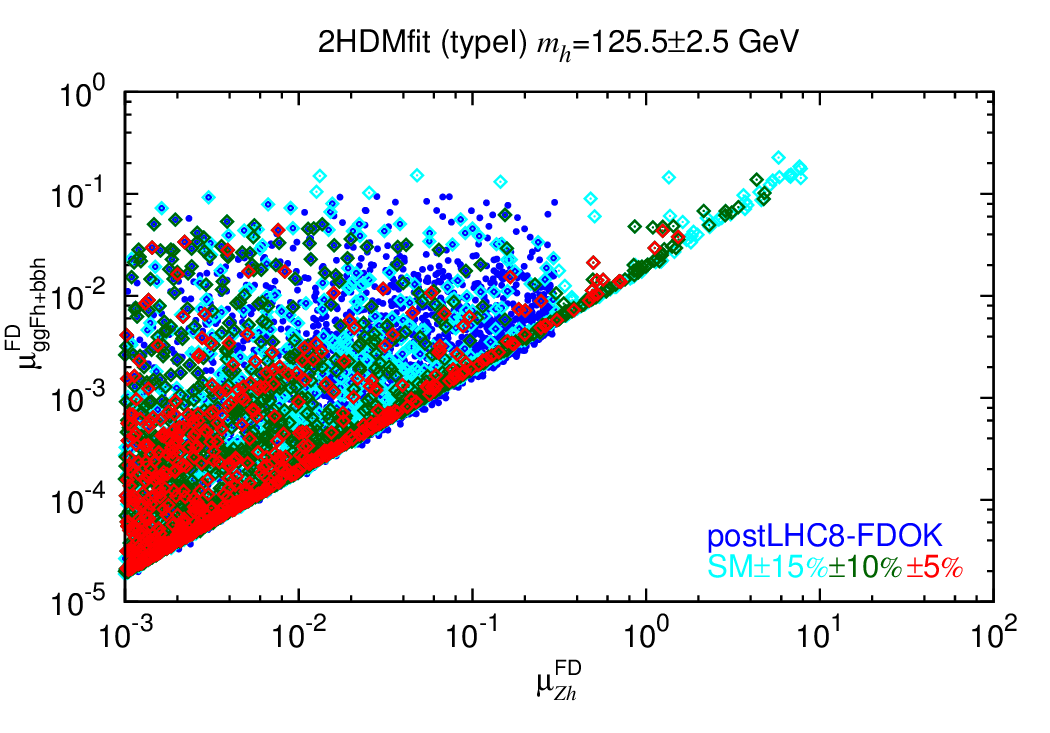}
\includegraphics[width=0.49\textwidth]{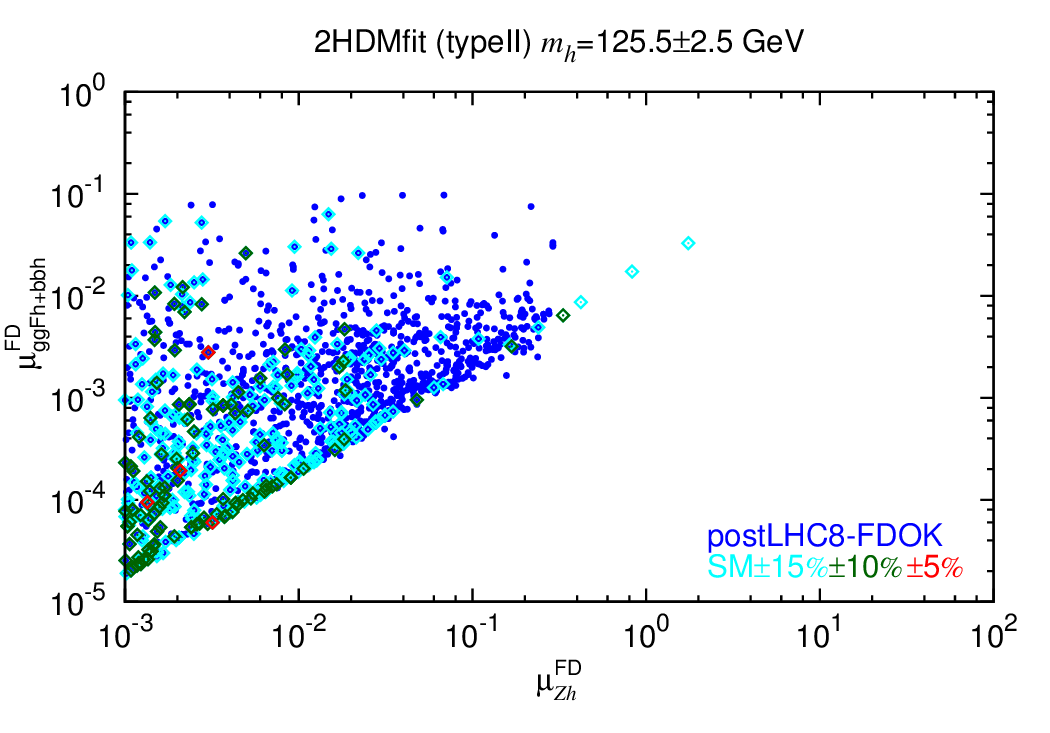}
\end{center}\vspace*{-5mm}
\caption{Scatterplots of $\mu_{{\rm ggF}h+{\rm bb}h}^{{\rm FD}}$ vs.\ $\mu_{Zh}^{\rm FD}$. Top plots illustrate how high the FD fractions can go for postLHC8 points (not imposing any FD limit).  The plots in the lower row illustrate how convergence of the $h$ properties to SM-like values would limit the maximum possible FDs. We display only points with $\mu_{Zh}^{\rm FD}\geq 10^{-3}$  --- there are many points with much lower values.
}
\label{ggfvszh}
\end{figure}

In Fig.~\ref{ggfvszh}, we exhibit a few features of $\mu_{Zh}^{\rm FD}$ and $\mu_{{\rm ggF}h+{\rm bb}h}^{{\rm FD}}$.
The upper two plots 
show the relative importance of $\mu_{Zh}^{\rm FD}$ compared to $\mu_{{\rm ggF}h+{\rm bb}h}^{{\rm FD}}$.  
We observe that the FD to the $Zh$ final state from  $A\to Zh$, $\mu_{Zh}^{\rm FD}$, is almost always the largest due to the large $gg \to A$ production rate compared to the $Z^*\to Zh$ rate that defines the $\mu^h_{V{\rm H}}(Zh)$ ratio that is the fundamental LHC measurement of interest.

We must now ask what amount of feed down is too large in the ggF$h$+bb$h$ and the $Zh$ cases.  This, of course, depends upon the accuracy with which the ggF$h$+bb$h$ and $Zh$ channels are measured. At LHC8, very roughly, ggF$h$+bb$h$ and $Zh$ channels are measured to accuracies of order 15\% and of order 50\%, respectively. Thus, we adopt the following criteria:
\begin{itemize}
\item In order for FD not to affect our fits we should have  $\mu_{{\rm ggF}h+{\rm bb}h}^{{\rm FD}}\leq 0.1$ and $\mu_{Zh}^{\rm FD}\leq 0.3$.  Points satisfying both criteria were already denoted as ``FDOK" above. 
\item We further define ``Low FD" as being cases such that FDOK criteria are violated by virtue of  $0.1 <  \mu_{{\rm ggF}h+{\rm bb}h}^{{\rm FD}}\leq 0.2$ and/or  $0.3<\mu_{Zh}^{\rm FD}\leq 0.5$.  
\item Finally, ``High FD" is defined as  $\mu_{{\rm ggF}h+{\rm bb}h}^{{\rm FD}}>0.2$ and/or  $\mu_{Zh}^{\rm FD}>0.5$.  
\end{itemize}

The lower two plots of Fig.~\ref{ggfvszh} illustrate the fact that as the $125.5\gev$ resonance is shown to be closer and closer to SM-like in all the various channels the maximum amount of FD that is possible is greatly reduced, becoming quite small for the SM$\pm 5\%$ case. Increased precision in the signal strength measurements thus reduces the ``danger" of FD contamination.  

\begin{figure}[t]
\begin{center}
\includegraphics[width=0.49\textwidth]{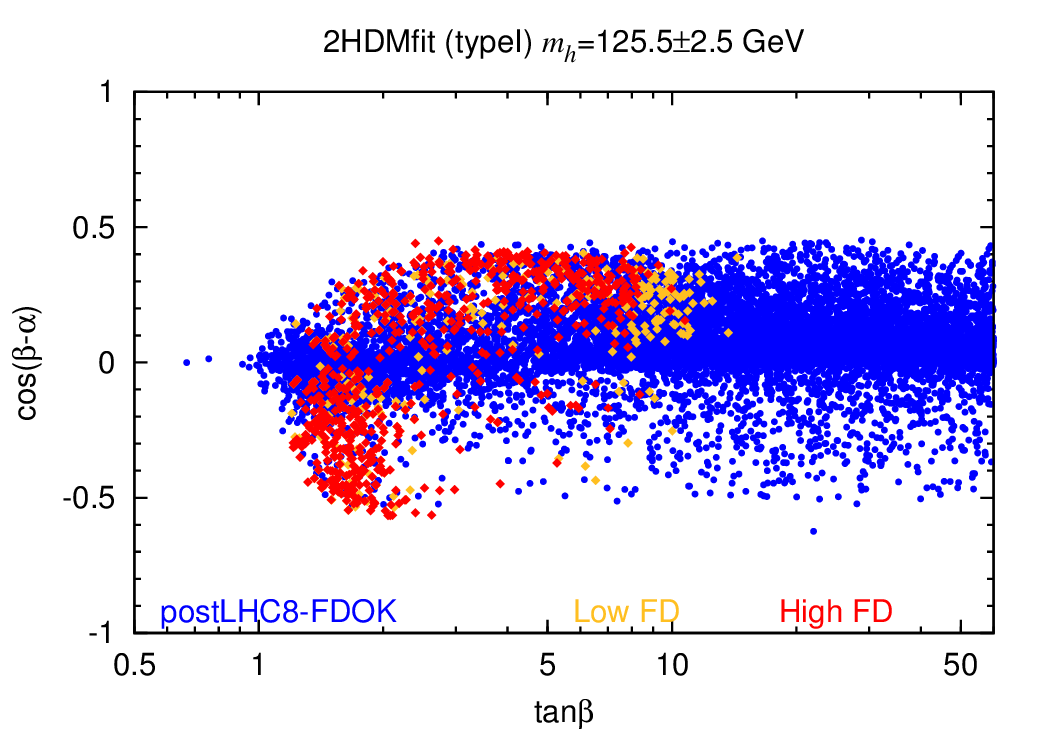}
\includegraphics[width=0.49\textwidth]{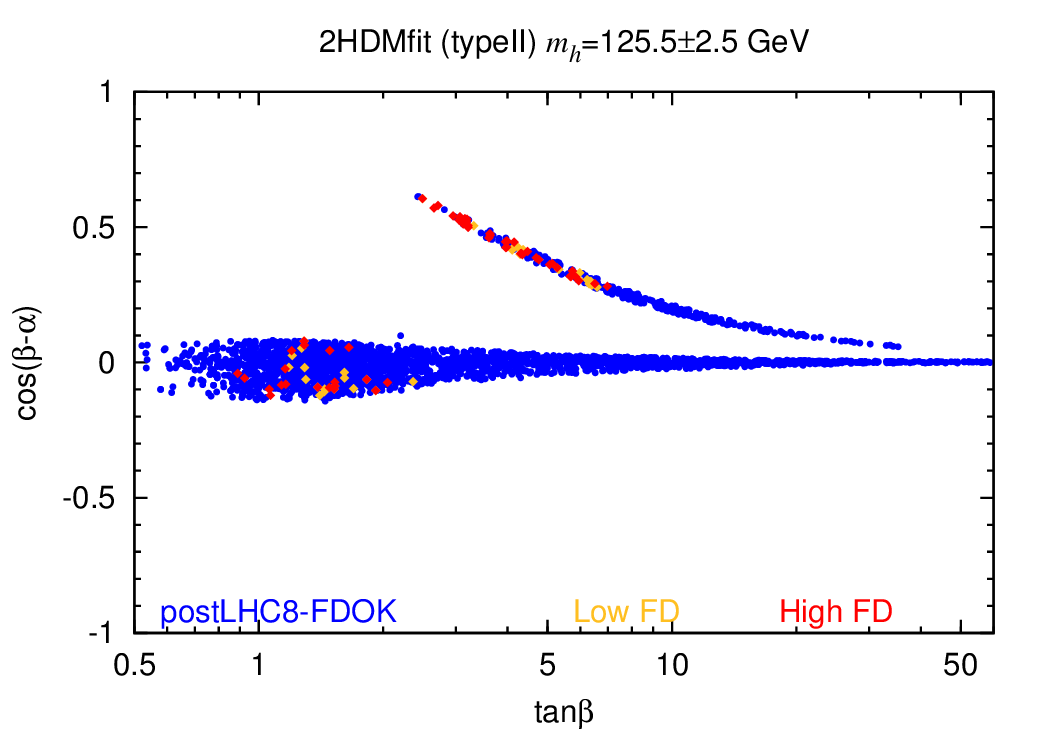}\\
\includegraphics[width=0.49\textwidth]{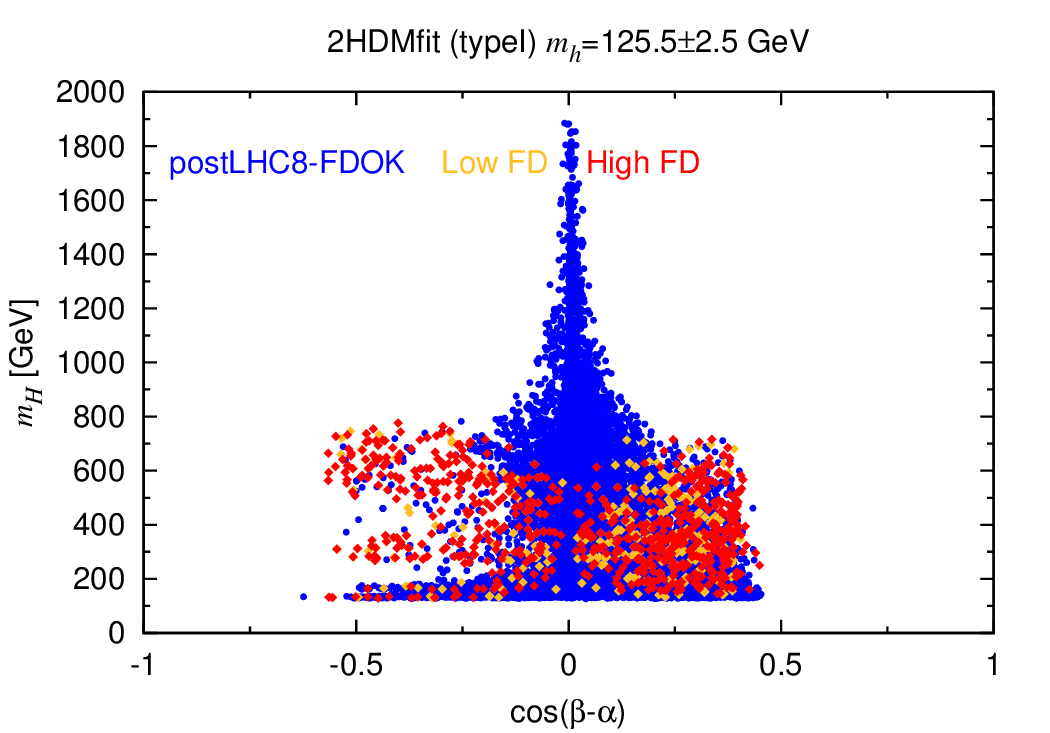}
\includegraphics[width=0.49\textwidth]{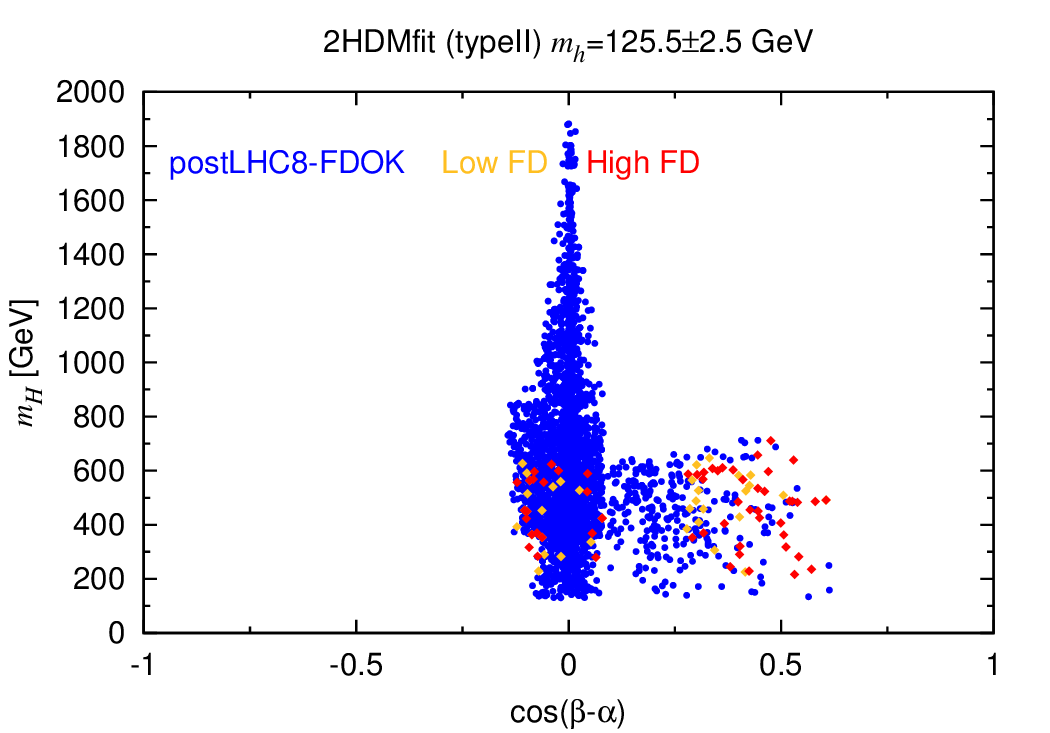}\\
\includegraphics[width=0.49\textwidth]{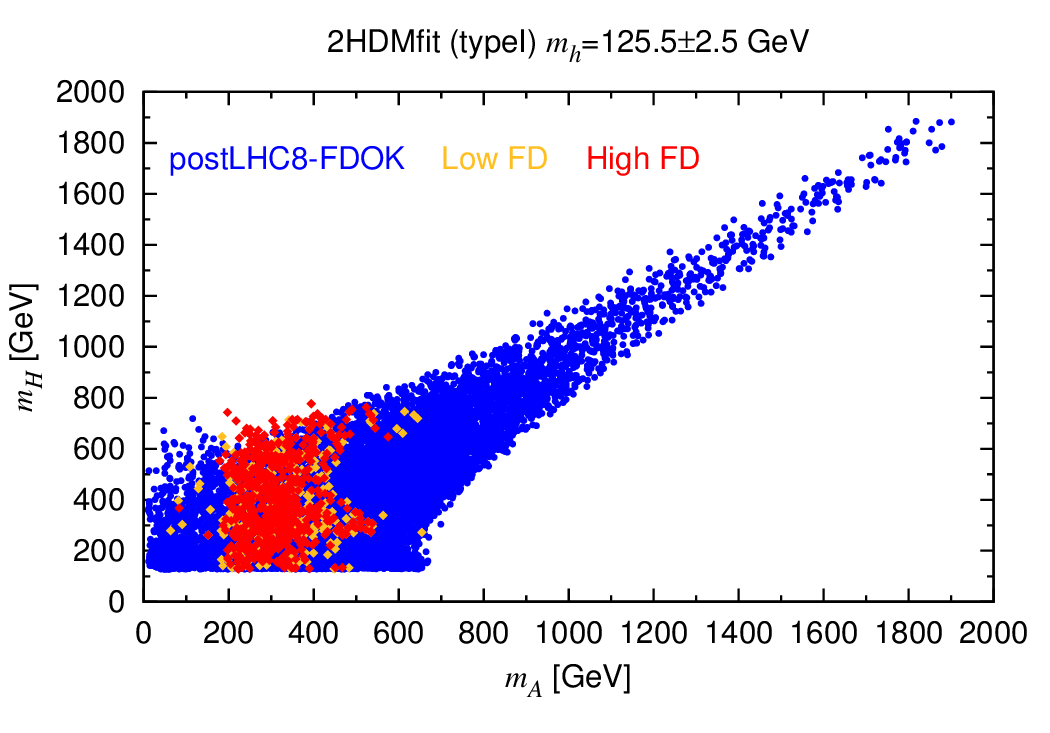}
\includegraphics[width=0.49\textwidth]{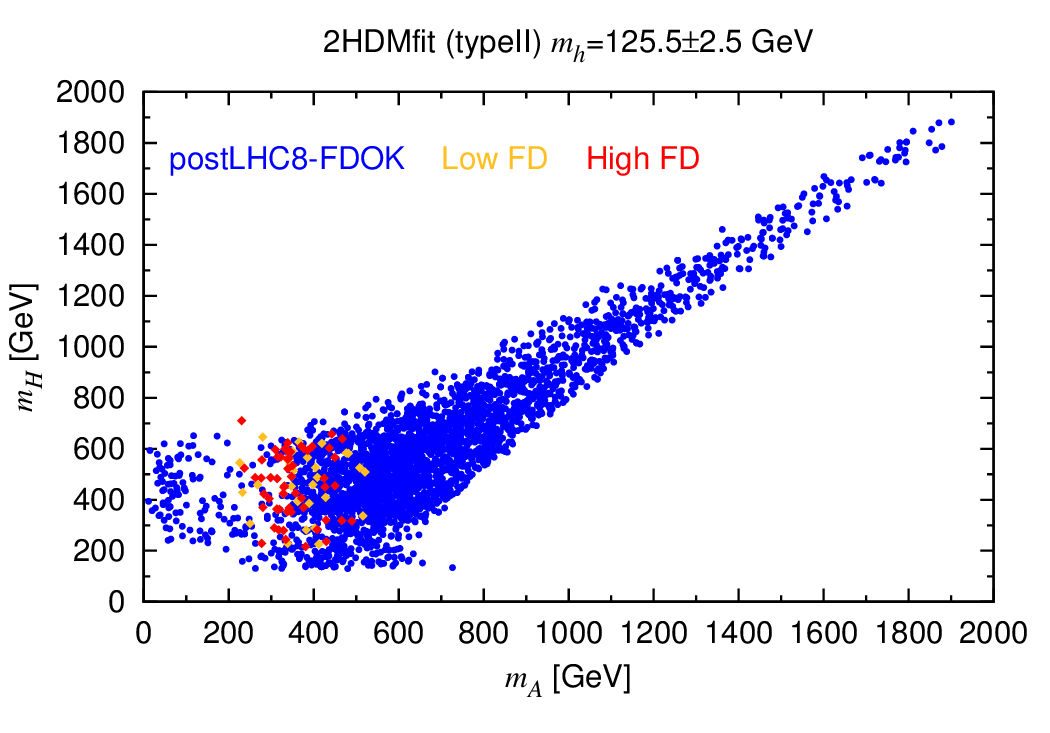}\\
\end{center}\vspace*{-5mm}
\caption{Constraints in the planes $\cbma$ vs.\ $\tanb$ (top row), $\mH$ vs.\ $\cbma$ (middle row) and $\mhh$ vs.\ $\mha$ (bottom row) 
for $\mh\sim 125.5\gev$. The blue points are the postLHC8-FDOK points. The orange (red) points are allowed postLHC8, but have Low FD (High FD). Many, but not all, Low FD and High FD points have postLHC8-FDOK points hidden below.  
Regarding the $\mH$ vs.\ $\cbma$ plots, replacing $\mH$ by $m_A$ or $m_{\hpm}$ gives essentially the same picture.
}
\label{bmavstbfd}
\end{figure}

To illustrate the effect of FD, we repeat the plots of Figs.~\ref{bmavstb} and \ref{mHvscba}  in Fig.~\ref{bmavstbfd} showing additional points that have Low FD or High FD  in at least one of the channels discussed above.  In these plots, the orange (red) points are those for which we have Low FD (High FD). Note that the points in the plots of Fig.~\ref{bmavstbfd} with Low FD or High FD all correspond to moderate $\tan \beta$ (up to $\sim 10$). However, these points overlap those having an acceptable FD level.  Thus, imposing the FD limits does not actually constrain the parameter spaces of the Type~I or Type~II models.

\clearpage

\section{\bf \boldmath $\mH\sim 125.5\gev$ scenarios}
\label{hhsection}

 Let us now turn to the case that the observed SM-like Higgs near
$125.5\gev$ is the heavier $CP$-even state of the 2HDM, $H$. In this case, the
lighter state, $h$, must have escaped LEP searches.  However, we will see that a signal for the $h$
could be hiding in the present data from the LHC and might be revealed in several focused analyses.  The pseudoscalar $\ha$
can be either lighter or heavier than the $H$. Since perturbativity for the quartic couplings prevents the $\ha$  from being 
heavier than about 1 TeV, it can also give interesting signatures at LHC14.

\begin{figure} [b!]
\begin{center}
\includegraphics[width=0.49\textwidth]{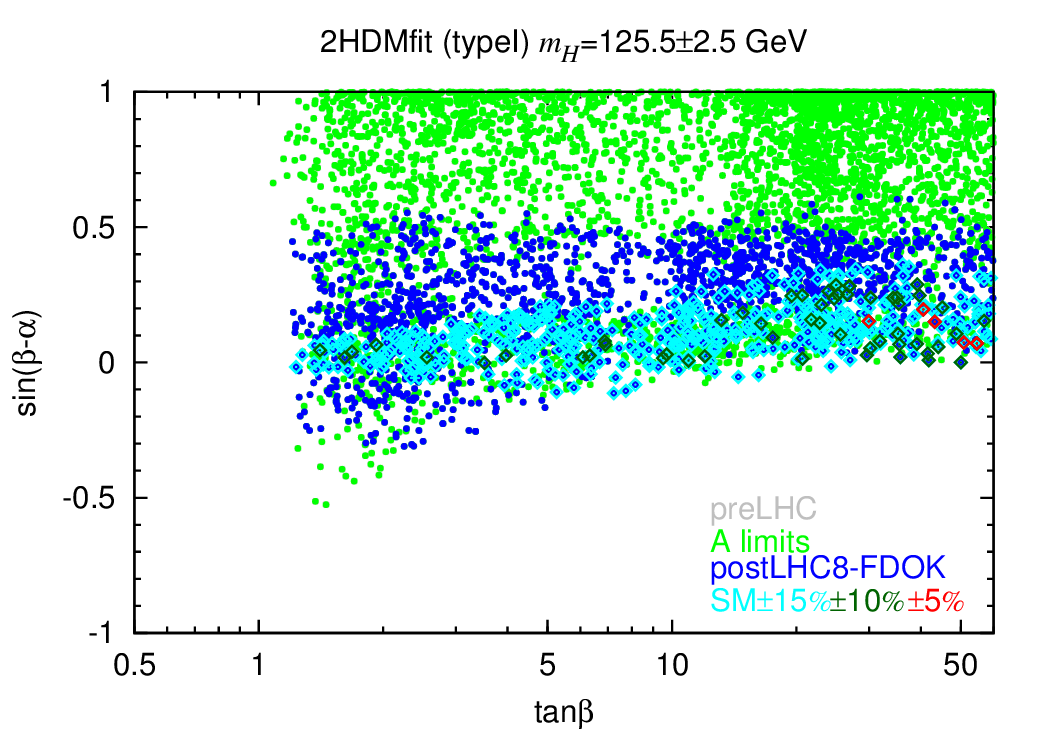}
\includegraphics[width=0.49\textwidth]{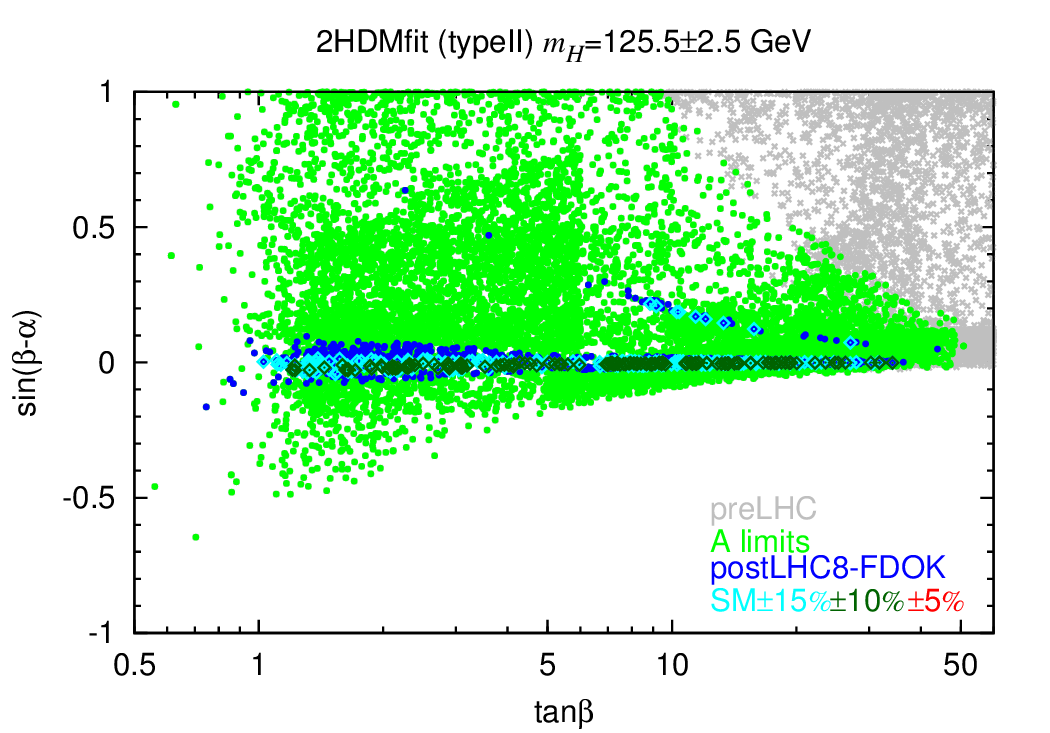}
\end{center}\vspace*{-5mm}
\caption{Constraints on the 2HDM of Type~I and Type~II in the $\sbma$ versus $\tanb$ plane for 
$\mH\sim 125.5\gev$.  We show points that survive at the preLHC (grey), A-limits (green), postLHC8-FDOK (blue), SM$\pm15\%$ (cyan), SM$\pm10\%$ (dark green), and SM$\pm 5\%$ (red) levels. There are no FDOK requirements imposed on the preLHC and A-limits points. The SM$\pm15\%,\pm10\%,\pm 5\%$ points {\it are}  subjected to FDOK requirements.}
\label{bmavstbH}
\end{figure}

We begin by presenting in Fig.~\ref{bmavstbH} plots for the $\mH\sim 125.5\gev$ scenarios analogous to those of Fig.~\ref{bmavstb} for the $\mh\sim 125.5\gev$ scenarios. Here, we have again required small FD (from $A$ production and decays); possible FD
contributions will be discussed later in this section. In the case of Type~I, consistency with the observed $125.5\gev$ signal restricts  $\sin(\beta-\alpha)$  less than was the case for $\cos(\beta-\alpha)$ in the $\mh\sim 125.5\gev$ case.  In contrast, for Type~II the constraints on $\sin(\beta-\alpha)$ are similar in nature to the limits on $\cos(\beta-\alpha)$ in the case of the $h$. There is, however, an important difference.  Namely, if $\pm 5\%$ agreement with the SM can be verified in all the channels listed in Eq.~(\ref{xxyychannels}), then $\mH=125.5\gev$ is eliminated in Type~II but not in Type~I. This can be traced to the fact that the charged-Higgs loop does not decouple at large $\mhpm$ and ends up suppressing the $H\gam\gam$ coupling and therefore the $\gam\gam$ final state rates.  More details regarding the nondecoupling of the charged Higgs loop contribution to the $H\gam\gam$ coupling are presented in Appendix \ref{nondecoup}.

\begin{figure} [t]
\begin{center}
\includegraphics[width=0.58\textwidth]{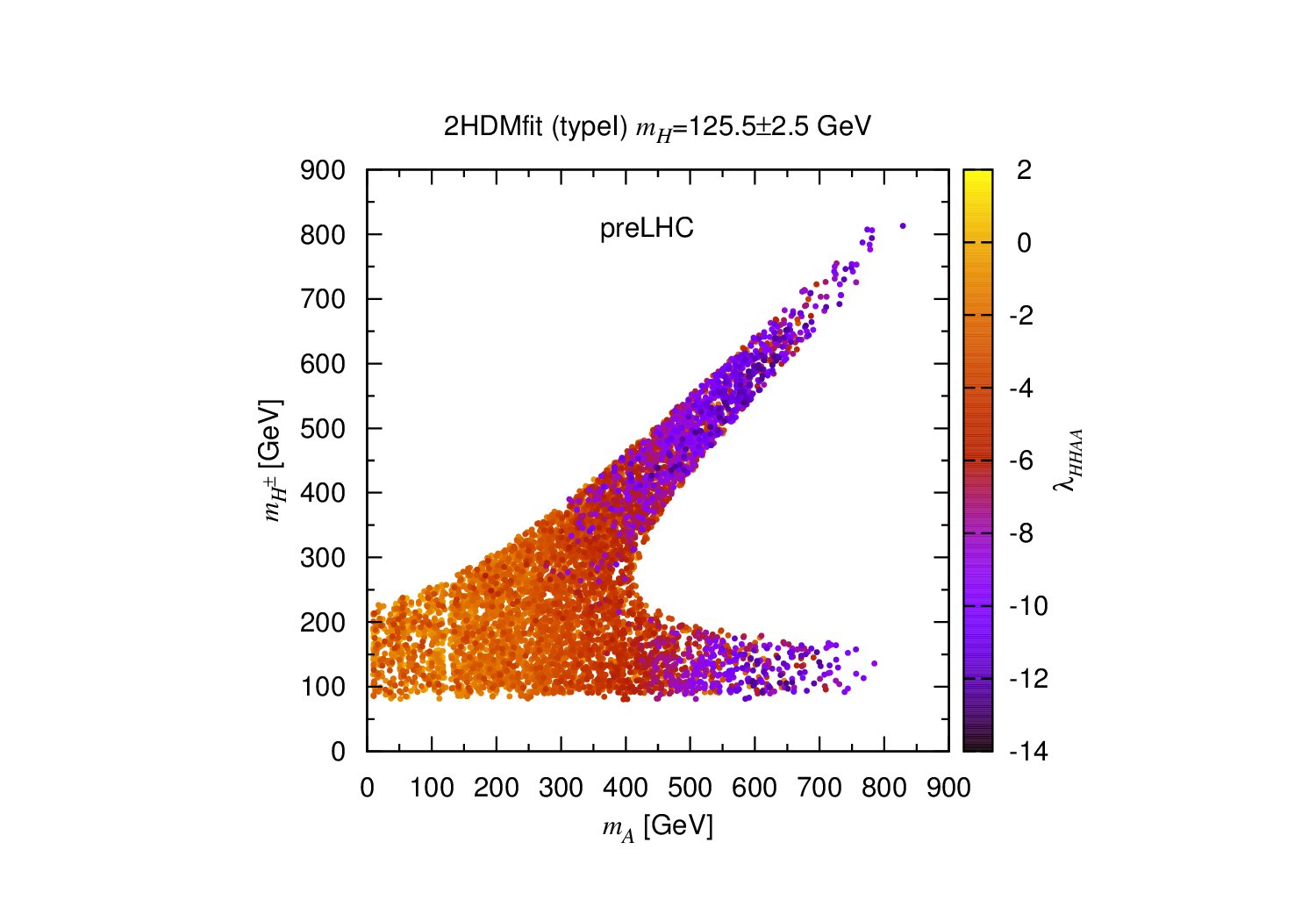}
\hspace{-30mm}
\includegraphics[width=0.58\textwidth]{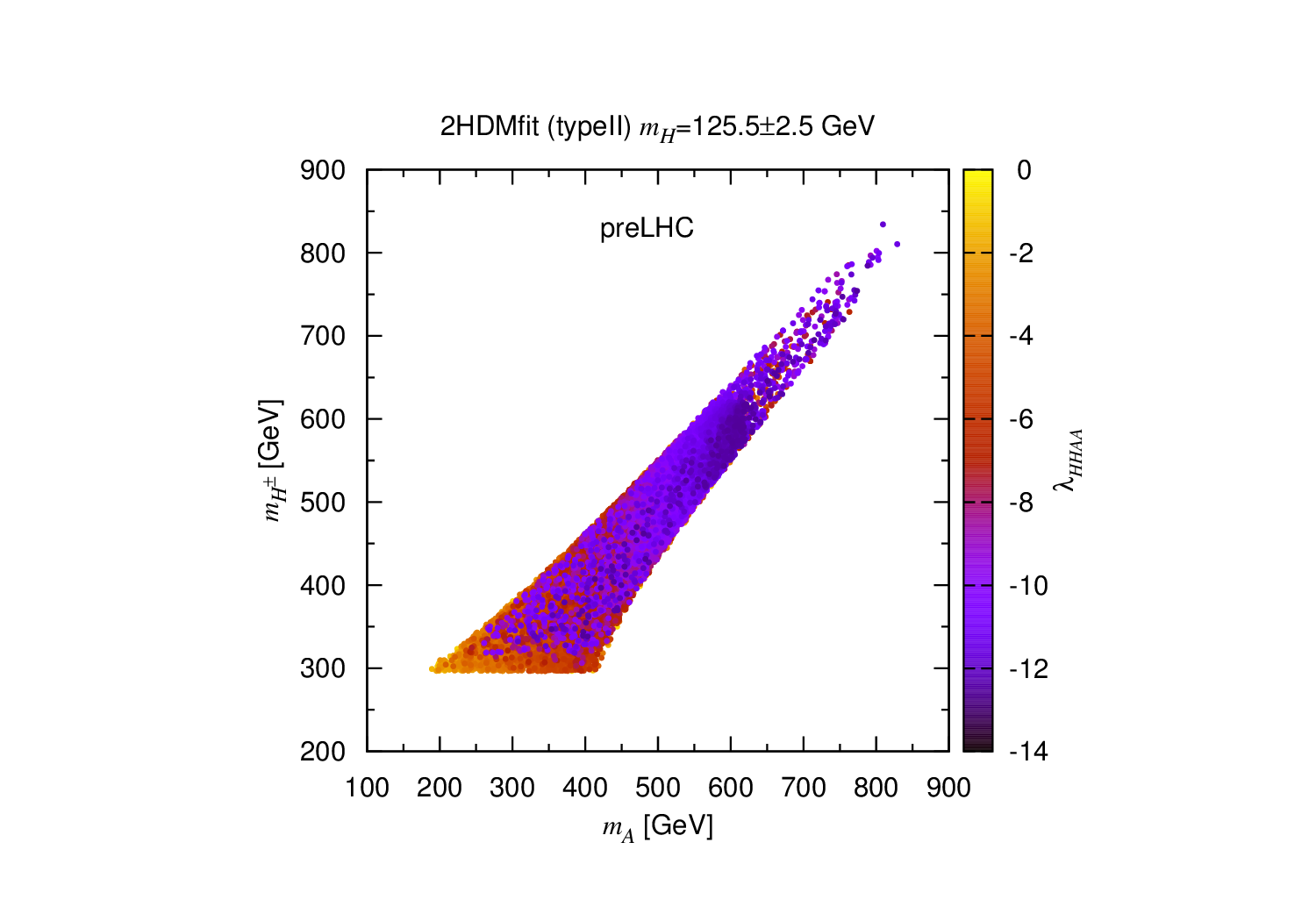}\\
\vspace{-5mm}
\includegraphics[width=0.58\textwidth]{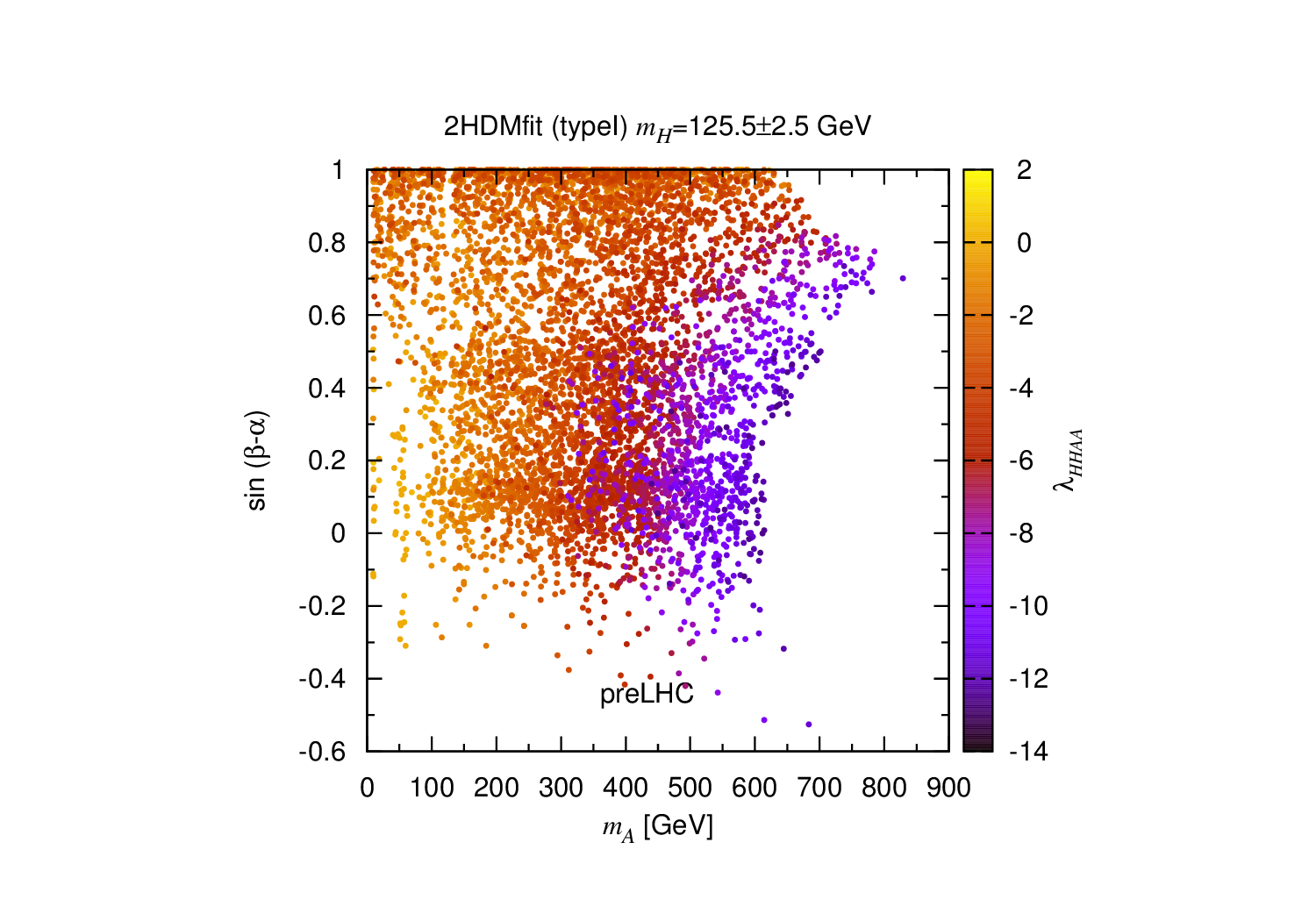}
\hspace{-30mm}
\includegraphics[width=0.58\textwidth]{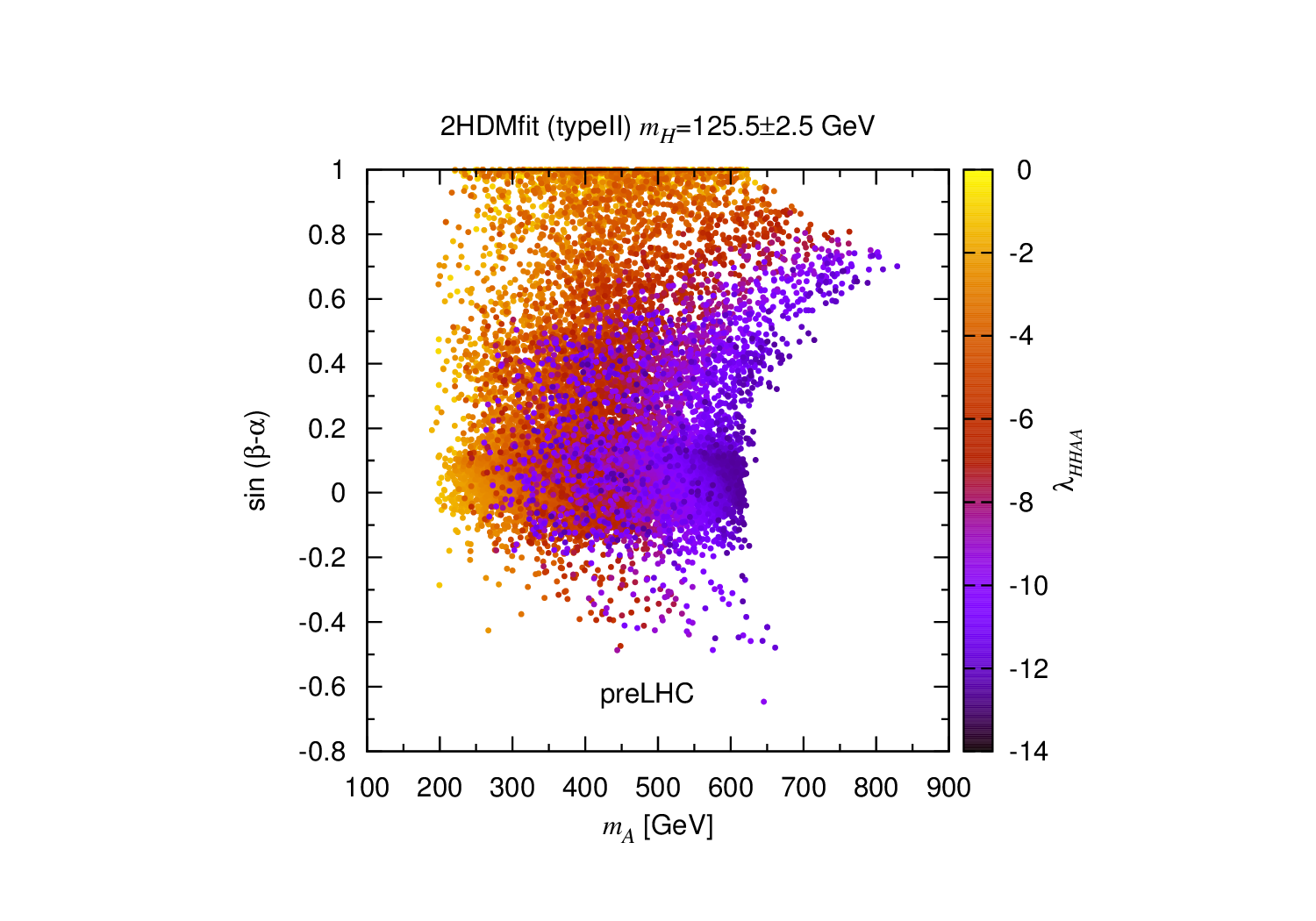}
\end{center}\vspace*{-10mm}
\caption{Scatterplots of $\lam_{HHAA}$ in the $\mhpm$ vs.\ $\mha$ and $\mha$ vs.\ $\sbma$ planes for the case of  
$\mH\sim 125.5\gev$.  The values of $\lam_{HHAA}$ are color coded as indicated by the scales on the right of the plots. The full set of preLHC conditions is satisfied for all points shown.}
\label{lamthreepert}
\end{figure}

There is a maximum $\mhpm$ that can be achieved before perturbativity is violated, but this maximum applies for all parameter choices, unlike the $\mh=125.5\gev$ case for which there is a true decoupling limit.  To illustrate this, we present in Fig.~\ref{lamthreepert} (upper row) ``temperature" plots showing the quartic coupling $\lam_{HHAA}$  in the plane of $\mhpm$ vs.\ $\mha$.  $\lam_{HHAA}$ is one of a few that most frequently encounter the perturbativity bound. We see that $\lam_{HHAA}$ hits its perturbativity bound of $\sim 4\pi$ at about $\mha\sim \mhpm\sim 800\gev$ for both Type~I and Type~II.  In the Type~I case the perturbativity limit is also reached at low $\mhpm$ if $\mha$ is as heavy as $\sim 800\gev$.  This wing of the $\mhpm$ vs.\ $\mha$ plot is not present for Type~II because of the lower bound of about $300\gev$ from $B$ physics constraints. In the bottom row of Fig.~\ref{lamthreepert} we present temperature plots of $\lam_{HHAA}$ in the $\mha$ vs.\ $\sbma$ plane, showing that at small to moderate $|\!\sbma|$ the perturbative bound is already exceeded by $\mha \sim 600\gev$. In any case, the bottom line is that there is no decoupling limit for the $\mhh=125.5\gev$ case and nondecoupling effects are inevitably of importance.

\begin{figure} [t]
\begin{center}
\includegraphics[width=0.49\textwidth]{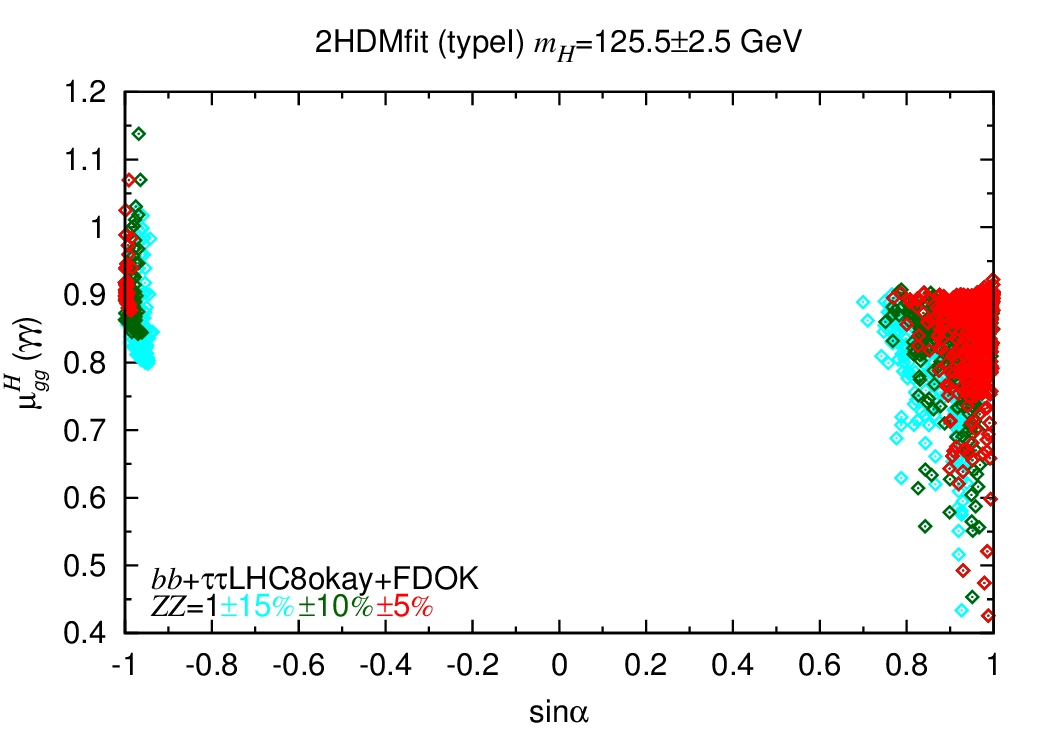}
\includegraphics[width=0.49\textwidth]{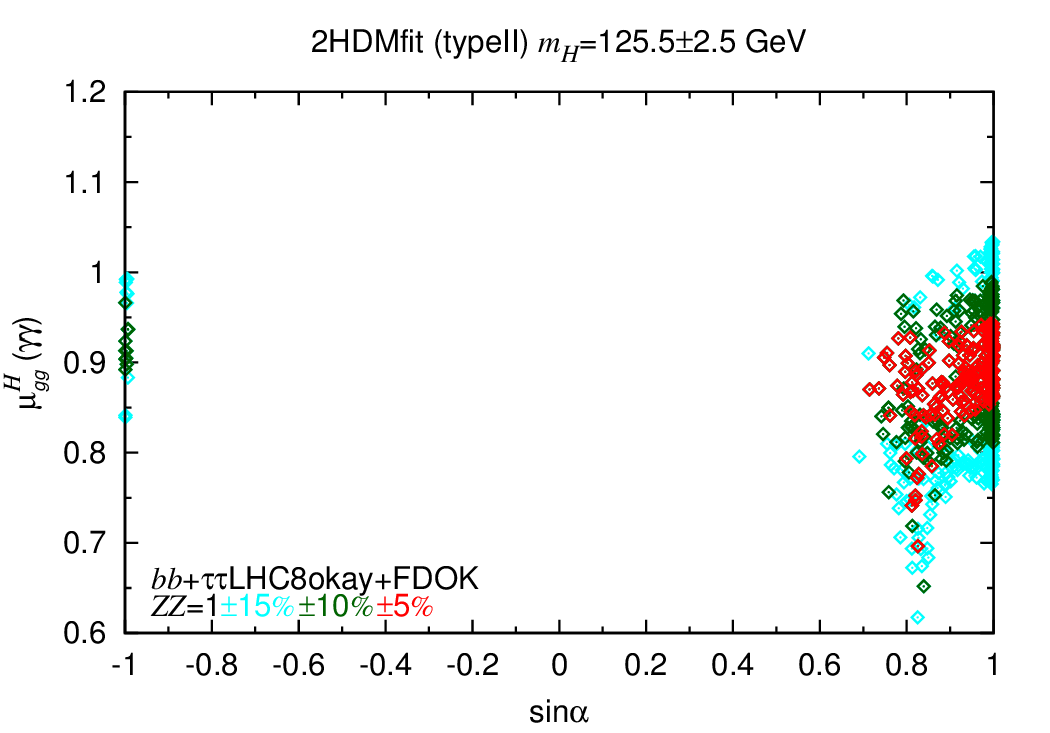}
\end{center}\vspace*{-5mm}
\caption{We plot $\mu_{gg}^\hh(\gam\gam)$ as a function of $\sina$ for points having $gg\to ZZ$ and $VV\to ZZ$ rates within $\pm15\%$, $\pm 10\%$ or $\pm 5\%$ of the SM predictions.}
\label{gamgamratesH}
\end{figure}

The resulting $\gam\gam$ final state rates are illustrated in Fig.~\ref{gamgamratesH}. There, we see that consistency with $\pm 5\%$ for the $ZZ$ final state rates and {\it simultaneously} for the $gg\to H\to \gam\gam$ rate is only possible on the $\sin\alpha<0$ (\ie\ $\cu^\hh<0$) branch in the Type~I model.  Most of this mismatch can, as said earlier, be traced to the nondecoupling of the charged-Higgs loop contribution to the $H \gam\gam$ coupling.  In the end only the few red points on the $\sina<0$ branch of the Type~I model having $\mu_{gg}^\hh(\gam\gam)\gsim 0.95$ can survive if $\leq \pm 5\%$ deviations from the SM are required for both the $ZZ$ and $\gam\gam$ final states.

\begin{figure} [t]
\begin{center}
\includegraphics[width=0.49\textwidth]{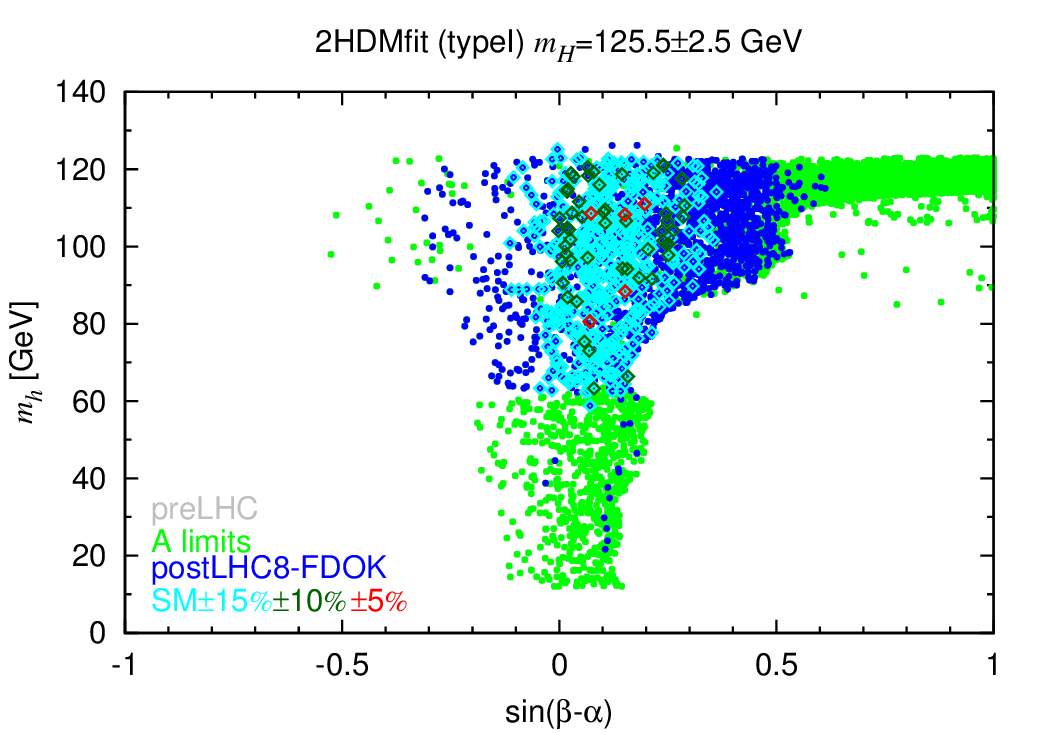}
\includegraphics[width=0.49\textwidth]{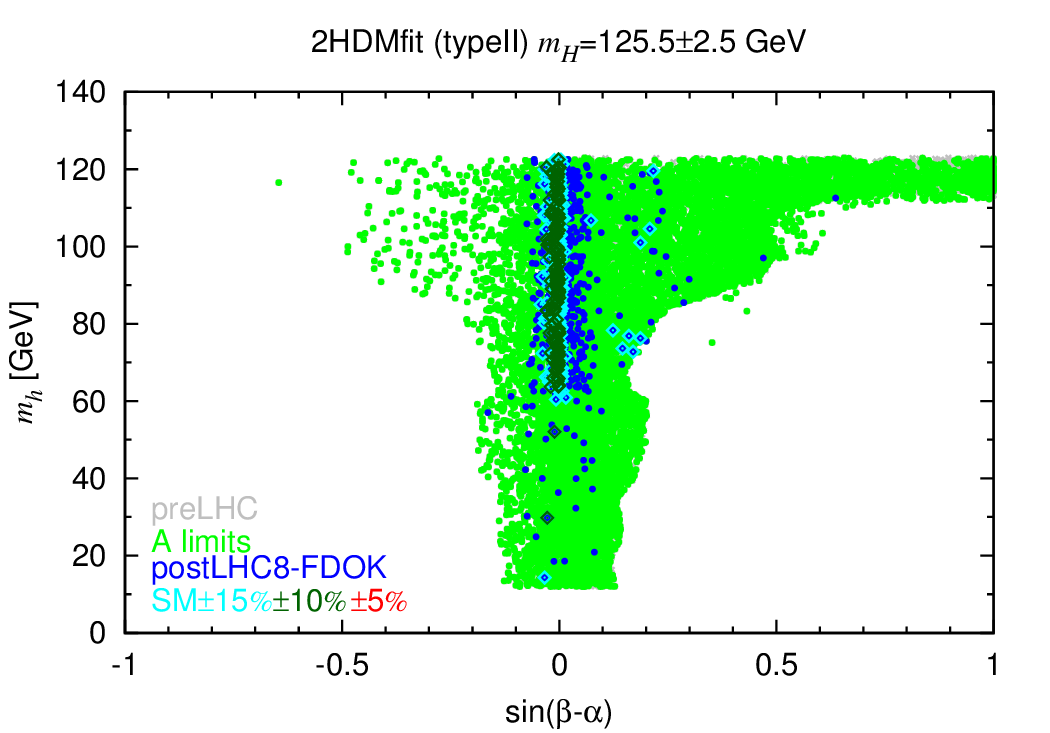}\\
\includegraphics[width=0.49\textwidth]{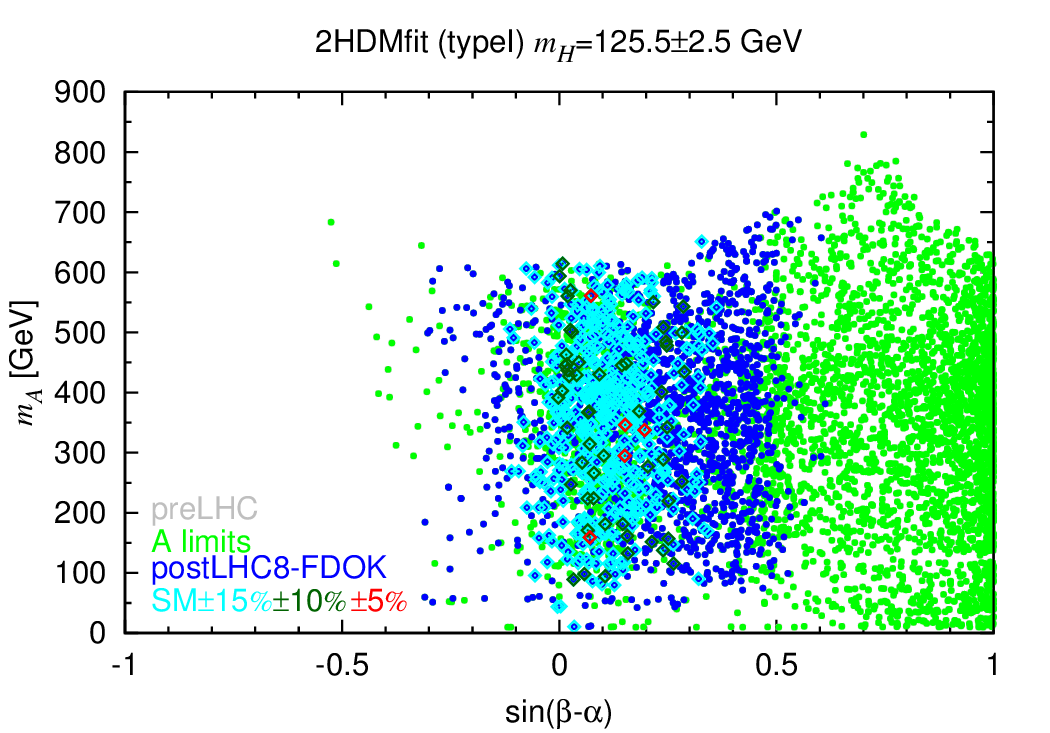}
\includegraphics[width=0.49\textwidth]{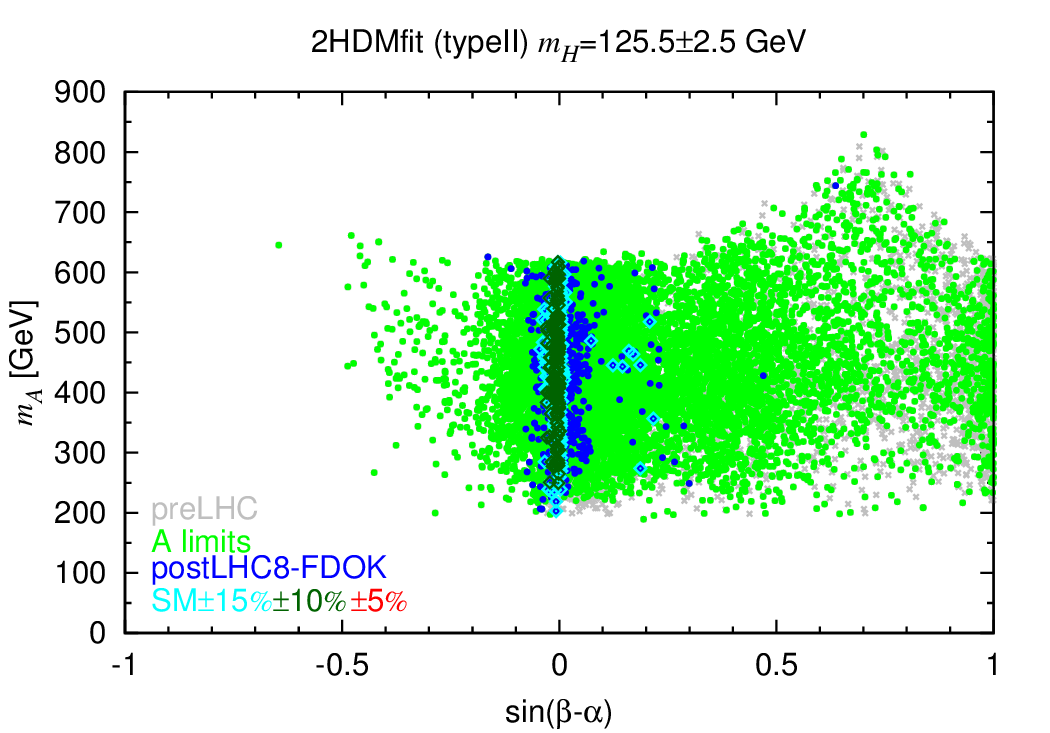}\\
\end{center}\vspace*{-5mm}
\caption{Constraints in the $\mh$ vs.\ $\sbma$ and $\mA$ vs.\ $\sbma$ planes for the $\mH\sim 125.5\gev$ scenarios. Results for $m_{\hpm}$ vs.\ $\sbma$ are very similar to those for $\mA$ vs.\ $\sbma$. There are no FDOK requirements imposed on the preLHC and A-limits points. The SM$\pm15\%,\pm10\%,\pm 5\%$ points {\it are}  subjected to FDOK requirements.}
\label{mhlmavssbma}
\end{figure}

As regards the $h$ and $A$ masses associated with a good fit by the $H$ to the LHC data and other limits we refer to Fig.~\ref{mhlmavssbma}.  There, we see that a proper fit at the postLHC8 level is easily achieved if $\mhl\gsim 60\gev$, for which $\hh\to \hl\hl$ decays are kinematically forbidden. However, there is also a scattering of points for which small values of $\mhl$ are possible.  Such points correspond to parameters for which the $H\hl\hl$ coupling is small. A very ``fine-tuned" scan is necessary to find these low-$\mhl$ points for which $\br(\hh\to \hl\hl)$ is small enough that the $H$ signals fit the LHC data at an adequate level.

\begin{figure} [t]
\begin{center}
\includegraphics[width=0.49\textwidth]{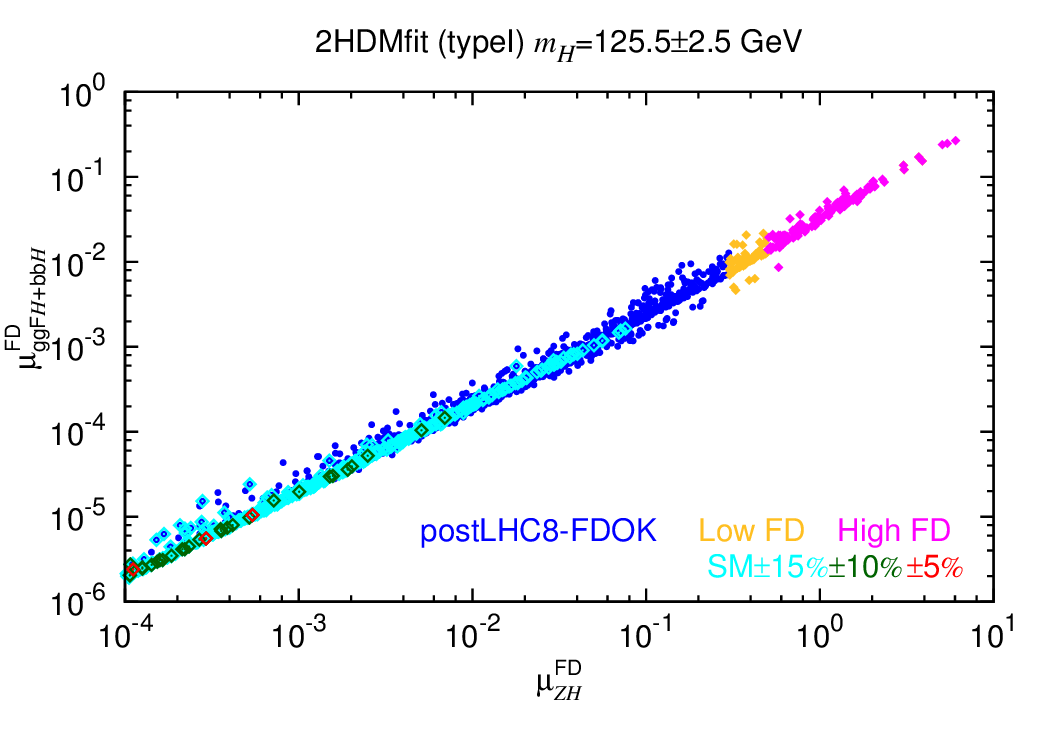}
\includegraphics[width=0.49\textwidth]{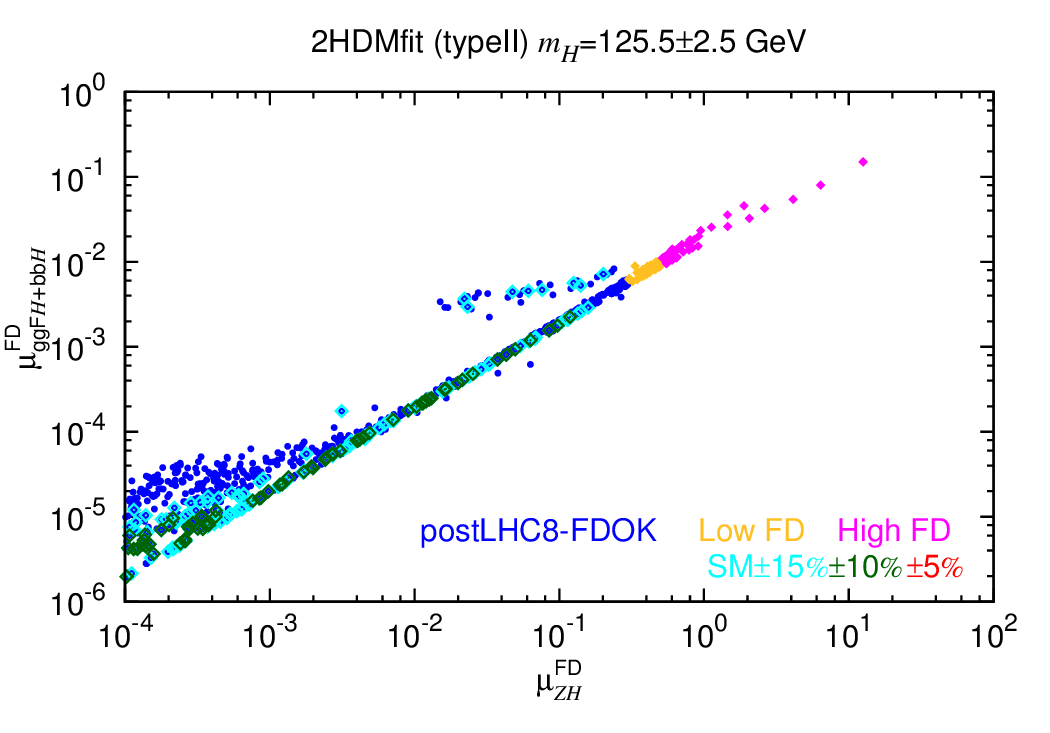}
\end{center}\vspace*{-5mm}
\caption{We plot $\mu_{{\rm ggF}H+{\rm bb}H}^{{\rm FD}}$ vs.\ $\mu_{ZH}^{\rm FD}$ illustrating how high FD fractions can go for postLHC8 points for the $\mH\sim 125.5\gev$ scenarios.  Also shown is how convergence of the $H$ properties to SM-like values would limit the maximum possible feed downs. We display only points with $\mu_{ZH}^{\rm FD}\geq 10^{-4}$  --- there are many points with much lower values. 
}
\label{ggfvszhh}
\end{figure}

Let us now address the issue of feed down. Given that $\mha$ can be quite large, there is certainly the possibility of
 $A\to ZH$ feed down contributions to the $H$ signals. The $HAZ$ coupling is proportional to $\sin(\beta-\alpha)$, which the fits require to be $\lsim 0.5$ in magnitude. What is important, however, is $\br(\ha\to Z\hh) = \Gamma(\ha\to Z\hh)/\Gamma_{\rm tot}(\ha)$, which can still be large. Figure~\ref{ggfvszhh} shows the FD $\mu$ values analogous to those considered for the $\mhl\sim 125.5\gev$ scenario.  To be precise, we consider
 \bea
 \mu_{{\rm ggF} H+{\rm bb}H}^{\rm FD}&\equiv&{\left( \sigma_{{\rm ggF}A}+\sigma_{{\rm bb}A}\right) P_{\rm FD}(A \to H+X) \over 
 \sigma_{{\rm ggF} H}+\sigma_{{\rm bb}H} }\, , \\
 \mu_{Z\hh}^{\rm FD}&\equiv&{\sigma_{{\rm ggF}A}\br(\ha \to Z\hh)\over \sigma_{ZH}}\,.
 \label{fdHdefs}
 \eea
We observe that substantial FD is indeed possible.  The ratio defining $\mu_{Z\hh}^{\rm FD}$ above, has the greatest potential for being large because of the large $gg\to A$ production rate in the numerator compared to the $Z^*\to ZH$ rate appearing in the denominator. In contrast,  in $\mu_{{\rm ggF} H+{\rm bb}H}^{\rm FD}$ both numerator and denominator are ggF-dominated. As in the $\mh=125.5\gev$ case, we exclude from subsequent plots those points which have FD levels that exceed $10\%$ relative to the $gg\to \hh+bb\hh$ production modes and $30\%$ in the $Z\hh$ associated production mode.

\begin{figure}[t]
\begin{center}
\includegraphics[width=0.49\textwidth]{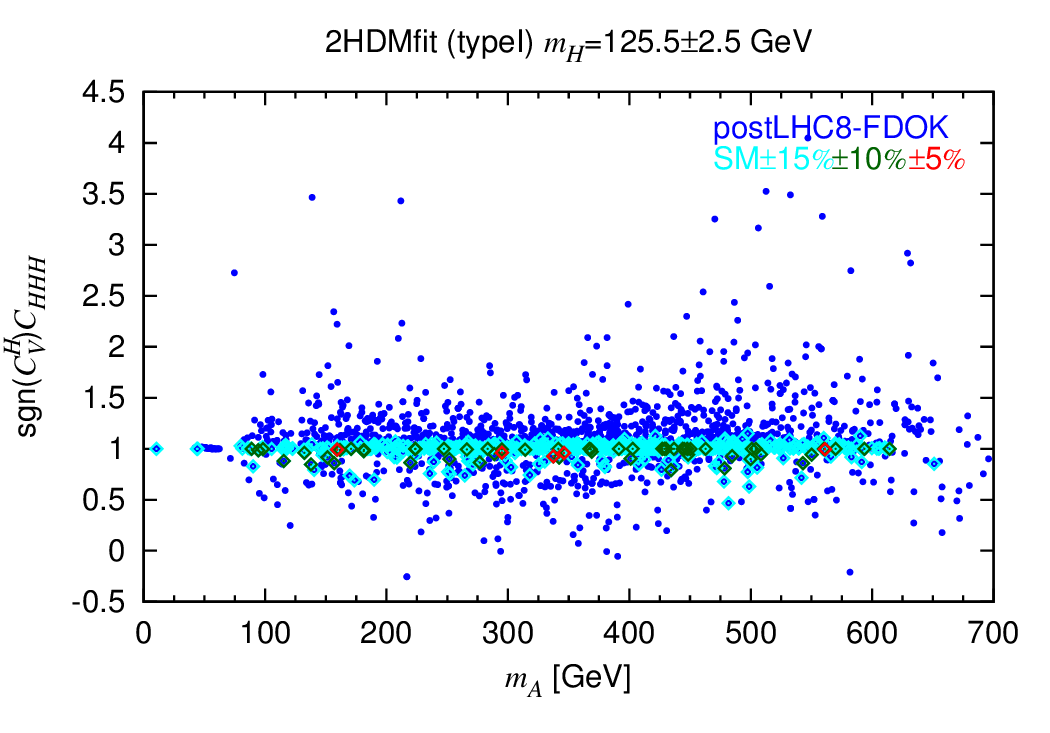}
\includegraphics[width=0.49\textwidth]{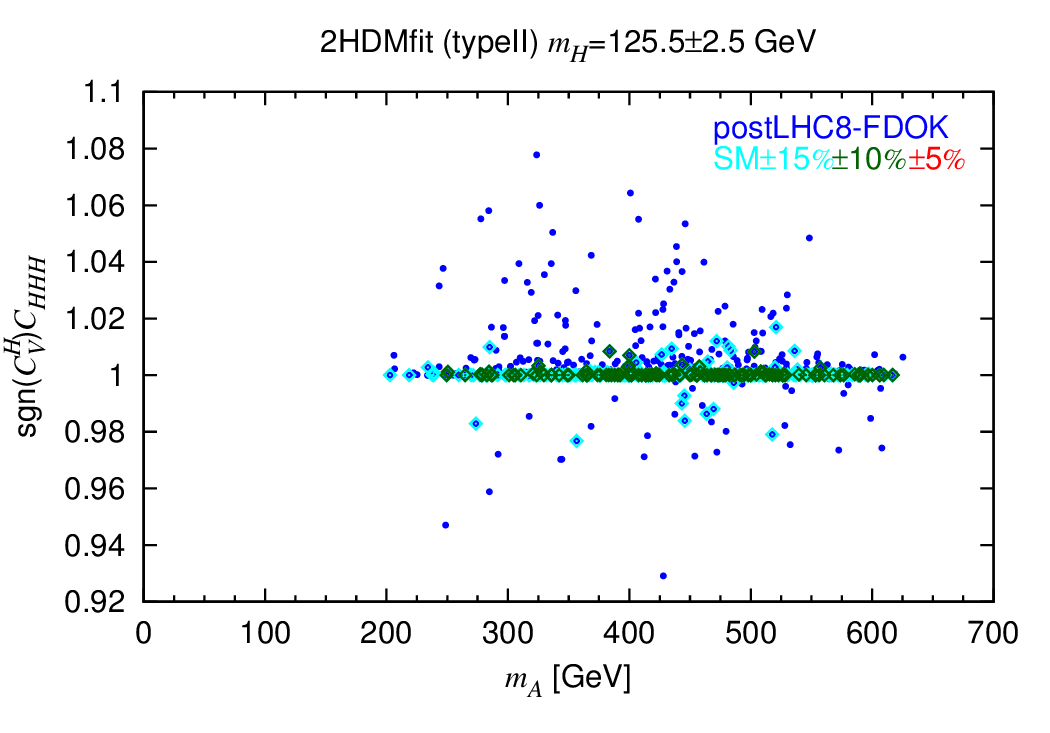}
\end{center}\vspace*{-5mm}
\caption{We display points in the $\sgn(C_V^H)C_{HHH}$ vs.\ $\mA$  plane for the $\mhh\sim 125.5\gev$ scenario comparing current $H$ fits  to the case where future measurements show that {\it all} channel rates   are within $\pm 15\%,\pm 10\%,\pm 5 \%$ of the SM Higgs prediction; FDOK is required in all cases. Color scheme is as for Fig.~\ref{bmavstbH}, except that preLHC and A-limits points are not displayed.}
\label{HHHcoup}
\end{figure}

As in the case of $\mhl\sim 125.5\gev$, it is interesting to assess the prospects for detecting a deviation in the triple-Higgs coupling as one goes from the current data set to  $\hh$ rates that are increasingly SM-like.
In Fig.~\ref{HHHcoup}, we plot $\sgn(C_V^H) C_{HHH}$, \ie\ the ratio of the triple-Higgs coupling $\lam_{HHH}$ to the value it should have in the SM limit, as a function of $\mha$.  [We  include $\sgn(C_V^H)$ because some of the points have  $C_V^H<0$ for our scanning procedure.] We observe that as the LHC signals become increasingly SM-like, the deviations of $\sgn(C_V^H)C_{HHH}$ from unity are even more tightly limited than in the case of $C_{hhh}$ for $\mhl\sim 125.5\gev$.  Of course, we also see (again) that very few (no) points survive the SM$\pm 5\%$ constraint in the case of Type~I (Type~II).

\begin{figure} [t]
\begin{center}
\includegraphics[width=0.49\textwidth]{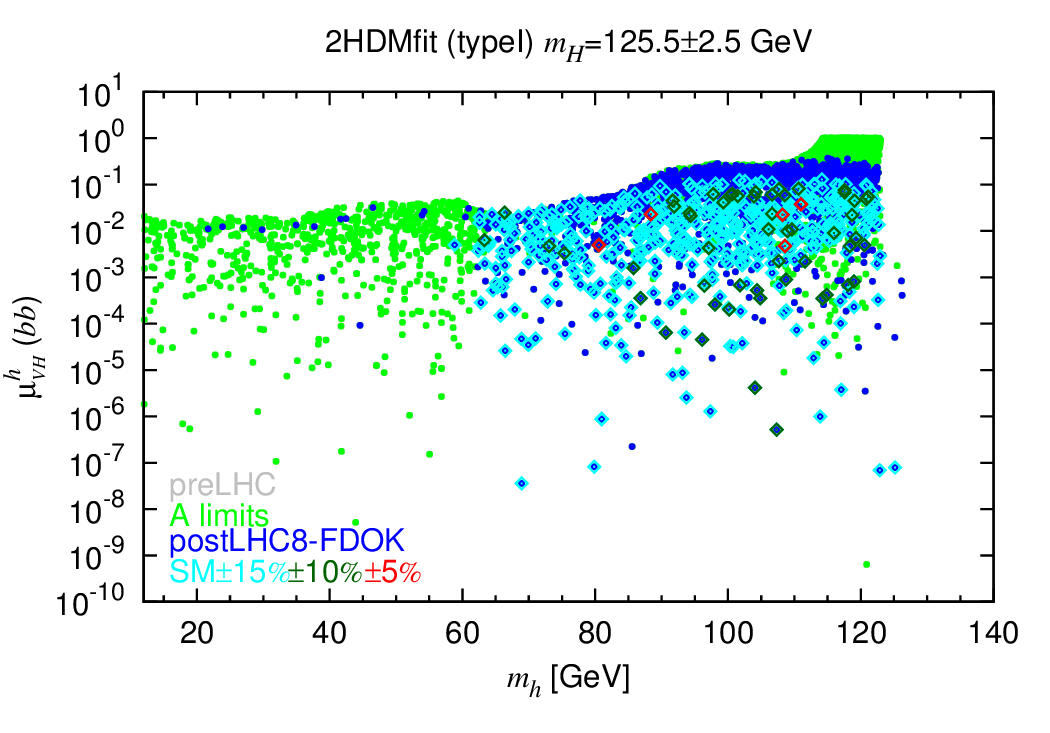}
\includegraphics[width=0.49\textwidth]{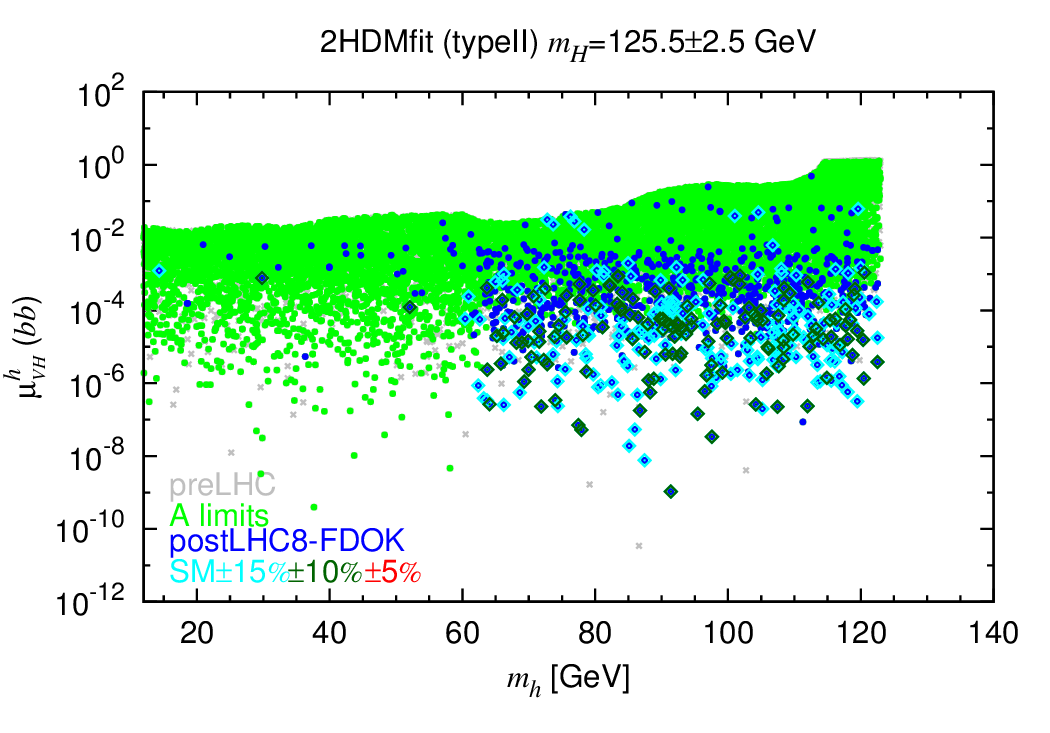}
\end{center}\vspace*{-5mm}
\caption{$\mu_{\rm VH}^h(b\anti b)$, \ie\ $V^*\to V\hl$ associated production with $\hl\to b\anti b$ relative to the SM, as a function of $\mhl$.  Note that $\mu_{\rm VH}^\hl(bb)$ is actually independent of energy and that the ratio also applies to any situation where the subprocess of interest is $V^*\to Vh$, including the LEP $Z^*\to Zh$ process. There are no FDOK requirements imposed on the preLHC and A-limits points. The SM$\pm15\%,\pm10\%,\pm 5\%$ points {\it are}  subjected to FDOK requirements.}
\label{vhbbvsmh}
\end{figure}

Let us next assess the feasibility for detecting the lighter $\hl$. As already noted, finding points with $\mhl\lsim 60\gev$ for which $\br(H\to hh)$ is small enough to still allow the $\hh$ rates in the various channels to fit the $125.5\gev$ signal is highly nontrivial and this scenario will be discussed in detail elsewhere. 
The most interesting modes for $\hl$ detection may be $gg\to h\to \gam\gam$ and   $V^* h$ with $h\to b\anti b$.  In the context of the current $8\tev$ data, only the latter mode is of interest --- expected signal strengths as a function of $m_h$ are plotted in Fig.~\ref{vhbbvsmh}.  While for many points the expected rates are obviously too small to have allowed detection of the $h$,  there also exist postLHC8-FDOK points for which detection in the $Vh(b\anti b)$ final state might be on the edge.  We speculate that for the $Vh(b\anti b)$  final state, a leptonic trigger on the $V$ might still allow the predicted signal to emerge for the higher $\mu_{\rm VH}^h(b\anti b)$ values. 

As an aside, it is easily inferred from Fig.~\ref{vhbbvsmh} that the postLHC8 and, even more so, the SM$\pm15\%,\pm 10\%, \pm 5\%$ requirements eliminate a large swath of the points that survive the $A$-limits constraint. It is also noteworthy that in the case of Type~I all preLHC points automatically satisfy the $A$-limits requirement, whereas some preLHC (grey) points get excluded by the A limits in the case of Type~II.

Considering Fig.~\ref{vhbbvsmh}, it is moreover interesting to ask whether the $\gsim 2\sigma$ LEP excess in the $Z b\anti b$ final state at $M_{b\anti b}\sim 98\gev$ could be explained by  $\mh\sim 98\gev$ and $\mu_{\rm VH}^\hl(b\bar b)\sim 0.1-0.3$.  We see that this is indeed possible in both the Type~I and Type~II models given current postLHC8 constraints on the $\hh$ properties.  Of course, the scatterplots suggest that this explanation is more fine-tuned in the Type~II case.  Furthermore, if   the $\hh$ rates are found to be within $\pm 15\%$ of the SM rates,  the value of  $\mu_{\rm VH}^\hl(b\anti b)$ is pushed well below the desired range in the case of Type~II and is at a marginal level in the case of Type~I.  At the SM$\pm 5\%$ level, the few surviving Type~I points have $\mu_{\rm VH}^\hl(b\anti b) \lsim 0.05$ (assuming that a more extensive scan would reveal red points with $\mhl\sim 98\gev$ that would have a signal level comparable to those around $90\gev$ and $108\gev$ plotted), a value that is not very consistent with the LEP $\sim 2.3 \sigma$ excess observed.

\begin{figure} [t]
\begin{center}
\includegraphics[width=0.49\textwidth]{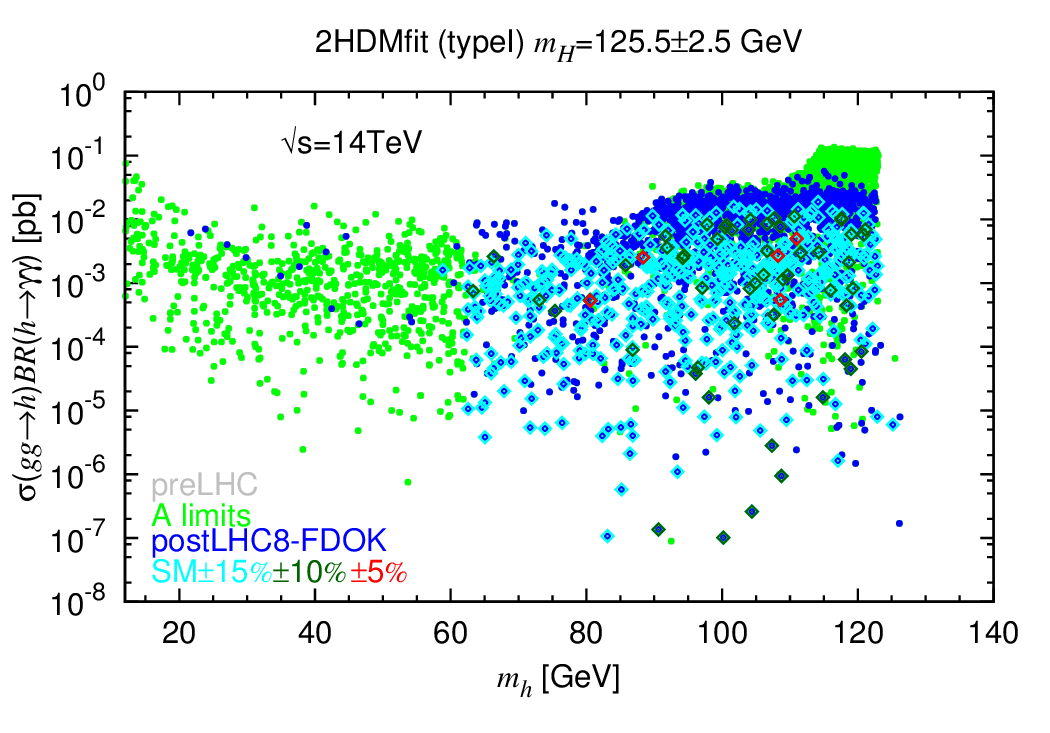}
\includegraphics[width=0.49\textwidth]{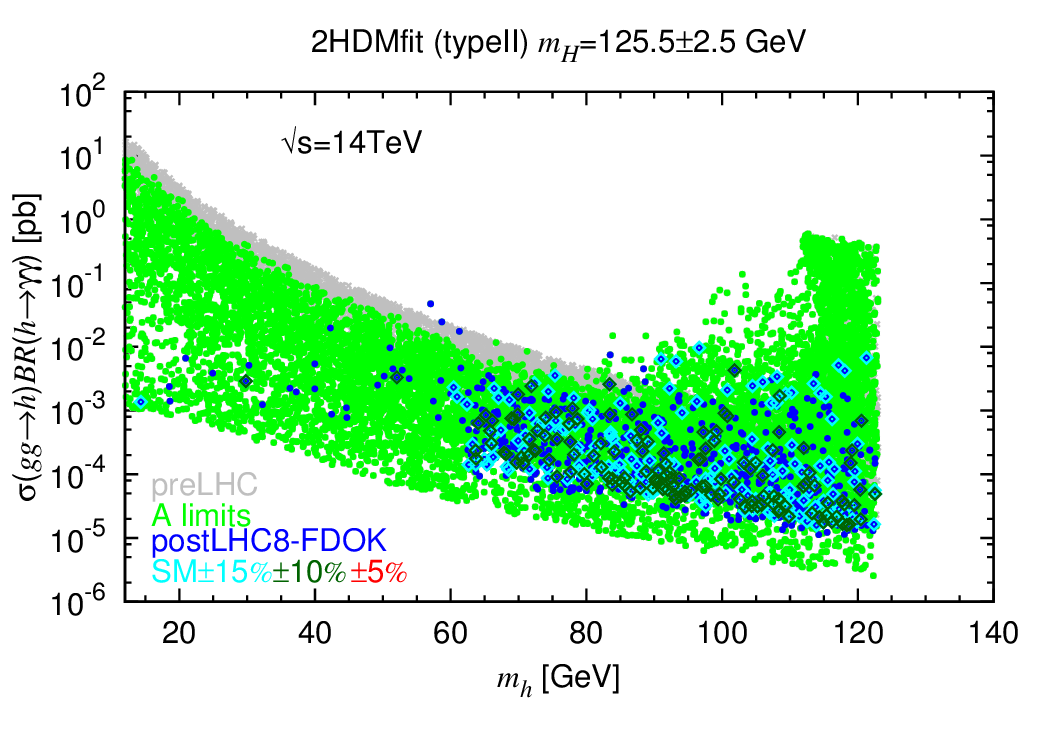}
\end{center}\vspace*{-5mm}
\caption{$\sig(gg\to h)\br(h\to \gam\gam)$ for $\rts=14\tev$ with postLHC8-FDOK constraints imposed as well as further limitations imposed by SM$\pm 15\%,\pm10\%,\pm5\%$ constraints. There are no FDOK requirements imposed on the preLHC and A-limits points. The SM$\pm15\%,\pm10\%,\pm 5\%$ points {\it are}  subjected to FDOK requirements.}
\label{siggghAA}
\end{figure}

At 14 TeV, there is also potential for detecting the $\hl$ in the  $gg\to h\to \gam\gam$ mode, as shown in Fig.~\ref{siggghAA}. Of course, while a significant event yield is possible for $L\geq 300\fbi$, the level of continuum irreducible and reducible backgrounds must be assessed and could prove too large for the blip at $\mhl$ to be observable.

\begin{figure} [h!]
\begin{center}
\includegraphics[width=0.49\textwidth]{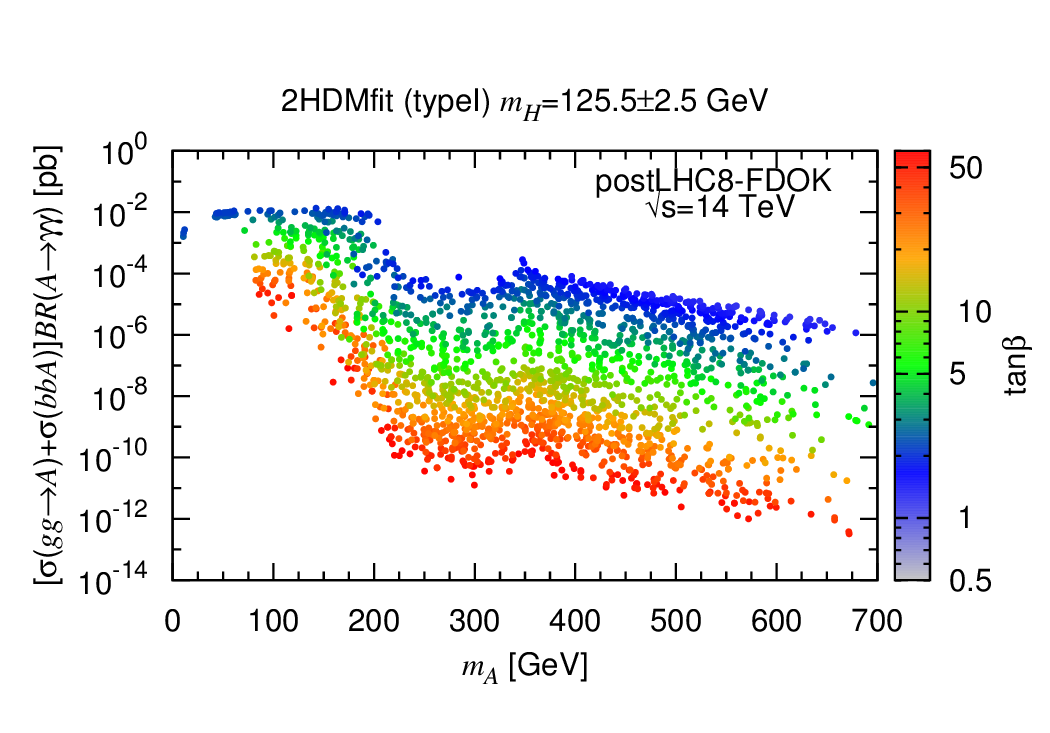}
\includegraphics[width=0.49\textwidth]{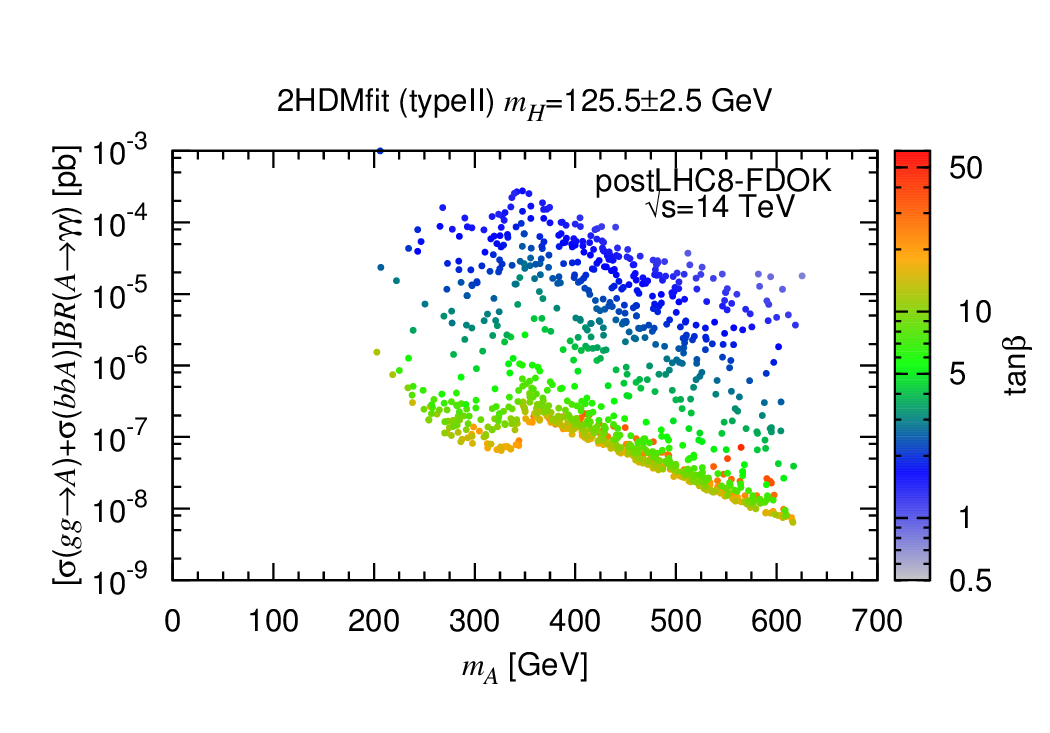}
\includegraphics[width=0.49\textwidth]{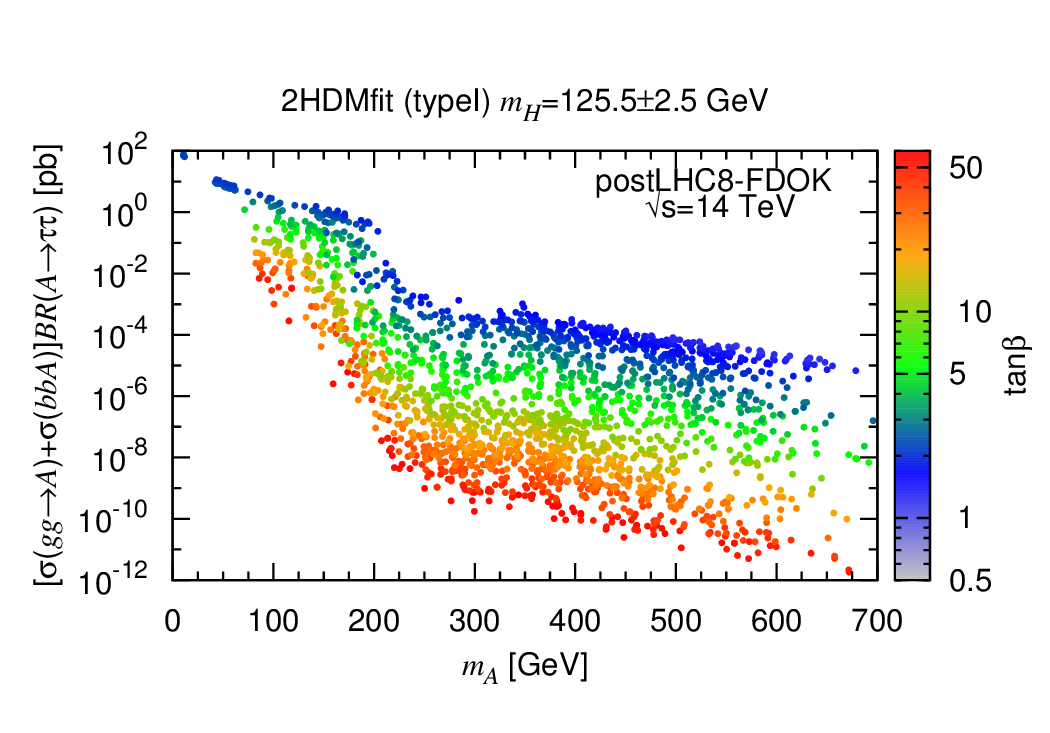}
\includegraphics[width=0.49\textwidth]{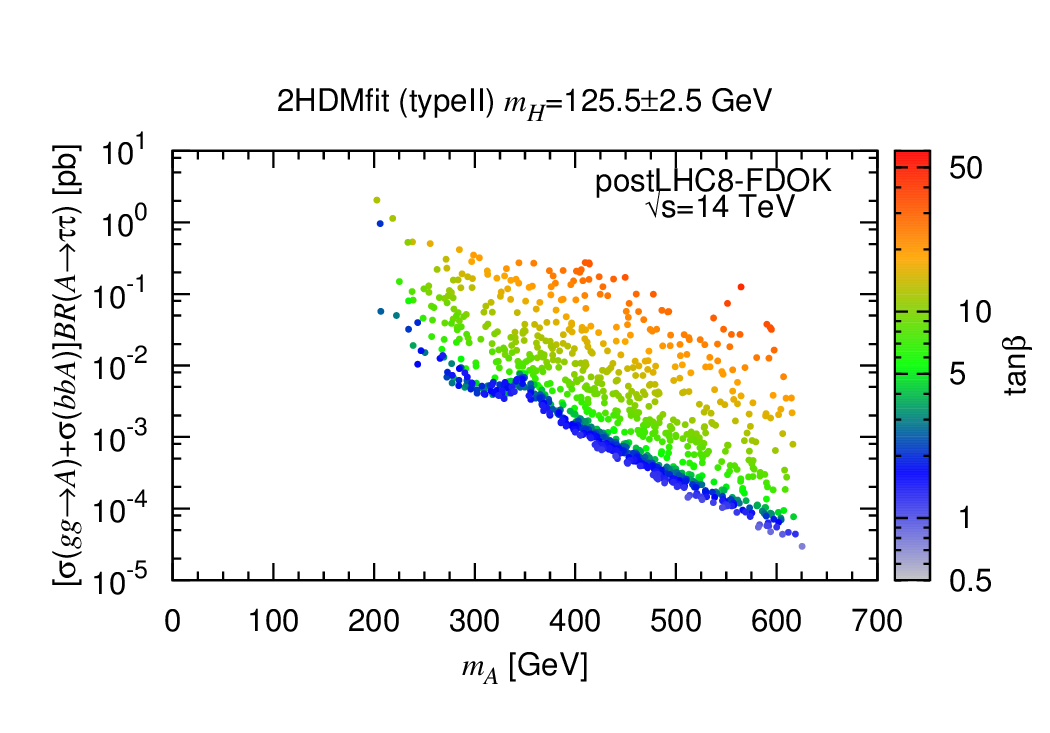}
\includegraphics[width=0.49\textwidth]{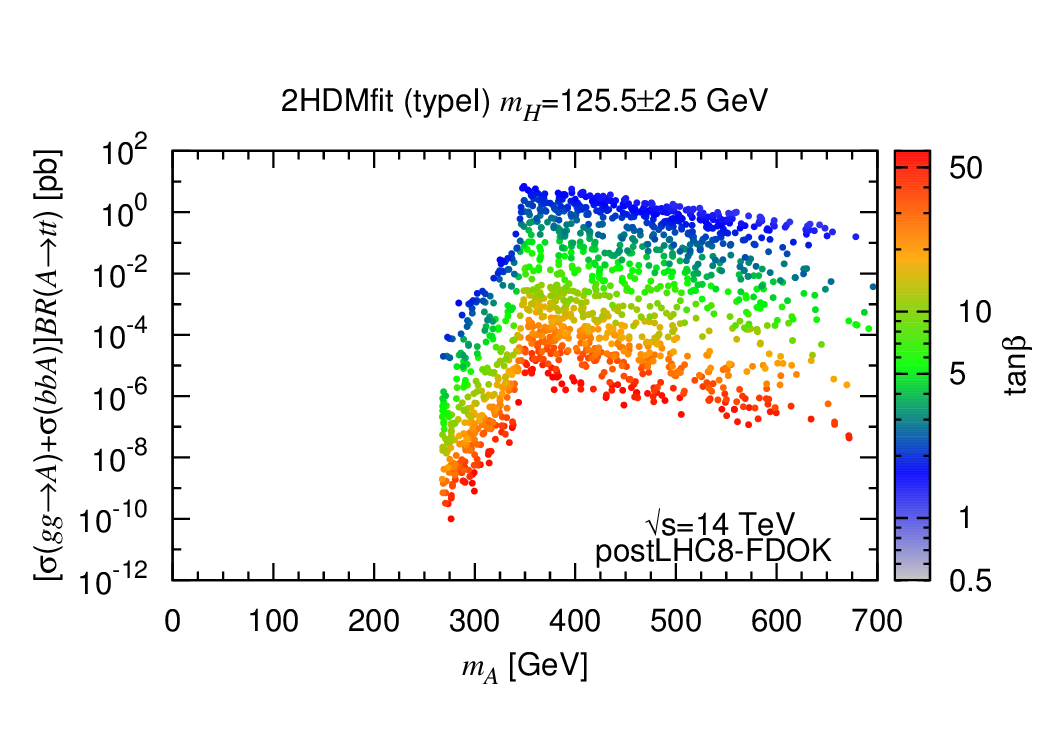}
\includegraphics[width=0.49\textwidth]{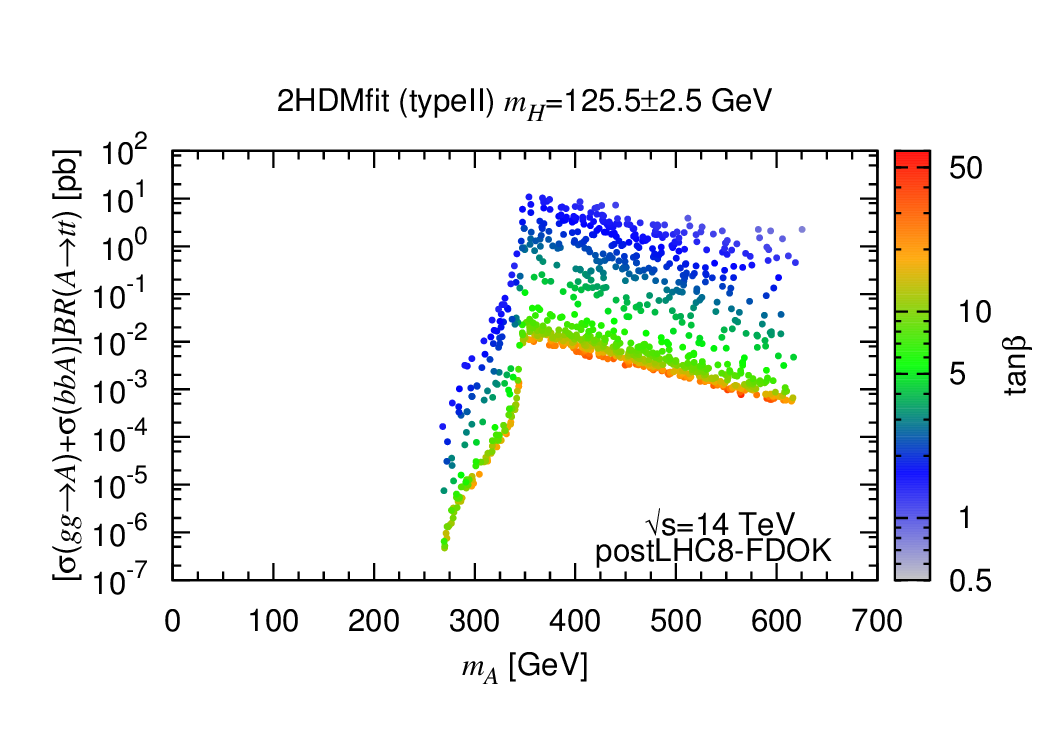}
\end{center}\vspace*{-5mm}
\caption{Rates (in pb) of pseudoscalar $A$ production at $\sqrt{s}=14$~TeV as a function of $\mA$ for the $m_H\sim 125.5$~GeV scenarios, separated into different $A$ decay modes: $A\to \gam\gam$ (top),  $A\to \tau\tau$ (middle) and $A\to t\bar t$ (bottom). In each case, we sum over $gg\to A$ and $gg\to bbA$ production. The values of $\tan\beta$ are color-coded as indicated by the scale on the right of the plots. 
}
\label{Aprod14tevmH125}
\end{figure}

Finally, let us turn to the question of detecting the pseudoscalar $A$.  
Figure~\ref{Aprod14tevmH125} shows cross sections for pseudoscalar $A$ production, concretely  $[\sigma(gg \to A)+\sigma({ bb}A))]\times \br (A \to \gam\gam)$ (top), $\times \, \br(A\to \tau\tau)$ (middle) and $\times \, \br(A\to t\anti t)$ (bottom)   at $14\tev$ as a function of $\mha$.\footnote {As commented in the last section, we plot the sum as this defines the inclusive production rate.  Of course, separating $gg\to A$ and $bbA$ production processes would eventually be possible.}
Again, there is a large range of possible cross section values at any given $\mha$, with the $\tan\beta$ dependence, of course, being the same as for the $\mh\sim 125.5\gev$ case. 
As already observed in Fig.~\ref{lamthreepert}, the possible range of $\mha$ is limited when $\mhh\sim125.5\gev$. 
In the case of Type~II models, $\mha\lesssim 200\gev$ is eliminated due to the $B$-physics  limit of $\mhpm\gsim 300\gev$ and the requirement of an acceptable $T$ parameter (which limits the $A-H^\pm$ mass difference). In the case of Type~I models,  $\mha\lsim 60\gev$ is possible but finding points with small enough $\hh\to\ha\ha$ to allow the $H$ to have reasonably SM-like properties requires significant fine-tuning. 
For most $\mha$, the Type~II maximal and minimal cross sections tend to be substantially (by a factor of $>1000$) larger than for Type~I. 
The lowest cross section values in Type~I models are really very small at the largest allowed $\mha$ values and would not allow the detection of the $A$ boson.  In contrast, in Type~II models, even the very lowest cross section value of $\sim 5\times 10^{-5}\pb$ at $\mha\sim 630\gev$ would imply a handful of events 
for $L=300\fbi$.
The maximum  Type~II values  imply a substantial number of events at all $\mha$, even at the largest masses,  $\mha\sim 630\gev$.  [We remind the reader that $630\gev$ is the upper limit allowed by perturbativity once the precision Higgs constraints, which limit $|\!\sbma|$  to smaller values,   have been included, cf.\ Fig.~\ref{lamthreepert}.]

As before, the rates for $A$ production in the $\mu\mu$  final state are simply obtained by the $\tanb$-independent rescaling factor $\br(A\to\mu\mu)/\br(A\to \tau\tau)\sim 3.5\times 10^{-3}$.  For $\mha\lsim 2m_t$, the  cross section values near the upper limit obtained from such rescaling of the $\tau\tau$ final state rates shown in Fig.~\ref{Aprod14tevmH125}  are likely to be observable given the relatively narrow nature of the mass peak (typically of order a few GeV) and the excellent $\mu\mu$ invariant mass resolution.   

Last but not least, an interesting question is whether the $\hl$, $\ha$ (and $\hpm$)  could {\it all} escape detection for some parameter choices when $\mhh\sim 125.5\gev$. Given the upper limit discussed above of $\mha\lsim 630\gev$  (after Higgs fitting constraints), a careful examination is required.

Consider first the $\ha$.   In the case of Type~I, Fig.~\ref{Aprod14tevmH125} shows that
for $\mha\in[200,300]\gev$ (and most probably all the way out to the maximum allowed $\mha$) and large $\tanb$ the $\gam\gam,\tau\tau,t\anti t$ rates are all very small and would not allow discovery of the $\ha$ even with $L=3000\fbi$. In the case of Type~II, the anticorrelation between the $\tau\tau$ and $t\anti t$ rates as $\tanb$ is varied, combined with the relatively substantial $\tau\tau$ rates for $\mha$ below the $t\anti t$ threshold combine to guarantee that the $\ha$ should be detectable with $L=3000\fbi$ in either the $\tau\tau$ or the $t\anti t$ mode, and perhaps in both.

As regards the $\hl$, it is first of all clear that if $|\!\sbma|$ is very small (as certainly both allowed and preferred by Higgs fitting) then the $V^*\to Vh$ rates will be very tiny, as illustrated in Fig.~\ref{vhbbvsmh} in the $\mhl>60\gev$ region for both Type~I and Type~II.   The other potentially viable mode is $gg\to\hl\to\gam\gam$.  However, Fig.~\ref{siggghAA} shows very small rates in this case as well for Type~I.  In contrast, the lowest cross sections in this channel in Type~II are of order $10^{-5}\pb$, a level which might be accessible with $L=3000\fbi$. 

\begin{figure} [t]
\begin{center}
\includegraphics[width=0.49\textwidth]{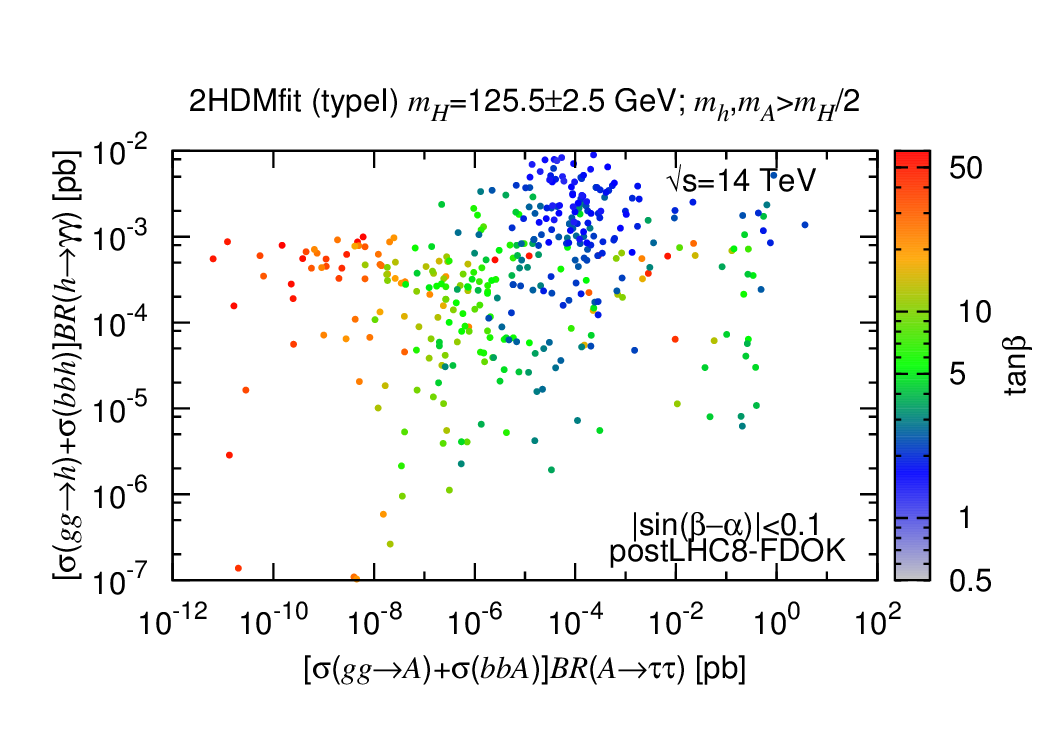}
\includegraphics[width=0.49\textwidth]{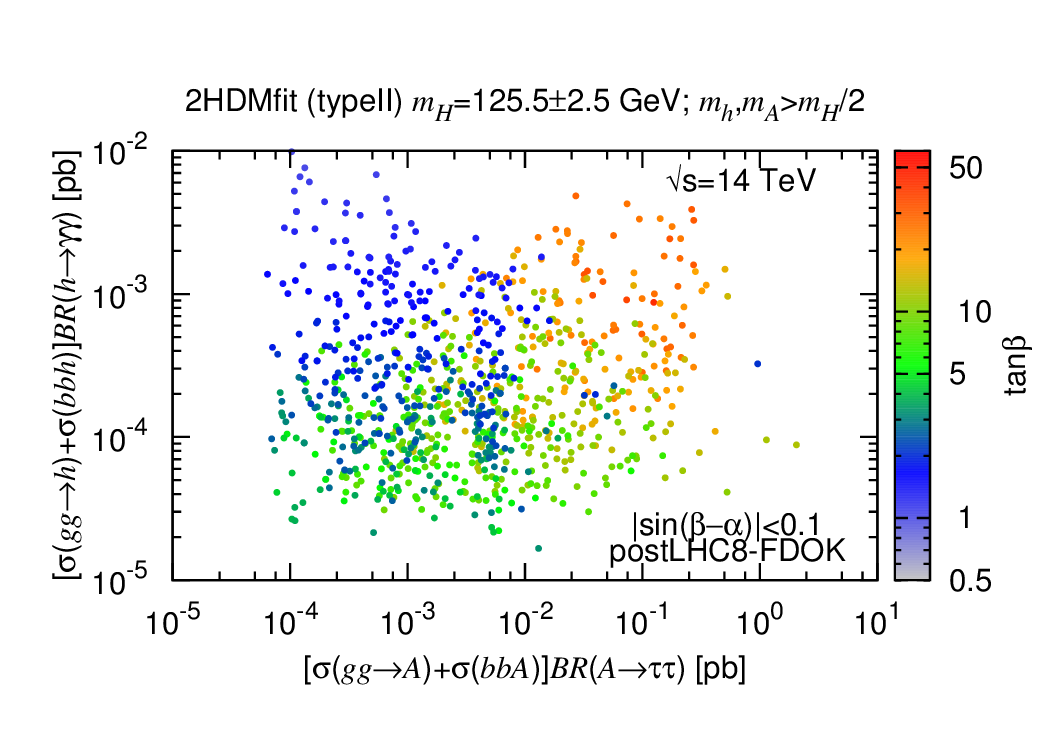}
\end{center}\vspace*{-5mm}
\caption{We plot $[\sig(gg\to h)+\sig(bbh)]\br(h\to \gam\gam)$ vs.\ $[\sig(gg\to A)+\sig(bbA)]\br(A\to \tau\tau)$ for $\rts=14\tev$ with postLHC8-FDOK constraints imposed. Colors indicate the value of $\tanb$.  We have required $|\!\sbma|<0.1$ so that the rate for $V^*\to Vh$ will be at most 0.01 times the SM rate at the same mass.}
\label{sigcorrplot}
\end{figure}

To summarize the above, in Type~II $\mhh\sim 125.5\gev$ scenarios, the prospects for the discovery of the $A$ boson are very good, and there is an excellent chance of finding the $\hl$ as well in the $\gam\gam$ final state.  In contrast, in Type~I there are clearly regions of parameter space for which no Higgs boson other than the SM-like $H$ will be discoverable without going to still higher luminosity or energy.  These results are summarized in Fig.~\ref{sigcorrplot} where we see from the right-hand plot that in Type~II both the $\hl$ and $\ha$ have an excellent chance of being detectable in the $\gam\gam$ and $\tau\tau$ modes, respectively. In contrast, the left-hand plot shows that there is clearly a corner of parameter space where detection of both the $\hl$ and the $\ha$ will be extremely challenging.  Of course, we have not explored prospects for $\hpm$ discovery in this paper, but its production cross sections are typically substantially below those for the $\ha$ since $\hpm$ production requires at least one top quark in the final state.

Finally, we note the regions with $\mha\sim 200-350\gev$ and low $\tanb$ of the $\gam\gam$ and $\tau\tau$ final state Type~I plots that have been depleted by the removal of points with excessive feed down.  These regions are, of course, the analogue of the FD-depleted regions discussed in the case of $\mh\sim 125.5\gev$.

\section{Conclusions}\label{conclusions}

The latest Higgs data from the LHC clearly favor a fairly SM-like Higgs boson with mass of about $125.5\gev$.  
In this paper we have quantified this in the context of  Type~I and Type~II 2HDMs, identifying either the $h$ 
or the $H$ as the $125.5\gev$ state.  Indeed, the vector boson pair coupling of the $125.5\gev$ state must be quite close to the SM Higgs coupling, but at low $\tanb$ there is significant dispersion about the SM values for the fermionic couplings.  Given the constraints on the $125.5\gev$ state, we have ascertained expectations for the other Higgs bosons $H$, $\hpm$ and $A$ in the case that it is the $h$ that gives the SM-like signal. We demonstrated that there are many parameter space points (satisfying all 95\% C.L. constraints on the $h$) that for $\rts=14\tev$ could allow for observation of the $H$ and $A$ in a  variety of modes, including not just the $ZZ$ (for $H$ only),  $\tau\tau$ and $t\bar t$ final states, but also $\mu\mu$ and $\gam\gam$. In the $\mH\sim 125.5\gev$ case, the $A$ can also be detected in these same modes (except $ZZ$).  In addition, there is good probability for viable signals for the lighter $h$.  In particular, the $V^*\to Vh$ with $h\to b\anti b$ channel might yield a detectable signal for at least a fraction of the points surviving the 95\% C.L. limits on the $H$ signal at $125.5\gev$.

Nondecoupling of the charged-Higgs loop contribution to the $h\gam\gam$ or $H\gam\gam$ coupling plays an important role once rates are required to be within $\pm 5\%$ of the SM predictions.  The result is elimination of all but a tiny fraction of the $\mhh\sim 125.5\gev$ Type~I scenarios and {\it all} of the $\mhh\sim 125.5\gev $ Type~II scenarios as well as {\it all} of the wrong-sign Yukawa Type~II scenarios in the $\mhl\sim 125.5\gev$ case.

Along the way, we clarified the important role played by constraints deriving from requirements of perturbativity in reducing the otherwise huge parameter space of the 2HDMs.  We also delineated the (fortunately) rather limited impact of removing points that suffer from a large amount of feed down from heavier Higgs bosons decaying either directly or via chain decay to a final state containing the $125.5\gev$ state.

An important general conclusion is that even  if the $125.5\gev$ signal rates converge to very SM-like  values, the 2HDMs predict ample opportunity for detecting the other Higgs bosons.  Still, it is also true that  in the case of $\mh\sim 125.5\gev$ in the decoupling limit of  very large $\mha\sim\mhh\sim \mhpm$ detection of even one of the other Higgs bosons at LHC14 would not be possible.  However, the case of $\mhh\sim 125.5\gev$ is different.  Because the maximum $\mha$ is limited, having the $\hl$, $\ha$ and $\hpm$ all escape detection is very unlikely in Type~II 2HDM.  However, in Type~I if $\mha\in[200,300]\gev$ and $\tanb$ is large, then the $\ha$ will not be detectable and $\hl$ detection will be very marginal.  

Finally, we make special note of the scenarios with low $\mha<100\gev$ that escape all LEP and (so far) LHC limits and yet have quite substantial $gg\to A$ and $bbA$ production cross sections. 
It will be interesting to probe these scenarios, which are possible for both Type~I and Type~II in the $\mhl\sim 125.5\gev$ case and for Type~I in the $\mhh\sim 125.5\gev$ case, in ongoing analyses of LHC 8 TeV data and in future LHC running at higher energy.

\section{Acknowledgements}

This work was supported in part by U.S. DOE Grant No. DE-SC-000999 and by IN2P3 under PICS FR--USA Contract No.~5872. 
Y.J.\ is also supported by  LHC-TI fellowship  US NSF Grant No. PHY-0969510. In addition, he thanks the  ``Investissements dÕavenir, Labex ENIGMASS''  for partial financial support for a research stay at LPSC Grenoble. 
J.F.G.\ and S.K.\  thank the Aspen Center for Physics, supported by the National Science Foundation under Grant No. PHYS-1066293, for hospitality.

\appendix
\clearpage

\section{Generic formalism for feed down processes}
\label{app:feeddown}

We have quantitatively computed feed down as follows.  Let us define the final state of interest by 
$Y=\gam\gam,VV,b\anti b,\tau\tau$ (the only final states quantitatively of relevance for fitting the $\sim 125.5\gev$ signal using the 7 + 8 TeV data set).  The production processes will be denoted by 
 $X{\rm H}=$ggFH (standing for $gg\to ${\rm H}), bbH (standing for associated production of the H with a $b\anti b$ pair),  ttH, VBFH (standing for $WW/ZZ\to {\rm H}$), and VH. (Note that the regular Roman H is generic and we will be using these production processes for various different choices of H.) In this paper, we will consider ggFH and bbH together and compute their combined feed down. We wish to consider the feed down into  ${\bf H}= h$ and $H$ where $\bf H$ is always the $125.5\gev$ state.  In the process, we will have to consider production of all Higgses that are heavier than the ${\bf H}$, which Higgses we will denote by ${\cal H}$.

 For all but VH, we can define the fraction of feed down contributing to production mode $X$ as 
 \beq
\mu_{X{\bf H}}^{\rm FD}\equiv  {\sum_{\cal H}  \sigma_{X{\cal H}}P_{\rm FD}({\cal H} \to {\bf H}+\mbox{anything}) \over \sigma_{X{\bf H}} }\,.
\eeq
Note that we have cancelled a common factor of $\br({\bf H}\to Y)$,  the branching ratio for $\bf H$ to decay to any final state $Y=\gam\gam,VV,b\anti b,\tau\tau$, that is common to the numerator and denominator.  In the above, we have to take into account the possibilities that the intermediating ${\cal H}$ can decay to two $\bf H$'s or only one $\bf H$.  Thus, we actually have
\beq
P_{\rm FD}({\cal H}\to {\bf H}+\mbox{anything})=2P_{{\cal H},2{\bf H}}+P_{{\cal H},1{\bf H}}\,,
\eeq
so that in the first case the ``anything" includes the 2nd ${\bf H}$.
For $2{\bf H}$ final states we include a multiplicative factor of 2 since both of the final ${\bf H}$'s will contribute. The formulas for the various $P_{{\cal H},2{\bf H}}$ and $P_{{\cal H},1{\bf H}}$'s cases can be found in \cite{Arhrib:2013oia} which we will repeat in the following context (including some corrections we found).

The VH process must be handled differently due to the fact that it is directly impacted by $gg\to A$ and $bbA$ production with $A \to Z{\bf H}$ in the case of $V=Z$ and by $gg\to t b \hpm\to tb \wpm {\bf H}$ in the case of $V=W$.  The largest of these fractional contaminations is that associated with  $gg\to A+bbA$ since the cross sections for $\hpm$ production are quite a bit smaller. In the $V=Z$ case, the leading contribution to the fractional feed down is
\beq
\mu_{Z{\bf H}}^{\rm FD}={[\sigma_{{\rm ggF}A}+\sigma_{{\rm bb}A}]\br(A\to Z {\bf H} )\over \sigma_{{\rm Z}{\bf H}}}\,,
\eeq
where the $\br({\bf H}\to Y)$ branching ratios in numerator and denominator have once again cancelled and $\sigma_{{\rm Z}{\bf H}}$ is that arising from the $Z^*\to Z{\bf H}$ subprocess.  In the above,
we have neglected (a good approximation) contributions to the $Z{\bf H}+{\rm anything}$ final state due to  the FD processes other than the direct $gg\to A\to Z{\bf H}$ production process.  In particular, we neglect the terms which lead to final states containing $Z{\bf H}$ plus additional particles.
Besides yielding small contributions to $\mu_{V{\bf H}}^{\rm FD}$, such terms yield final states that are much more complicated than the simple $Z{\bf H}$ final state, containing, for example, more than one $Z$ or a $Z+W$ pair. The selection criteria for the $Z{\bf H}$ final state are likely to be such that these more complicated final states were discarded or strongly discriminated against. In the following subsections, we apply these general formulas to the two 2HDM scenarios discussed in this paper, showing how to determine the magnitude of feed down associated with a given point in parameter space.

\subsection*{\boldmath Example I: Feed down of heavier Higgs bosons to a $125.5\gev$ $h$} 

In this case, the ${\bf H}=h$ can be produced from the chain decay of heavier Higgs bosons ${\cal H}=H,A,\hpm$. Since the cross section for $gg\to tb \hpm$ is relatively smaller than those for $H$ and $A$ (because of the $t$ in the $tb\hpm$ final state), we will not include $\hpm$ as an intermediate source for the FD mechanism.  In practice, important FD contributions arise from (ggF+bb)$H$, (ggF+bb)$A$ or VBF$H$.  The most important feed down decays for ggF+bb are $H\to hh$ and $A\to Zh$ (or $A\to ZH$ for the case of $\mH\sim 125.5\gev$) and for VBF $H\to hh$. However, the approximations of
\beq
P_{H,2h}\simeq \br(H\to hh)\,,\quad P_{A,1h}\simeq \br(A\to Zh)\,,
\eeq
are not generally adequate. 

It is convenient to consider separately the chains that can yield $2h$ in the final state vs.\ those that yield only $1h$ in the final state.  Some sample net branching ratios, denoted (following \cite{Arhrib:2013oia}) by $P$ are the following.  First, if $H$ is the primary then $H\to AA$ and $H\to \hp\hm$ are possible for some 2HDM points in addition to $H\to hh$.  Thus, we have
\begin{align}
P_{H,2h}&=\br(H\to hh)+\br(H\to\hp\hm)\bigl[\br(\hp\to \wp h)^2 \label{H2h} \\
&+ \br(\hp\to \wp A)^2\br(A\to Zh)^2 \nonumber \\
&+ 2 \br(\hp\to \wp h)\br(\hp\to \wp A) \br(A\to Zh)\bigr] \nonumber \\
&+ \br(H\to AA)\bigl[\br(A\to Zh)^2+ { 4} \br(A\to \wm \hp)^2\br(\hp\to \wp h)^2 \nonumber \\
&+ { 4} \br(A\to Zh)\br(A\to \wm \hp)\br(\hp\to \wp h)\bigr] \,, \nonumber
\end{align}
\begin{align}
P_{H,1h}&=\br(H\to ZA)\br(A\to Zh)+2\br(H\to \wm\hp)\br(\hp\to \wp h) \label{H1h} \\
&+ 2\br(H\to\hp\hm)\bigl[\br(\hp\to \wp h) \nonumber \\
&+ \br(\hp\to \wp A)\br(A\to Zh)\bigr]p(\hp\not\to h) \nonumber \\
&+ \br(H\to AA)\bigl[2 \br(A\to Zh)+{ 4}\br(A\to \wm \hp)\br(\hp\to \wp h)\bigr]p(A\not\to h) \,, \nonumber
\end{align}
where
\begin{align}
p(\hp\not\to h)&= 1-\br(\hp \to \wp h)-\br(\hp\to \wp A ) \br(A\to Z h) \,,\\
p(A\not\to h)&= 1-\br(A\to Zh)-{ 2} \br(A\to\wm \hp)\br(\hp\to \wp h) \,. \nonumber
\end{align}
If $A$ is the primary then possibly accessible modes at the first stage are $A\to ZH$, and $A\to \wp\hm,\wm\hp$, resulting in
\begin{align}
P_{A,2h}&=\br(A\to ZH)\overline P_{H,2h}+2\br(A\to \wm\hp)\br(\hp\to \wp H)\br(H\to hh) \,, \\
P_{A,1h}&=\br(A\to Zh)+\br(A\to ZH) \overline  P_{H,1h} \nonumber \\
&+ 2\br(A\to\wm\hp)\br(\hp\to \wp h) \nonumber \,,
\label{A12h}
\end{align}
where the appropriate expressions for the two $P$'s are the same as given earlier with the $A$ channels eliminated since $H$ being a secondary particle must be lighter than the $A$.
\begin{align}
\overline P_{H,2h}&=\br(H\to hh)+\br(H\to \hp\hm)\br(\hp\to\wp h)^2 \,, \\
\overline P_{H,1h}&=2\br(H\to \wm\hp)\br(\hp\to \wp h) \nonumber \\
&+ 2\br(H\to\hp\hm)\br(\hp\to \wp h) p(\hp\not\to h) \nonumber \,.
\end{align}
For the sake of completeness, we also provide the formula for the case that $\hp$ or $\hm$ is produced as the primary Higgs. In this case, the first level decays of relevance are $\hp\to \wp h$ and $\hp\to \wp H$, leading to
\begin{align}
P_{\hp,2h}&=\br(\hp \to \wp H)P_{H,2h}+\br(\hp\to \wp A)\br(A\to ZH)\br(H\to hh) \,, \\
P_{\hp,1h}&= \br(\hp\to \wp h)+\br(\hp\to \wp H)P_{H,1h}+\br(\hp\to \wp A)\br(A\to Zh) \,, \nonumber
\end{align}
where the $P_{H,2h}$ and $P_{H,1h}$ are as given earlier in Eqs.~(\ref{H2h}) and (\ref{H1h}), respectively, with kinematics for the $\hp$ primary situation eliminating the terms involving $H\to \wm \hp$ and $H\to \hp\hm$.

\begin{figure}[h]
\begin{center}
\includegraphics[width=0.55\textwidth]{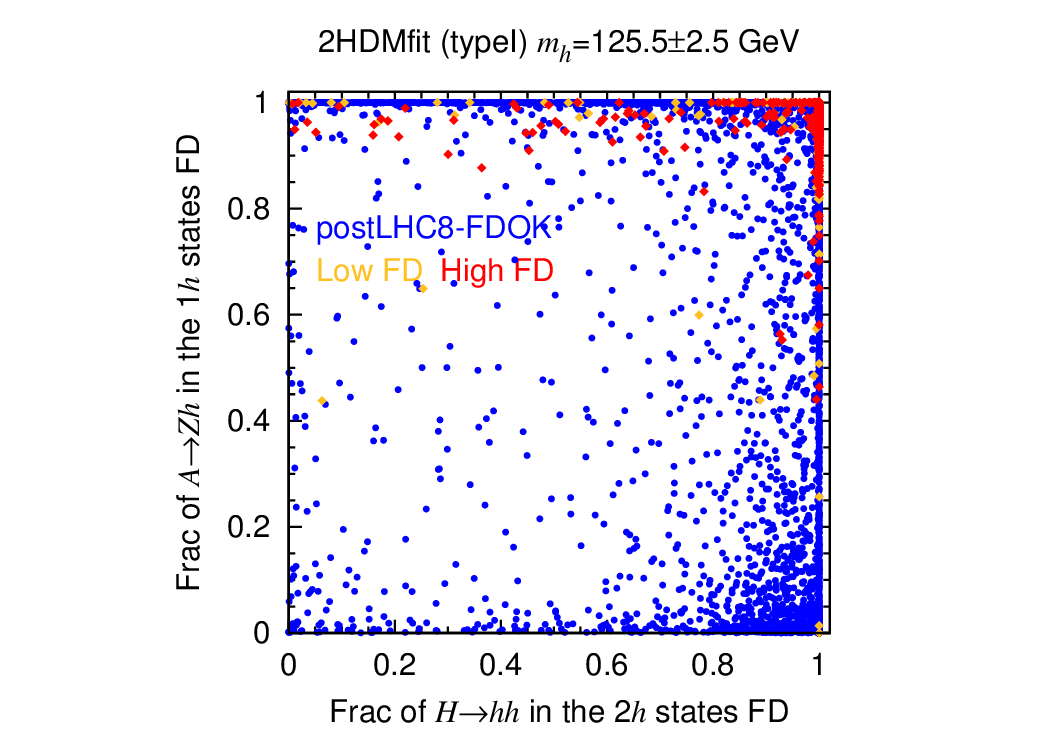}
\hspace*{-25mm}
\includegraphics[width=0.55\textwidth]{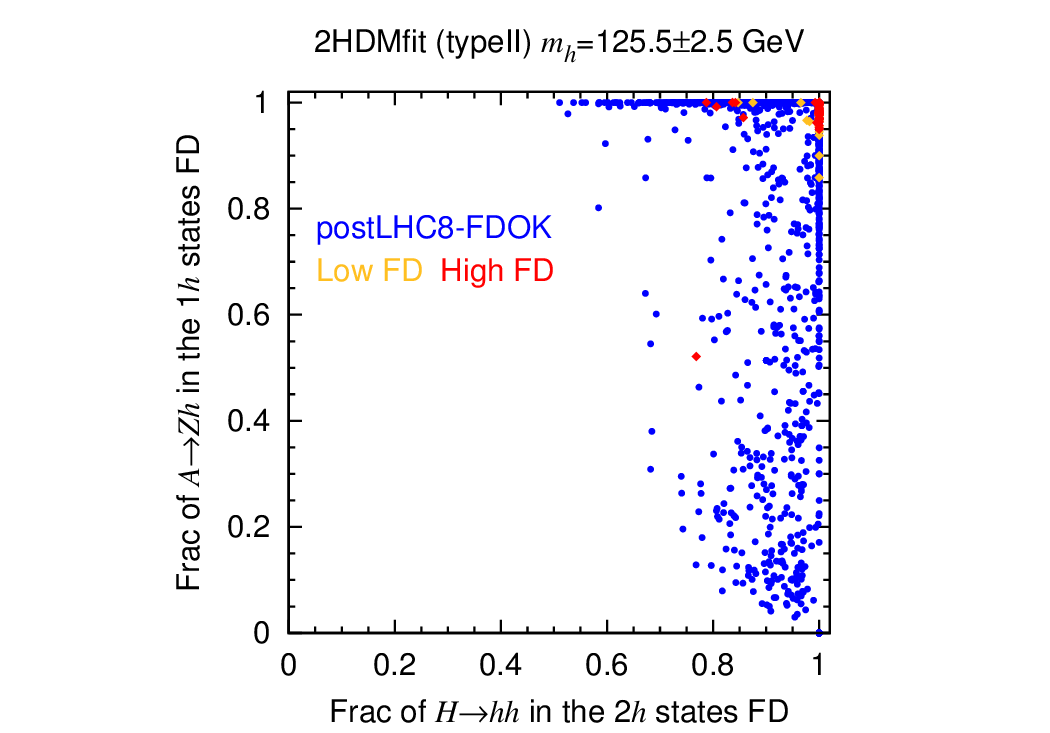}
\end{center}\vspace*{-5mm}
\caption{We plot  $[\sigma(gg\to H+bbH)2\br(H\to hh)]/[\sigma(gg\to H+bbH)2P_{H,2h}+\sigma(gg\to A+bbA)2P_{A,2h}]$ on the $x$ axis and $[\sigma(gg\to A+bbA)\br(A\to Zh)]/[\sigma(gg\to H+bbH)P_{H,1h}+\sigma(gg\to A+bbA)P_{A,1h}]$ on the $y$ axis. The postLHC8-FDOK points are displayed in blue. The cyan, green and red points obey FDOK constraints and have all the channel rates of Eq.~(\ref{xxyychannels}) within SM$\pm15\%$, SM$\pm 10\%$ and SM$\pm 5\%$, respectively; cf.  Fig.~\ref{cbmavstbfuture}.}
\label{fractionfd}
\end{figure}

Furthermore, in order to illustrate the importance of keeping the full formulas we show in Fig.~\ref{fractionfd} the fraction of $1h$ final states coming directly from $A\to Zh$ vs.\ the fraction of $2h$ final states coming directly from $H\to hh$.

\subsection*{\boldmath Example II: Feed down of heavier Higgs to a $125.5\gev$ $H$}

The case of ${\bf H}=H$ in the 2HDM is much simpler. Ignoring the $\hpm$ as explained above, the only heavier Higgs which can feed the $H$ signal is ${\cal H}=A$. Therefore, the most important FD processes are  $gg \to A+bbA$ with $A\to ZH$. However, the approximations of
\beq
P_{A,1H}\simeq \br(A\to ZH)\,,
\eeq
are not generally applicable. In order to measure the FD precisely, one should use the full expression for $P_{A,2H}$ and $P_{A,1H}$ as follows
\begin{align}
P_{A,2H}&=\br(A\to \hp\hm) \br(\hp\to \wp H)^2 \,, \\
P_{A,1H}&=\br(A\to ZH)+2\br(A\to\wm\hp)\br(\hp\to \wp H) \nonumber \\
&+ 2\br(A\to \hp\hm) \br(\hp\to \wp H) p(\hp\not\to H) \nonumber \,,
\end{align}
where
\beq
p(\hp\not\to H)=1-\br(\hp \to \wp H)\,.
\eeq

\section{\boldmath Nondecoupling of the $\hpm$ loop contribution to the $H \gam\gam$ coupling and wrong-sign Yukawa effects on the $ggH$ coupling.}
\label{nondecoup}

We have claimed that the $\mhh=125.5\gev$ Type~II scenarios can be either eliminated or confirmed when the LHC measurements reach a precision such that the rates in the various initial$\times$final state channels can be measured to 5\% accuracy.  As we have said, this is because for this scenario the charged Higgs loop does not decouple in the $H \gam\gam$ coupling calculation and results in at least a 5\% reduction in the coupling  relative to the SM prediction, \ie\ in $\cp^H$.  
This can easily be explained using the results of Appendix B of \cite{Ferreira:2014naa} as translated to the current situation.  

We first emphasize that in all the fits, whether Type~I or Type~II, the LHC data require that the sign of the top-quark Yukawa, \ie\ $\cu^H$, be the same as the sign of the $\hh VV$ coupling $\cv^H$ in order that the $\gam\gam$ final state rates not be too enhanced. Referring to Table~\ref{tab:couplings},  we see that a SM-like $\hh$ results in $\cbma=\pm 1$, for which $\cu^H=\pm 1$, respectively.  In our scan range of $|\alpha|\leq \pi/2$ both signs are found. For Type~I, $\cu^H$ and $\cd^H$ have the same sign.  For Type~II, $\cd^H$ is always positive within the $\alpha$ scan range. Thus, in Type~I we always have the same relative signs between all the couplings as for a SM Higgs boson, whereas in Type~II there will be scan points for which $\cd^H$ has the ``wrong" sign relative to $\cv^H$ and $\cu^H$.  This will impact the $gg\hh$ 1-loop coupling because the interference between the top-quark and bottom-quark loops will change sign, thereby increasing the $gg\hh$ coupling by a factor of $\cg^H\sim 1.12$ relative to the SM value --- see Appendix B of \cite{Ferreira:2014naa} --- a level that can be probed in a future LHC run from the observation of the decay products of Higgs bosons originating from gluon fusion.  In contrast, the change in sign of the bottom-quark loop contribution to the $H \gam\gam$ coupling yields a $\lsim 1\%$ decrease in $\cp^H$ due to the dominance of the $W$- and $t$-loop contributions.  However, it turns out that for {\it all} of the Type~II scenarios the charged-Higgs loop contribution to the $H \gam\gam$ coupling does not decouple as $\mhpm$ becomes large.  As a result, all Type~II scenarios with $\mhh\sim 125.5\gev$ will be eliminated if  future LHC data determine that $\cp^H$ is within $5\%$ of the SM prediction of unity. This happens regardless of the common sign of $\cv^H$ and $\cu^H$, denoted by $\sgn(\cv^H)$.

\begin{figure}[h!]
\begin{center}
\includegraphics[width=0.49\textwidth]{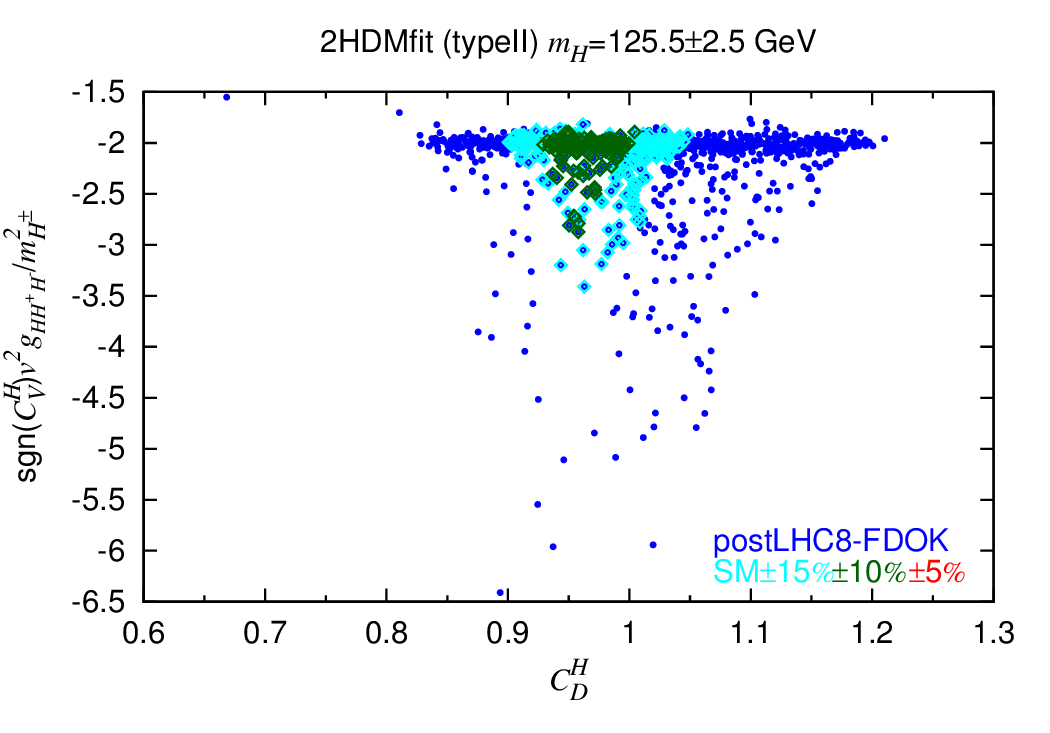}
\includegraphics[width=0.49\textwidth]{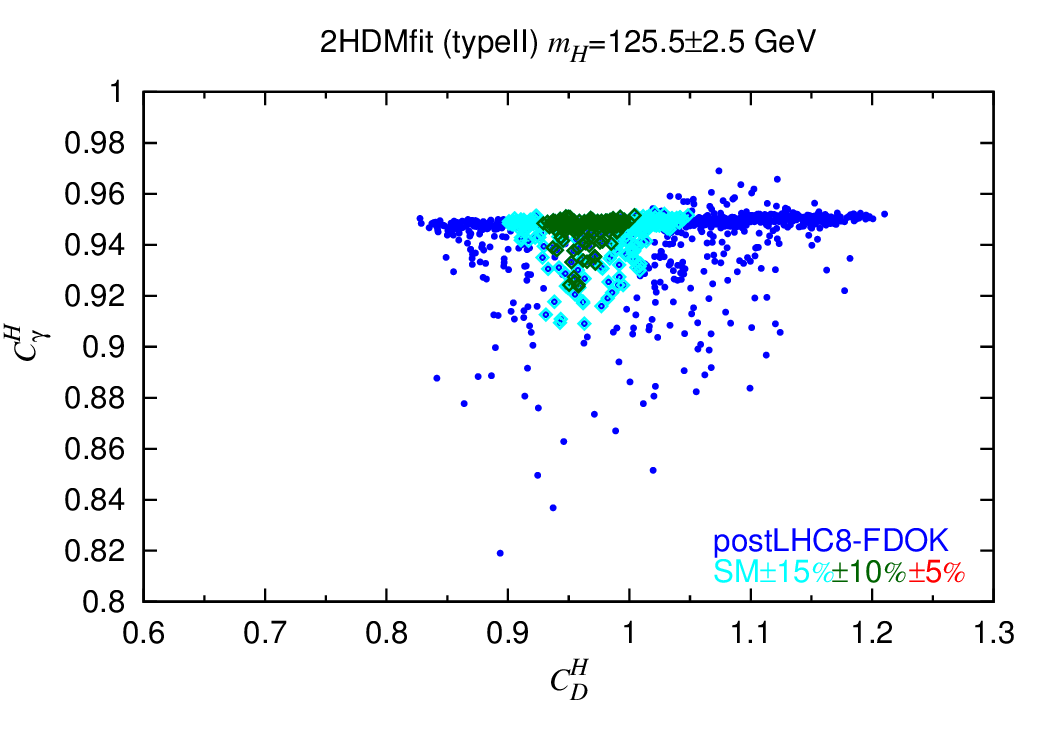}
\end{center}\vspace*{-5mm}
\caption{For Type~II 2HDM, we plot: $v^2 g_{\hh\hp\hm} \sgn(\cv^H) / \mhpm^2$ vs.\ $\cd^H$ (left) and $\cp^H$ vs.\ $\cd^H$ (right). We show FDOK points consistent at the postLHC8 level (blue) and after presumed LHC measurements showing SM consistency at the $\pm15\%$ (cyan) and  $\pm 10\%$ (dark green) level.  No points survive if SM consistency is demonstrated at the $\pm5\%$ level.  }
\label{cgamplot}
\end{figure}

To gain intuition, we employ the results of \cite{Gunion:2002zf} to find that in the limit where $\cbma\to \pm 1$ the dimensionless $\hh \hp\hm$ coupling takes the form
\beq
g_{\hh\hp\hm}=-{\sgn(\cv^H) \over v^2}(\mhh^2+2\mhpm^2-2\mhat^2)\,.
\label{ghhhphmform}
\eeq
Further, perturbativity requires that $\mhat^2$ be of modest size, and, of course, $\mhh$ is fixed at $\sim 125.5\gev$.
Thus, we have the asymptotic result that
\beq
{v^2 g_{\hh\hp\hm} \sgn(\cv^H) \over \mhpm^2} \to -2\,.
\eeq
As shown in Appendix B of \cite{Ferreira:2014naa} for the case of $\cv^H>0$ (but the same arguments apply with all signs changed to the $\cv^H<0$ case\footnote{The $b$-quark loop, which does not change sign when $\cv^H<0$, is negligible relative to other contributions.}), this means that $\cp^H$ is reduced by 5\% relative to the SM value of unity.
Of course, $\mhat$ is not exactly zero, but one finds that $\mhat^2$ lies below $(100\gev)^2$ while being able to reach large negative values for a large range of $\mhpm$. The result is that  $v^2 g_{\hh\hp\hm} \sgn(\cv^H)$ is always {\it at most} $-2 \mhpm^2$, implying that the decrease in $\cp^H$ is at least 5\%. This is illustrated in Fig.~\ref{cgamplot}.  

\begin{figure}[h!]
\begin{center}
\includegraphics[width=0.49\textwidth]{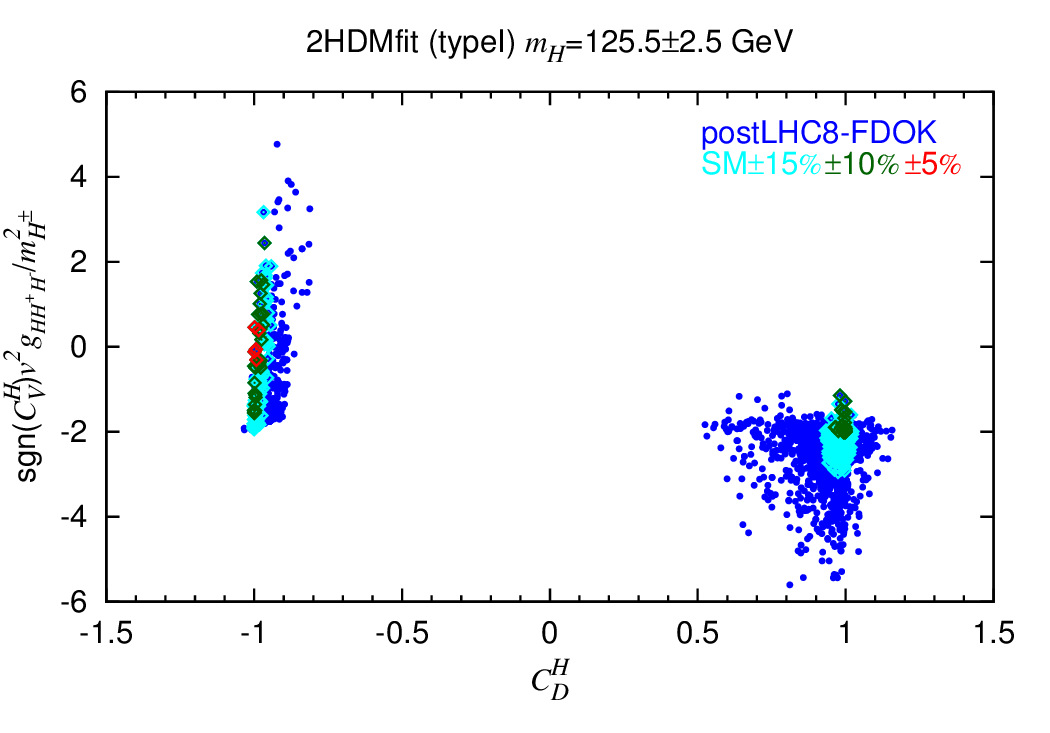}
\includegraphics[width=0.49\textwidth]{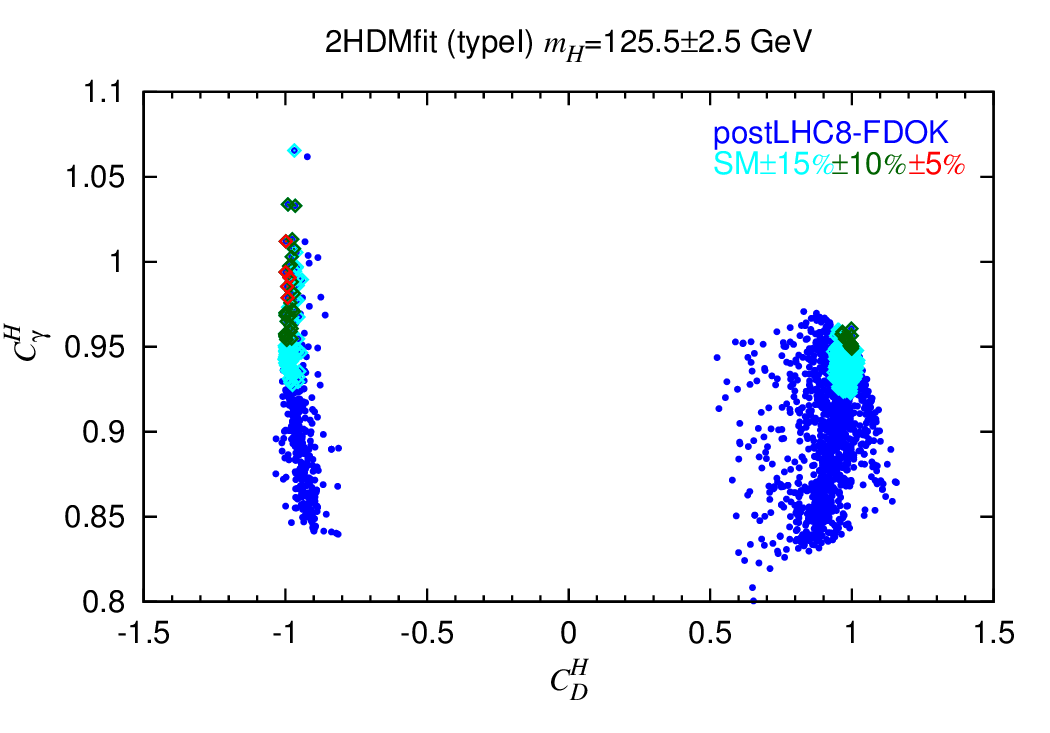}
\end{center}\vspace*{-5mm}
\caption{For Type~I 2HDM, we plot: Left, $v^2 g_{\hh\hp\hm} \sgn(\cv^H) / \mhpm^2$ vs.\ $\cd^H$ (left) and $\cp^H$ vs.\ $\cd^H$ (right), showing FDOK points consistent at the postLHC8 level (blue) and after presumed LHC measurements showing SM consistency at the $\pm15\%$ (cyan),  $\pm 10\%$ (dark green) and $\pm 5\%$ (red) levels.  }
\label{cgamplottypeI}
\end{figure}

This Type~II situation can be contrasted with Type~I. As stated earlier, in Type~I $\cv^H$, $\cu^H$ and  $\cd^H$ all have the same sign, which can be either plus or minus with respect to the SM convention of $\cv^{\hsm}=+1$.  We plot in Fig.~\ref{cgamplottypeI} the same quantities as in Fig.~\ref{cgamplot}. We observe that the bulk of the parameter space, namely that portion with $\cd^H>0$,  has $v^2 g_{\hh\hp\hm} \sgn(\cv^H)  \lsim -2 \mhpm^2$  leading to at least a $5\%$ suppression of $\cp$. 
In the other branch, we have $\cd^H<0$.  In the end, various effects are competing and $v^2 g_{\hh\hp\hm} \sgn(\cv^H) \gsim -2 \mhpm^2$, with values near zero and positive values as well being possible.  The result is that $\cp^H\sim 1$ is achievable and a good fit to LHC data is possible, although quite rare within our scans.

\begin{figure}[h!]
\begin{center}
\includegraphics[width=0.49\textwidth]{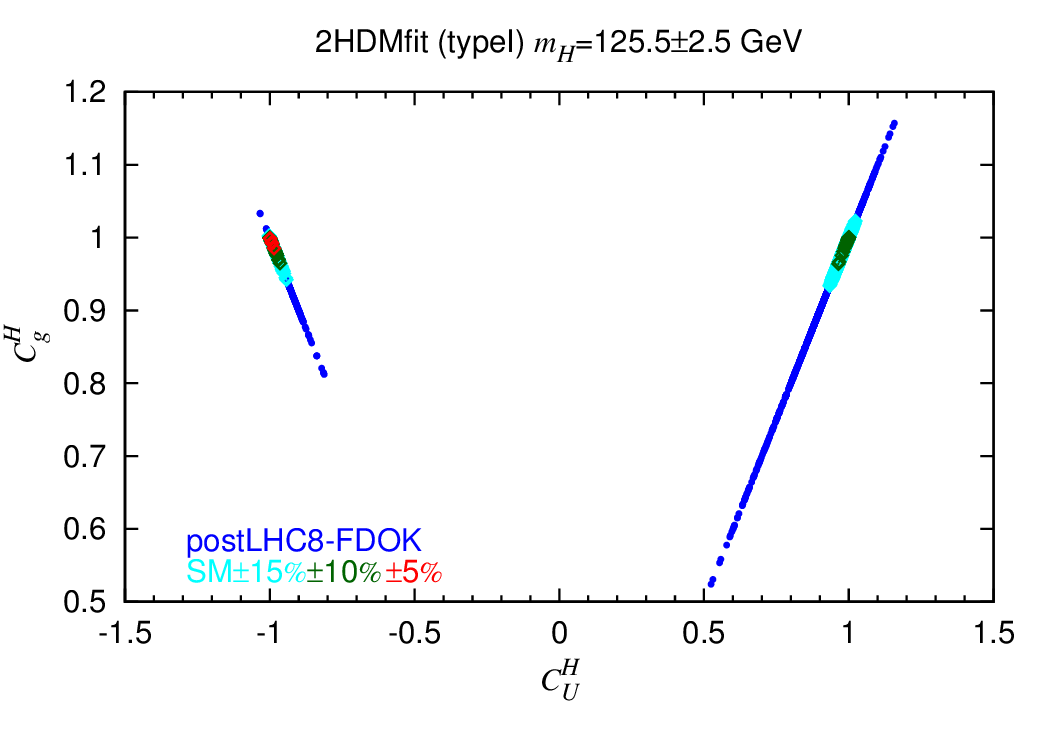}
\includegraphics[width=0.49\textwidth]{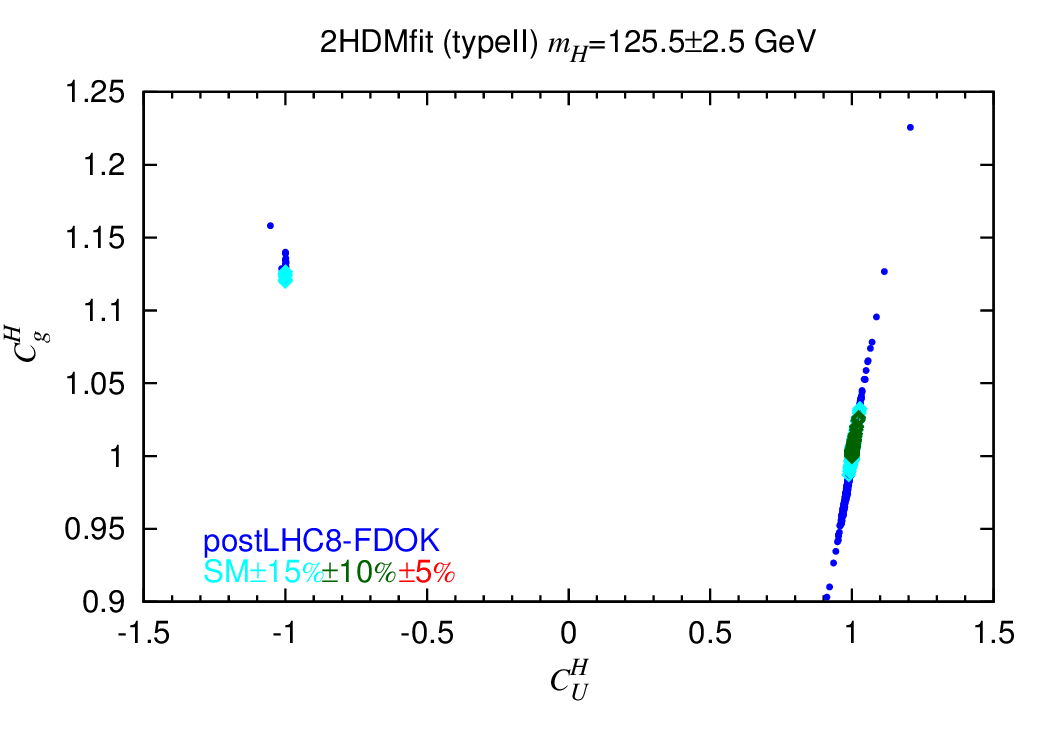}
\end{center}\vspace*{-5mm}
\caption{We plot $\cg^\hh$ vs.\ $\cu^\hh$ for Type~I (right) and Type~II (left).  We show FDOK points consistent at the postLHC8 level (blue) and after presumed LHC measurements showing SM consistency at the $\pm15\%$ (cyan) and  $\pm 10\%$ (dark green) levels.  Points consistent with the  SM (red points) at the $\pm5\%$ level are only present in the $\cu^\hh<0$ branch of the Type~I model. }
\label{cgcufig}
\end{figure}

A second means of discrimination relative to the SM arises in the context of the $gg\hh$ coupling in the case of a Type~II model.  As noted above, in Type~II $\cd^H$ always has the same sign as in the SM, but $\cu^H$ can have either the SM sign or a sign opposite the SM, \ie\ $\cu^H\sim -1$.  In this latter case, the sign of the interference terms between the top-quark and bottom-quark loop contributions to the $gg\hh$ coupling changes. This results in a significant increase of about $12\%$ in the $gg\hh$ coupling, as detailed in \cite{Ferreira:2014naa}. As also discussed there, $\cg^H$ can eventually be measured to an accuracy of about 5\% in future LHC runs as well as at a future linear collider, which would allow either confirmation or conflict with the $\cu^\hh\sim -1$ Type~II case. In Fig.~\ref{cgcufig}, we plot $\cg^\hh$ as a function of $\cu^\hh$ for both Type~I and Type~II models.  There, we see that $\cg^H$ is not generally a useful discriminator in the case of  the Type~I model nor on the $\cu^\hh\sim +1$ branch in the case of a Type~II model.

\clearpage
\bibliographystyle{JHEP}
\bibliography{2hdmfit_v2}

\providecommand{\href}[2]{#2}\begingroup\raggedright\begin{thebibliography}{10}

\bibitem{Aad:2012tfa}
{\bf ATLAS} Collaboration, G.~Aad et~al., {\it {Observation of a new particle
  in the search for the Standard Model Higgs boson with the ATLAS detector at
  the LHC}},  {\em Phys.Lett.} {\bf B716} (2012) 1--29,
  [\href{http://xxx.lanl.gov/abs/1207.7214}{{\tt arXiv:1207.7214}}].

\bibitem{Chatrchyan:2012ufa}
{\bf CMS} Collaboration, S.~Chatrchyan et~al., {\it {Observation of a new boson
  at a mass of 125 GeV with the CMS experiment at the LHC}},  {\em Phys.Lett.}
  {\bf B716} (2012) 30--61, [\href{http://xxx.lanl.gov/abs/1207.7235}{{\tt
  arXiv:1207.7235}}].

\bibitem{ATLASnew}
{\bf ATLAS} Collaboration, G.~Aad et~al., {\it {Updated coupling measurements
  of the Higgs boson with the ATLAS detector using up to 25/fb of proton-proton
  collision data}},  2014.
\newblock ATLAS-CONF-2014-009.

\bibitem{CMS:new}
{\bf CMS} Collaboration, {\it {Combination of standard model Higgs boson
  searches and measurements of the properties of the new boson with a mass near
  125 GeV}},  2013.
\newblock CMS PAS HIG-13-005.

\bibitem{Aad:2013wqa}
{\bf ATLAS} Collaboration, G.~Aad et~al., {\it {Measurements of Higgs boson
  production and couplings in diboson final states with the ATLAS detector at
  the LHC}},  {\em Phys.Lett.} {\bf B726} (2013) 88--119,
  [\href{http://xxx.lanl.gov/abs/1307.1427}{{\tt arXiv:1307.1427}}].

\bibitem{Gunion:1989we}
J.~F. Gunion, H.~E. Haber, G.~L. Kane, and S.~Dawson, {\it {THE HIGGS HUNTER'S
  GUIDE}},  {\em Front.Phys.} {\bf 80} (2000) 1--448.

\bibitem{Gunion:2002zf}
J.~F. Gunion and H.~E. Haber, {\it {The CP conserving two Higgs doublet model:
  The Approach to the decoupling limit}},  {\em Phys.Rev.} {\bf D67} (2003)
  075019, [\href{http://xxx.lanl.gov/abs/hep-ph/0207010}{{\tt
  hep-ph/0207010}}].

\bibitem{Branco:2011iw}
G.~Branco, P.~Ferreira, L.~Lavoura, M.~Rebelo, M.~Sher, et~al., {\it {Theory
  and phenomenology of two-Higgs-doublet models}},  {\em Phys.Rept.} {\bf 516}
  (2012) 1--102, [\href{http://xxx.lanl.gov/abs/1106.0034}{{\tt
  arXiv:1106.0034}}].

\bibitem{Chiang:2013ixa}
C.-W. Chiang and K.~Yagyu, {\it {Implications of Higgs boson search data on the
  two-Higgs doublet models with a softly broken $Z_2$ symmetry}},  {\em JHEP}
  {\bf 1307} (2013) 160, [\href{http://xxx.lanl.gov/abs/1303.0168}{{\tt
  arXiv:1303.0168}}].

\bibitem{Grinstein:2013npa}
B.~Grinstein and P.~Uttayarat, {\it {Carving Out Parameter Space in Type-II Two
  Higgs Doublets Model}},  {\em JHEP} {\bf 1306} (2013) 094,
  [\href{http://xxx.lanl.gov/abs/1304.0028}{{\tt arXiv:1304.0028}}].

\bibitem{Coleppa:2013dya}
B.~Coleppa, F.~Kling, and S.~Su, {\it {Constraining Type II 2HDM in Light of
  LHC Higgs Searches}},  {\em JHEP} {\bf 1401} (2014) 161,
  [\href{http://xxx.lanl.gov/abs/1305.0002}{{\tt arXiv:1305.0002}}].

\bibitem{Eberhardt:2013uba}
O.~Eberhardt, U.~Nierste, and M.~Wiebusch, {\it {Status of the
  two-Higgs-doublet model of type~II}},  {\em JHEP} {\bf 1307} (2013) 118,
  [\href{http://xxx.lanl.gov/abs/1305.1649}{{\tt arXiv:1305.1649}}].

\bibitem{Chang:2013ona}
S.~Chang, S.~K. Kang, J.-P. Lee, K.~Y. Lee, S.~C. Park, et~al., {\it {Two Higgs
  doublet models for the LHC Higgs boson data at $\sqrt{s}=$ 7 and 8 TeV}},
  {\em JHEP} {\bf 1409} (2014) 101,
  [\href{http://xxx.lanl.gov/abs/1310.3374}{{\tt arXiv:1310.3374}}].

\bibitem{Cheung:2013rva}
K.~Cheung, J.~S. Lee, and P.-Y. Tseng, {\it {Higgcision in the Two-Higgs
  Doublet Models}},  {\em JHEP} {\bf 1401} (2014) 085,
  [\href{http://xxx.lanl.gov/abs/1310.3937}{{\tt arXiv:1310.3937}}].

\bibitem{Celis:2013ixa}
A.~Celis, V.~Ilisie, and A.~Pich, {\it {Towards a general analysis of LHC data
  within two-Higgs-doublet models}},  {\em JHEP} {\bf 1312} (2013) 095,
  [\href{http://xxx.lanl.gov/abs/1310.7941}{{\tt arXiv:1310.7941}}].

\bibitem{Wang:2013sha}
L.~Wang and X.-F. Han, {\it {Status of the aligned two-Higgs-doublet model
  confronted with the Higgs data}},  {\em JHEP} {\bf 1404} (2014) 128,
  [\href{http://xxx.lanl.gov/abs/1312.4759}{{\tt arXiv:1312.4759}}].

\bibitem{Baglio:2014nea}
J.~Baglio, O.~Eberhardt, U.~Nierste, and M.~Wiebusch, {\it {Benchmarks for
  Higgs Pair Production and Heavy Higgs Searches in the Two-Higgs-Doublet Model
  of Type II}},  {\em Phys.Rev.} {\bf D90} (2014) 015008,
  [\href{http://xxx.lanl.gov/abs/1403.1264}{{\tt arXiv:1403.1264}}].

\bibitem{Inoue:2014nva}
S.~Inoue, M.~J. Ramsey-Musolf, and Y.~Zhang, {\it {CPV Phenomenology of Flavor
  Conserving Two Higgs Doublet Models}},  {\em Phys.Rev.} {\bf D89} (2014)
  115023, [\href{http://xxx.lanl.gov/abs/1403.4257}{{\tt arXiv:1403.4257}}].

\bibitem{Craig:2013hca}
N.~Craig, J.~Galloway, and S.~Thomas, {\it {Searching for Signs of the Second
  Higgs Doublet}},  \href{http://xxx.lanl.gov/abs/1305.2424}{{\tt
  arXiv:1305.2424}}.

\bibitem{Barger:2013ofa}
V.~Barger, L.~L. Everett, H.~E. Logan, and G.~Shaughnessy, {\it {Scrutinizing
  the 125 GeV Higgs boson in two Higgs doublet models at the LHC, ILC, and Muon
  Collider}},  {\em Phys.Rev.} {\bf D88} (2013) 115003,
  [\href{http://xxx.lanl.gov/abs/1308.0052}{{\tt arXiv:1308.0052}}].

\bibitem{Carena:2013ooa}
M.~Carena, I.~Low, N.~R. Shah, and C.~E. Wagner, {\it {Impersonating the
  Standard Model Higgs Boson: Alignment without Decoupling}},  {\em JHEP} {\bf
  1404} (2014) 015, [\href{http://xxx.lanl.gov/abs/1310.2248}{{\tt
  arXiv:1310.2248}}].

\bibitem{Chen:2013rba}
C.-Y. Chen, S.~Dawson, and M.~Sher, {\it {Heavy Higgs Searches and Constraints
  on Two Higgs Doublet Models}},  {\em Phys.Rev.} {\bf D88} (2013) 015018,
  [\href{http://xxx.lanl.gov/abs/1305.1624}{{\tt arXiv:1305.1624}}].

\bibitem{Kanemura:2014dea}
S.~Kanemura, H.~Yokoya, and Y.-J. Zheng, {\it {Complementarity in direct
  searches for additional Higgs bosons at the LHC and the International Linear
  Collider}},  {\em Nucl.Phys.} {\bf B886} (2014) 524--553,
  [\href{http://xxx.lanl.gov/abs/1404.5835}{{\tt arXiv:1404.5835}}].

\bibitem{Wang:2014lta}
L.~Wang and X.-F. Han, {\it {Study of the heavy CP-even Higgs with mass 125 GeV
  in two-Higgs-doublet models at the LHC and ILC}},  {\em JHEP} {\bf 1411}
  (2014) 085, [\href{http://xxx.lanl.gov/abs/1404.7437}{{\tt
  arXiv:1404.7437}}].

\bibitem{Dorsch:2013wja}
G.~Dorsch, S.~Huber, and J.~No, {\it {A strong electroweak phase transition in
  the 2HDM after LHC8}},  {\em JHEP} {\bf 1310} (2013) 029,
  [\href{http://xxx.lanl.gov/abs/1305.6610}{{\tt arXiv:1305.6610}}].

\bibitem{Shu:2013uua}
J.~Shu and Y.~Zhang, {\it {Impact of a CP Violating Higgs Sector: From LHC to
  Baryogenesis}},  {\em Phys.Rev.Lett.} {\bf 111} (2013) 091801,
  [\href{http://xxx.lanl.gov/abs/1304.0773}{{\tt arXiv:1304.0773}}].

\bibitem{Ahmadvand:2013sna}
M.~Ahmadvand, {\it {Baryogenesis within the two-Higgs-doublet model in the
  Electroweak scale}},  {\em Int.J.Mod.Phys.} {\bf A29} (2014) 1450090,
  [\href{http://xxx.lanl.gov/abs/1308.3767}{{\tt arXiv:1308.3767}}].

\bibitem{Dorsch:2014qja}
G.~Dorsch, S.~Huber, K.~Mimasu, and J.~No, {\it {Echoes of the Electroweak
  Phase Transition: Discovering a second Higgs doublet through $A_0 \rightarrow
  ZH_0$}},  {\em Phys.Rev.Lett.} {\bf 113} (2014), no.~21 211802,
  [\href{http://xxx.lanl.gov/abs/1405.5537}{{\tt arXiv:1405.5537}}].

\bibitem{Drozd:2012vf}
A.~Drozd, B.~Grzadkowski, J.~F. Gunion, and Y.~Jiang, {\it {Two-Higgs-Doublet
  Models and Enhanced Rates for a 125 GeV Higgs}},  {\em JHEP} {\bf 1305}
  (2013) 072, [\href{http://xxx.lanl.gov/abs/1211.3580}{{\tt
  arXiv:1211.3580}}].

\bibitem{Belanger:2013xza}
G.~Belanger, B.~Dumont, U.~Ellwanger, J.~Gunion, and S.~Kraml, {\it {Global fit
  to Higgs signal strengths and couplings and implications for extended Higgs
  sectors}},  {\em Phys.Rev.} {\bf D88} (2013) 075008,
  [\href{http://xxx.lanl.gov/abs/1306.2941}{{\tt arXiv:1306.2941}}].

\bibitem{Arhrib:2013oia}
A.~Arhrib, P.~Ferreira, and R.~Santos, {\it {Are There Hidden Scalars in LHC
  Higgs Results?}},  {\em JHEP} {\bf 1403} (2014) 053,
  [\href{http://xxx.lanl.gov/abs/1311.1520}{{\tt arXiv:1311.1520}}].

\bibitem{HIG13025}
{\bf CMS} Collaboration, {\it {Search for extended Higgs sectors in the $H\to
  hh$ and $A \to Zh$ channels in $\rts = 8 \tev$ pp collisions with
  multileptons and photons final states, CMS-PAS-HIG-13-025}}, .

\bibitem{Eriksson:2009ws}
D.~Eriksson, J.~Rathsman, and O.~Stal, {\it {2HDMC: Two-Higgs-Doublet Model
  Calculator Physics and Manual}},  {\em Comput.Phys.Commun.} {\bf 181} (2010)
  189--205.

\bibitem{Eriksson:2010zzb}
D.~Eriksson, J.~Rathsman, and O.~Stal, {\it {2HDMC: Two-Higgs-doublet model
  calculator}},  {\em Comput.Phys.Commun.} {\bf 181} (2010) 833--834.

\bibitem{htautau}
{\bf CMS} Collaboration, {\it {Higgs to tau tau (MSSM), CMS-PAS-HIG-13-021}}, .

\bibitem{ATLAS:2013nma}
{\bf ATLAS} Collaboration, {\it {Measurements of the properties of the
  Higgs-like boson in the four lepton decay channel with the ATLAS detector
  using 25 fb¿1 of proton-proton collision data, ATLAS-CONF-2013-013}}, .

\bibitem{CMS-PAS-HIG-13-002}
{\it {Properties of the Higgs-like boson in the decay $H\to ZZ\to 4\ell$ in
  $pp$ collisions at $\sqrt s=7$ and 8 TeV}},  Tech. Rep. CMS-PAS-HIG-13-002,
  CERN, Geneva, 2013.

\bibitem{cms2l2tau}
\url{https://twiki.cern.ch/twiki/bin/view/CMSPublic/Hig13002TWiki\#Limit\_Plot}.

\bibitem{CMS-PAS-HIG-13-014}
{\it {Search for a heavy Higgs boson in the $H \to ZZ \to 2\ell 2nu$ channel in
  $pp$ collisions at $\sqrt s= 7$ and 8 TeV}},  Tech. Rep. CMS-PAS-HIG-13-014,
  CERN, Geneva, 2013.

\bibitem{korytov:privcom}
A.~Korytov.
\newblock private communication.

\bibitem{Ferreira:2014naa}
P.~Ferreira, J.~F. Gunion, H.~E. Haber, and R.~Santos, {\it {Probing wrong-sign
  Yukawa couplings at the LHC and a future linear collider}},  {\em Phys.Rev.}
  {\bf D89} (2014) 115003, [\href{http://xxx.lanl.gov/abs/1403.4736}{{\tt
  arXiv:1403.4736}}].

\bibitem{Efrati:2014uta}
A.~Efrati and Y.~Nir, {\it {What if $\lambda_{hhh}\neq 3m_h^2/v$}},
  \href{http://xxx.lanl.gov/abs/1401.0935}{{\tt arXiv:1401.0935}}.

\bibitem{Dawson:2013bba}
S.~Dawson, A.~Gritsan, H.~Logan, J.~Qian, C.~Tully, et~al., {\it {Higgs Working
  Group Report of the Snowmass 2013 Community Planning Study}},
  \href{http://xxx.lanl.gov/abs/1310.8361}{{\tt arXiv:1310.8361}}.

\bibitem{Coleppa:2014hxa}
B.~Coleppa, F.~Kling, and S.~Su, {\it {Exotic Decays Of A Heavy Neutral Higgs
  Through HZ/AZ Channel}},  {\em JHEP} {\bf 1409} (2014) 161,
  [\href{http://xxx.lanl.gov/abs/1404.1922}{{\tt arXiv:1404.1922}}].

\end{thebibliography}\endgroup

\end{document}